\documentclass[oneside,onecolumn,11pt,a4paper]{article}

\usepackage[hmarginratio=1:1,left=21mm,top=32mm,columnsep=20pt]{geometry} % margins

\usepackage[T1]{fontenc} % Use 8-bit encoding that has 256 glyphs

\usepackage{lmodern}
\usepackage{fancyhdr} % Headers and footers
\usepackage{titling} % Customizing the title section

\usepackage{abstract} % abstract customization

\usepackage{titlesec} % customization of titles
\titleformat{\section}[block]{\large \scshape \bfseries}{\thesection.}{0.5em}{} % section titles look
\titleformat{\subsection}[block]{\large \bfseries}{\thesubsection.}{0.2em}{} % subsection titles look

\providecommand{\keywords}[1] % keywords command
{\small	\textit{Keywords:} #1}

\usepackage[nottoc]{tocbibind} % add bibliography to tableofcontents

\usepackage[small,labelfont=bf,up,textfont=up]{caption} % custom captions under/above floats in tables or figures. [hang]

\usepackage[backend=biber,citestyle=numeric,bibstyle=authoryear]{biblatex}  % ,style=numeric is the default, alphabetic for keynames
\AtEveryBibitem{[\printfield{labelnumber}]\addspace} %ref nums
\usepackage{booktabs} % horizontal rules in tables
\usepackage{diagbox}  % diagonal lines in tables
\usepackage{tabularx}

\usepackage{csvsimple} % import csv to table

\usepackage{float} % force table to stay here via H

\usepackage[final]{graphicx}

\usepackage{enumitem} % customized lists
\setlist[itemize]{noitemsep} % make itemize lists more compact

\usepackage{hyperref} % enables hyperlinks in pdf

\usepackage[USenglish]{babel} % hyphenation and typographical rules

\usepackage{blindtext} % to generate dummy text

\usepackage{amsmath} % enhancements for improving mathematical formulas
\usepackage{amssymb} % extended symbol collection
\pretitle{\begin{center}\LARGE\bfseries} %  title formatting
\posttitle{\end{center}} %  title closing formatting

\title{Defining and finding lifelike entities with a lazy filter} 

\author{
\textrm{Mario Martinez-Saito} \thanks{\footnotesize\href{mailto:mmartinezsaito@gmail.com}{mmartinezsaito@gmail.com}} \\[1ex]
\normalsize Institute of Cognitive Neuroscience, HSE University, Russian Federation \\
}

\date{\today} % Leave empty to omit a date

\renewcommand{\maketitlehookd}{%
\begin{abstract}
\noindent 
Our binary intuitive understanding of life and lifelikeness is good enough for daily life, but not for research in the natural sciences.
Here we propose an operational definition of lifeness of a particular entity as a scalar, product of its defining information (algorithmic complexity) integrated over its lifetime.
We provide a parameter-free dynamical filtering algorithm that can efficiently and on-the-fly fit configurations of entities constituted by moving particles to a tree structure that can model a wide range of hierarchical system entities, while extracting and calculating the complexity, lifespan, and lifeness of the entity and all of its constituting subtrees or subentities.
We simulated 41 interacting particle worlds and found preliminary evidence suggesting that the lifeness of entities is associated with the distance to criticality, as roughly measured by the range of the pairwise interaction forces of the elemental particles of the worlds they inhabit.
This study is a proof of concept that defining, measuring, and quantifying lifeness is (1) feasible, useful for (2) simplifying theoretical discussions, for (3) hierarchically assessing biotic properties such as number of hierarchical levels and predicted longevity and for (4) classifying both artificial and biological entities, respectively via computer simulations and a combination of evolutionary and molecular biology approaches.
\end{abstract}
\keywords{hierarchical clustering; autopoiesis; non-reciprocal forces; Bayesian inference; interacting particle systems}
}

\begin{document}

% Print the title and the abstract
\maketitle

\tableofcontents

%-----------------------------------------------/-----------------------------------------
%	ARTICLE CONTENTS
%----------------------------------------------------------------------------------------

\section{Introduction}

Living beings persist while surrounded by potentially destructive fluctuations. But not all persistent beings seem to be biotic.
The \textit{intuitive} notions of life and non-life are sufficient for most mundane affairs, but as soon as one drills down to the microscopic scale their distinction starts to blur.
This is because our binary intuitive understanding of life is good enough for survival purposes \cite{Martinez-Saito2023}.
There are many definitions of life \cite{Trifonov2011}, but no consensus on which one is the best. This disagreement may be caused by conceiving life as a binary category, whereby things are exclusively either alive or dead. 
Here we propose a scheme for computing the ``lifeness'' of any subsystem in a system of interacting particles, and give a definition of lifeness as a continuum attribute.

Despite our intuitive physical perception \cite{Martinez-Saito2023} of matter as lumps, blobs, jets, or puffs, the world can be largely explained as a jumble of interacting tiny particles, so here we attempt to define and find life-like entities via a simple dynamical filtering algorithm in some of the simplest of conceivable worlds: baths of mutually interacting particles; and relate their associated lifeness to the properties of its environment.
Our definition is based on two simple observations.
First, living entities are (entities): they persist, or exist for an extended time span (lifespan).
Second, they persist while preserving a particular or idiosyncratic structure, that mirrors its environment (good regulator theorem) \cite{Conant1970,Friston2006}.
This structure is generally a complex adaptive system: a multiscale hierarchical entity where at each level and between levels an interplay between non-equilibrium dynamics and selective processes continuously reshape all subentities \cite{Cziko1997,Levin2003,Friston2015c,Ramstead2018}.
These straightforward observations have a myriad of implications that strongly depend on and condition the environment where the living entity dwells, and determine the dimensions of lifeness: a product of time and information.

The complexity of biotic processes suggests that mesoscopic objects self-organize in ways unlike anything we know at very large or very small scales \cite{Laughlin2000}.
``The miracles of nature revealed by modern molecular biology are no less astonishing than those found by physicists in macroscopic matter. Their existence leads one to question whether as-yet-undiscovered organizing principles might be at work at the mesoscopic scale, at least in living things.'' \cite{Laughlin2000} 
The selective process that enables the existence of only persistent entities is likely to underlie the organizing principles of the mesoscopic scale.

In the next section, we lay out the features of a world that enable self-organization (which in turn is the foundation of lifelike behavior), and then propose a definition of lifeness.
In Section \ref{sec:automata}, we describe the interacting particle system (IPS) that constitutes a model environment.
In Section \ref{sec:lifefinder}, we specify Bisect-Unite Nodes Clustered Hierarchically (BUNCH), a dynamical filtering algorithm conceived to detect lifelike entities and enable the computation of lifeness.
Then, in Section \ref{sec:simul} we run IPS simulations, apply the BUNCH algorithm, and relate the resulting lifeness to the parameters of its environment.
Finally, we conclude with a discussion of the results, caveats, and prospective research avenues.

\section{Lifeness: A quantitative definition of life}

How to identify ``live entities'' or even ``entities'' in general? 
They seem to be characterized by their persistence, but how long is enough to ``be'', and be what exactly?
Most of the entities or beings that we encounter belong to the type of entities that persist the longest: plainly, in a volatile state of affairs where new entities appear and old ones disappear intermittently, entities with long lifespans have the highest chance of being observed.
Can the persistence of entities be quantified?
Our reasoning roughly goes along the following lines: (1) lifelike entities are persistent, i.e. preserve their structure across time (diachronically); in other words, they occupy space in a specific configuration and possess some form of temporal invariance; (2) this can be achieved trivially in a static world, but only by having mechanisms to counteract destructive stimuli in a dynamic world; (3) a persisting entity must encompass a model complex enough to counteract the typical complexity of environmental fluctuations; hence (4) the entity's algorithmic complexity or information multiplied by its lifespan is a rough measure of the capacity to survive in complex environments, which we take to be a reasonably good definition of lifeness; (5) the so defined lifeness will be higher for entities capable of dwelling for long times in complex dynamical environments (those poised on a near-critical state).

\subsection{Why is our world complex and near-critical? \label{sec:critworld}}

Our surrounding world ---Earth's surface--- is a dissipative system sustained by a continuous stream of available energy (exergy) mostly coming from the Sun that exhibits biotic diversity across multiple spatio-temporal scales.
Lifelike behavior already arises at least from the nanoscopic scale of molecules. For example, a light source can be used to on-off turn the self-organized clustering of photocatalytic synthetic colloids in a suspension with fuel \cite{Palacci2013} \footnote{This is an expression diffusiophoresis, i.e. the propelling forces induced by the colloid and fuel gradients, that turns the colloids into microswimmers. The resulting crystals self-assemble and break dynamically resembling biotic behavior.}.

This scenario is unlikely to be fortuitous: a non-equilibrum steady state, characterized by persistent flows of matter and energy, but also by time invariant probabilities for the relevant physical variables.
Crucially, such systems spontaneously tend to roughly discretize their characteristic spatio-temporal scales \cite{Haken1983}, which manifests as discrete scales of description: i.e. geological (hundreds of millions of years), phylogenetic (millions of years), developmental (years), behavioral (days), neurophysiological or biochemical (milliseconds), etc.
This separation of temporal scales together with an exogenous driving source \cite{Vespignani1998} and a conservation law \cite{Bonachela2009} result in a physical system autonomously poised near a phase transition, i.e. a self-organized critical (SOC) system \cite{Bak1987} (or quasi-critical for non-conservative systems, SOqC \cite{Bonachela2009}).
Being poised near a critical phase does not even require much precision for some systems, that possess a stretched critical region (Griffiths phase) by virtue of their glass-like structural disorder; this eases self-organization \cite{Moretti2013}.
Spatio-temporal invariance is manifested in multiple physical phenomena as the power laws characterizing SOC.
But systems designed or evolved to minimize fluctuations while satisfying limited resource constraints (e.g.\ free energy), also called highly optimized tolerance (HOT) systems \cite{Carlson1999}, may also exhibit power laws.
The archetypal example is living systems, which evince higher performance and robustness but only with respect to the uncertainty  accounted for in their generative models; their high performance (fitness) has the downside of resulting in extreme vulnerability to \textit{non-modeled} fluctuations.
An archetype of HOT design at multiple hierarchy levels of neurons, ganglia, and brain, is the placement of entities under the constraint of minimizing connection cost \cite{Cherniak1994,Banavar2000}.

Systems ---such as Earth's surface--- composed of many nonlinear interacting subunits slowly driven by an exogenous force that evolve into a non-equilibrium steady state (NESS) are typically characterized by self-organized criticality \cite{Bak1987}.
Critical phenomena entail spatio-temporally correlated fluctuations over several orders of magnitude, near-diverging correlation length, power-law distributed quantities, and scaling invariance with exponents that are universal across a large class of systems \cite{Goldenfeld1992,Sornette2004,Martinez-Saito2022c}.
The cerebral cortex, a subsystem of Earth's surface, seems to operate near a critical regime \cite{Beggs2003}, which affords optimal computational \cite{Beggs2008,Bertschinger2004} and more generally fitness advantages associated with dynamic adaptability \cite{Friston2000g,Ay2008,Friston2012a,Chialvo2010,Hidalgo2014,Munoz2018}.
This is one of the reasons that life is typically poised at criticality \cite{Mora2011}.

Another reason is that (as we will discuss soon) in order to survive living creatures (persistent entities) must have a good enough model of its environment \cite{Conant1970,Friston2006}, so the multilayered architecture of the brain must also be a reflection of the multiple spatio-temporal scales of its environment \cite{Chialvo2006}.
Although there also exist equilibrium systems poised at criticality (e.g. closed systems at a phase transition), no entropy is produced in them.
Whatever persistent entities inhabit them, they do not need any memory or model of their environment because the complexity of their internal model (and the amount of information they need to store) is bounded by the entropy production of their environment \cite{Bo2015}.
But this is unlike anything we consider to be alive.
Living beings typically inhabit open dissipative or NESS systems, where energy available for performing computations or storing information is costly, which imposes a trade-off between performance in terms of representational accuracy and (metabolic) power consumption \cite{Hasenstaub2010,Martinez-Saito2022c} or in other words between information storage capacity and computational efficiency \cite{Hartich2016}.
The kind of persistent entities that can inhabit equilibrium and non-equilibrium environments are qualitatively different, and their distinction is analogous to the distinction between non-autonomous and autonomous entities\footnote{Which also correspond to mere active inference and adaptive active inference \cite{Kirchhoff2018}.} (cf.\ Table \ref{tab:entity_examples}).

Another related and conspicuous attribute of our world is its (algorithmic) complexity.
Complexity is associated with ``rich forms of intermittency with the recurrent and self-limiting expression of stereotyped transient-like dynamics that give the illusion of a dynamically changing attractor'' \cite{Friston1997a}.
Complexity is also associated with the information required to describe a system at scales larger than a given characteristic length defining the range of interest of a phenomenological description \cite{Goldenfeld1992}, below which fluctuations are considered irrelevant noise \footnote{This is analogous to how blurring the phase space trajectories at an atomic scale entails a coarser description of the system where entropy increases spontaneous and monotonously, as illustrated by Gibbs' H-theorem \cite{Jaynes1965}.
Hence given a reasonably wide separation between the spatio-temporal scales of (uninteresting) fast small variables and (interesting) slow large variables, complexity can simply be taken to be the entropy or information of the macroscopic variables of interest.}.
Here the spatio-temporal scale of the threshold beyond which variables are of interest is defined by the smallest scales at which the relevant entity is not invariant anymore: e.g. for a yeast cell at a given temperature compatible with life, the scales associated with thermal fluctuations are below this threshold because the yeast is invariant under the thermal fluctuations of its constituting molecules. These fluctuations can then be simply described as a (Boltzmann) probability distribution, thereby enormously reducing the information needed to describe it\footnote{This is equivalent to computing the joint entropy of all the system variables except the contribution from noise or ``non-interesting'' variables, which are typically the fast small fluctuations. This corresponds to subtracting the entropy of the smallest components  in the $\Phi$ measure of complexity \cite{Tononi1994,Friston1995d} or discarding the highest Fourier modes of a large sequence of the system's dynamical trajectory coursing on its manifold \cite{Friston1997a}.}.
In our case, this means focusing on the information required to describe entities composed of atoms\footnote{Here ``atom'' refers to its original etymology of ``indivisible particle'', not the basic particles of chemical elements.}, while ignoring the information required to describe the constituent atoms---a form of ``organized complexity'' \cite{Weaver1948}.
This idea of discarding the information of the building blocks or elemental constituents is common to most measures of complexity \cite{Friston1995d}.
Hence complexity is an index of the disparity between any given configuration and the entropy-maximizing configuration, that increases with spatial scale. This is related to the fact that the mixing or uniformizing effects of entropic forces become stronger with increasing spatial scale.

Crucially, complexity is maximized at criticality \cite{Goldenfeld1992,Friston1995d,Lotfi2021}.
One of the telltale signs of criticality is a diverging correlation length, which in turn entails a diverging  susceptibility (the variation of an order parameter in response to an applied field)  \cite{Goldenfeld1992}.
In this respect, it has been shown in simulated ecosystems that ``for a family of individuals with similar parameters, the fittest possible agent sits exactly at criticality, and it is best able to encapsulate a wide variety of distributions'' as indicated by its higher susceptibility to environmental configurations \cite{Hidalgo2014}.
More generally, many more distinguishable outputs can be reproduced by inferential procedures or models poised at the peak of susceptibility, i.e.\ at criticality \cite{Mastromatteo2011}, which affords a source of diversity that can be leveraged via selective pressure to glean good enough models \cite{Friston2000g}, which  e.g. in neuroscience can be evinced e.g. as ``neural noise'' \cite{Deco2009a}, perceptual instability \cite{Friston2012a} or ``diachronic sampling'' \cite{Martinez-Saito2023}.  
In summary, self-organized systems poised near criticality (pericriticality) exhibit a rich stock of dynamic patterns, which promotes the survival of complex entities, such as those we call living beings, including ourselves (cf.\ anthropic principle \cite{Carter1983}).

\subsection{How (biotic) complexity may accrue synergetically and hierarchically in a world governed by local interactions \label{sec:synmepfep}}

In our world, persistent and complex enough entities typically require a continuous influx of usable energy to exist, by actively preempting destructive exogenous events in a NESS environment.
This is at least true from the microscopic scale of cellular life \footnote{And indirectly also for viruses, who depend on cell hosts for survival.} and upwards. 
Conversely, their building blocks or molecules, which are composed of atoms, are (passively) stable in cold enough baths ---they are held together by the energy barrier of chemical bonds--- and can thus exist in thermal equilibrium.

In an isolated system, the thermodynamic arrow of time is directed such that a step down in thermodynamic free energy is (on average) an irreversible step forward in time.
This (Second Law of Thermodynamics) characterizes the evolution of isolated systems via a trade-off where in each subsystem energy tends to decrease and entropy tends to increase, while the total energy remains constant.

Increasing entropy seems to be odds with the persistence of complex entities, which preserve their integrity. 
To do so they must counteract the Second Law's tendency to jumble everything thus precluding particular structures from remaining unchanged and from becoming mixed with their surroundings, when they cease to exist.

If compounds or entities are seen as metastable configurations of elemental particles that decay spontaneously and require energy to couple, then they require energy to persist, akin to biotic beings \cite{Sharma2007}.
Crucially, the amount of free energy they require to persist (live) is roughly proportional to their algorithmic complexity, which is the sum of the complexities of their constituting particles or subentities, because any effecting action to offset destructive fluctuations consumes free energy (commensurate with the fluctuation size).
In other words, the effect of entropic forces increases with the scale.
This is why while elementary particles are eternal, large entities must be well designed, be lucky, and ``strive'' \cite{Friston2013} to survive in order the persist.
The result is that entropy increase in non-equilibrium systems favors the selection and persistence of more complex entities (``dissipative structures'') \cite{Nicolis1977,Prigogine1978} because their very existence precludes the existence of simpler entities via selective evolution processes \cite{Cziko1997}. In synergetics, this is construed as the takeover of available energy gradients by the self-organized structures that consume them most efficiently \cite{Haken1983,Tschacher2007}, which are also the most complex persisting structures.
In principle complexity may thus accrue without bound by stacking step by step increasingly higher levels on organization.
Similarly, life is typically realized as interwoven hierarchies of entities that actively oppose environmental fluctuations, with faster entropy production associated with more complex persistent entities that require a larger influx of available energy.
Thus, ecosystems develop such that they systematically increase their ability to degrade the incoming (on Earth, solar) energy \cite{Schneider1994} and thus augment the rate of entropy production over what it would be in the absence of biota 
\cite{Ulanowicz1987}.
If the whole world were a NESS, this would entail the existence of a highest level of complexity, the ``omnientity'' (which can be modeled as an ``omnicluster'', cf. Section \ref{sec:omniclus}).

This view strongly resonate with the Maximum Entropy Production principle (MEP), a predictive, if somewhat ambiguous, principle for non-equilibrium systems, akin to the Second Law for equilibrium systems, which states that the steady state of open thermodynamic systems with sufficient degrees of freedom is kept in a state at which the production of entropy is maximized given the constraints of the system \cite{Dewar2003,Virgo2011}.
However its applicability is unsettled because typically it is unclear how to determine the correct constraints \cite{Virgo2010} since non-equilibrium systems are not uniformly governed by detailed balance constraints.

Therefore, if we envision life as sustained complexity drawn out in time, then both synergetics and MEP predict spontaneous evolution toward lifelike entities.
It has even been proposed that the rate of entropy production is  \textit{the} fitness criterion of natural selection \cite{Sharma2007}.

\subsection{Self-organizing persistent blobs in a biotic broth of short-range interactions \label{sec:socblobs}}

So far we have examined what sort of entities are likely to exist in a NESS system driven by a continuous influx of free energy.
Now we turn around this approach by asking what are the attributes that entities must possess to persist beyond a given timespan, to which the free energy principle (FEP) \cite{Friston2013} furnishes an answer while leading to a similar conclusion as synergetics and MEP.

The pattern of interactions between elementary entities or particles define the character of the world or IPS: interaction rules governing motion create space and time correlations, precluding mutual independences\footnote{Similarly to how on a graph it is the dependencies between nodes (typically instantiated as parameters) that define the structure of the graph by constraining it\cite{Bishop2007}.}.
Short-range (and sparse \cite{Friston2000g}) interactions is an essential feature of a world that accommodates persistent complex dynamical entities because it induces rich or complex behavior.
Infinite-range interactions induce strong coupling and coherence, whereas too short or weak interactions entails decoupling and hence independent fluctuations or noise; these regimes exhibit shorter correlation lengths, and hence less complex behavior, than the intermediate regime of short-range interactions \cite{Sornette2000}.

In a world governed by local and short-range interactions, particles or entities composed of them will necessarily be pairwise conditionally independent of each other, given the other entities standing between any pair (the Markov blanket) and mediating their interactions ---since ``action at a distance'' is not allowed, at least for far away distances.
In the case of cellular life, the stable structure mediating between the persistent entity or cell and the environment can be readily identified as its membrane \cite{Varela1974} (although in general the boundary constituting the Markov blanket may not be identified with a unique fixed structure).

How do we know that an entity is an entity? 
Clearly, entities or systems that do not minimize their entropy (when it exceeds the entropy of their own model of themselves) are unlikely to exist because they become mixed with their environment (die) \cite{Friston2012}. 
That is, persistent entities occupy with a high probability a small and bounded set of states (``viability set'') within the total set of possible states (phase space) that it might occupy \cite{Ramstead2018}. 
This notion can be formalized as an attractor, a minimal invariant set that attracts an initial open set of initial conditions \cite{Strogatz2000}, which is itself non-deterministic (random attractor) \cite{Friston2012,Friston2012d}.
During the process of natural selection, persistent entities accumulate information about the environment \cite{Frank2012b,Hartich2016}: to resist entropic erosion and maintain itself in a bounded set of states (homeostasis), they must be entangled with their surroundings in some sort of steady and mutual dynamical relationship. 
This is done by instantiating a causal, statistical model of its eco-niche relation by not just encoding but embodying a ``model of the world that has been sculpted by reciprocal interactions between self-organization and selection over time'' \cite{Ramstead2018}.
This is manifested as a sort of spatio-temporal invariance, which is another phrase for persistence: living beings are spatio-temporally invariant if they can move around unchanged in space-time trajectories, i.e.\ stay the same in different locations and times \cite{Martinez-Saito2023}.

This suggests a brute force scheme to find persistent beings: the peaks of a multidimensional (space-time) autocorrelation function of subsets of different sizes (``autocorrelation kernels'') of the dynamically evolving spatial configuration of atoms would indicate the existence of persistent entities.
Applying this scheme over all the subsets of the world and with spatio-temporal autocorrelation kernels would indicate whether there exists life at different spatial and temporal scales\footnote{But clearly from our subjective perspective, we only emphasize the temporal component of the spatio-temporal autocorrelation: we say survival but not traveled distance because from our deictic perspective, we are always here.} and whether the correlation length diverges.
Crucially, in searching for lifelike entities we are only interested in ``macroscopic'' autocorrelation kernels, beyond the atomic scale.
At large enough scales in complex enough world, simple interatomic interactions are not enough to warrant survival, and a model of the world becomes mandatory.
For example, in the real world, life stands on a hierarchy of at least five levels: quarks, baryons, atoms, molecules, cells, pluricellular organisms, societies, etc.
With each step up in the hierarchy, both the spatial and temporal scale is correspondingly increased \textit{discretely} \cite{Kiebel2008,Martinez-Saito2022c}.
In the context of FEP, this is construed as Markov blankets of Markov blankets: particles, cells \cite{Friston2015c} or subsystems at one scale (each comprising a Markov blanket that enshrouds internal states) constitute an ensemble of states with a sparse dependency structure \cite{Friston2000g}, which induce another layer of Markov blankets at the supraordinate scale \cite{Kirchhoff2018,Ramstead2018}. For consistency, this scheme ought to be extended to a multiscale framework \cite{Froese2014} e.g.\ via variational neuroethology from the molecular via ontogenetic and phylogenetic to the evolutionary spatio-temporal scales \cite{Ramstead2018}.
At evolutionary and phylogenetic scales we have species, races and strains that slowly transform via changes in the DNA codebook, which also determines, jointly with environmental factors, the ontogenetic and individual features.

Roughly, the blanket or membrane of an ``invariant blob'' is its spatial boundary and birth-death is the temporal boundary.
For example, atoms (as point-particles) are extended in time (eternal), but they lack spatial attributes (extension) other than position (barring a multiplicity of atom types).
But, according to FEP, from this ``not mixing'' property, we can already infer behavior: any ergodic random dynamical system wearing a Markov blanket will appear to self-organize to actively maintain its structural and dynamical integrity, where this semblance is almost inevitable \cite{Friston2013}.
In other, loose, words: persistent entities in a world governed by short-range interactions have a boundary and are lifelike.
For each persistent entity, this boundary induces a bipartition of the world into internal and external states and an associated circular causality: it will try to preserve its existence by stabilizing both their environmental and bodily states\footnote{There is a loose analogy with electrostatics. Closed electrical conductors have the property of keeping their interior at a constant electric potential: they shield their inside from outside by redistributing their moving surface charges so as to cancel out any external electric fields. Here the cavity is analogous to the entity body, the surface with moving charges to the generative internal model and the Markov blanket, and the external electric field to environmental stimuli.}.

It can be shown \cite{Friston2012,Friston2013} that if an entity survived or persistent for a long enough time (ergodic) then the internal states dynamically reconfigure themselves with the result that they appear to be \textit{reasonably good enough} at inferring the state of the external world and selecting actions (active inference) conducive to their survival (autopoietic).
Here reasonably good enough simply means that although internal states cannot in general perfectly represent (Bayesian beliefs about) external states, this is not a fatal problem: an approximate Bayesian inference scheme, typically gradient descent on a variational free energy function of internal and blanket states \cite{Friston2003,Friston2007}, that yields a reasonably high survival chance is good enough.

Finally, a persistent entity can be recursively composed of smaller persistent entities (each wearing its own Markov blanket) so that the behavior of nested persistent entities is conditioned by both bottom-up and top-down effects, where mixtures of Markov blanket states are the slow (unstable) modes that order parameters in synergetics \cite{Ramstead2018}.
If we assume the existence of an omnientity encompassing all other blankets (cf. previous section), it would also possess an ``omniblanket''\footnote{But note that if we assume the world to be an NESS then it would be an open system, which implies matter and energy exchange with its ``outside''. On the flip side, if the world is isolated then its blanket might degenerate into a static ``wall'' (e.g. a world or manifold with boundaries) or simply not exist (e.g. an infinite world or a closed manifold).}. 

In summary, synergetics and MEP predict spontaneous evolution toward complex self-organized (lifelike) entities (SOC) displaying a (hierarchical) separation of spatio-temporal scales, whereas FEP provides a mechanistic and parsimonious account of how such entities, only by assuming their protracted existence, accomplish lifelike behavior ---i.e. by performing approximate perceptual and active inference via a variational free energy function of its own states. 
Synergetics and MEP are closely related to FEP: the behavior and evolution of lifelike entities predicted by FEP typically leads to hoarding more available energy and to becoming (through selective processes) increasingly more complex.
Another way in which MEP and synergetics and FEP are closely related is that in a world governed by short-range interactions the processes that drive entropy production follow a local gradient on spatial configurations of particles, that locally suppresses free energy.
Analogously, the mechanism by which persistent entities reduce variational free energy is likely to be (a local) gradient descent on internal states (that parameterize external states), for reasons of computational tractability, efficiency, and expediency \cite{Friston2003,Friston2007,Friston2017}.

Overall, the existence of animate beings is actually both a struggle both for free energy or negative entropy \cite{Boltzmann1895,Friston2006} ---because they must stay confined within the limits of their life-compatible domains--- \textit{and} for positive entropy \cite{Sharma2007} ---because to the degree that more energy becomes available, more complex persistent entities occupy larger domains of the phase space, driven by the larger fluctuations associated with carrying more energy, which boost exploration.
In other words, it is competition for hoarding the limited available energy \textit{and} staying self-sustainable, i.e. alive. 
Living systems do not just destroy energy gradients by inhabiting variational free energy minima, they also create and maintain them by carving out the probability gradients that surround the minima \cite{Ramstead2018} or ``a living system must both distinguish itself from its environment and, at the same time, maintain its energetic coupling to its environment to remain alive'' \cite{Varela1997}.
In brief and coarsely, living beings are what persists in a complex system fed by an energy inflow, which tends to be the most complex and dissipative entity that the system can support.

\subsection{Is life a categorical or a continuous attribute?}

Portraying life as a binary category unavoidably leads to an impasse at a small enough spatio-temporal scales. For example, are viruses and/or giant viruses alive? \cite{Koonin2016} It may not be sensible to set a hard threshold for categorizing systems into biotic and non-biotic.

But then why do we think of life as a binary attribute?
Perhaps because typically we only need to deal with big and complex enough macroscopic forms of life, and such entities evince a sharp distinction or bifurcation that separates ``live'' and ``dead'' states, where the latter are characterized by an inability to preserve the structure of the live states followed by an irreversible break down and sundering of its constituents \footnote{Note that since a dead or inert state simply keeps breaking down and decaying, its lifeness (cf. Section \ref{sec:lifeness}) is negligible despite its initially high complexity.}.
This distinction becomes increasingly fuzzier at smaller scales.

We suggest that life is analogous to a phase of matter: just like the e.g. distinction between water and steam phases, it only makes sense beyond a certain spatial scale: in general you cannot tell apart ``steam molecules'' from water molecules.
In the same manner that water and steam can become indistinguishable at the liquid-vapor phase transition \footnote{\textit{Even} below the critical point.} and at nanoscopic scales, the distinction between live and dead phases is also ambiguous at their phase transitions (death and even birth) and at nanoscopic scales.
Hence life can be better understood as a continuous attribute.

\subsection{Lifeness as persistent complexity: long-lived and complex things look ``alive'' \label{sec:lifeness}}

The concept of lifeness mainly bespeaks two qualities: (1) persisting, carrying on, remaining, or surviving, i.e. continuing to exist for a finite timespan; and (2) steadiness, endurance, or constancy, in the sense of preserving a specific structure or entity.
These notions are roughly the answers to the questions \textit{how long?} and \textit{what?}, which can be measured with time $t$ and information $I$ (required to describe the what) units, such as seconds and bits.
Given an entity $E$ of complexity $I$ that persists for $t$ time, we simply propose that the product of these two variables is a reasonably good measure of lifeness: $L = t \cdot I$.
Thus if we ignore location (i.e. entities are assumed equivalent up to spatial translations) lifeness is proportional to both lifespan and complexity.
Note that the lifeness of atoms (elementary particles) is zero because being elementary they lack (moving) parts or children.
In general a system may enclose multiple nested entities; typically, smaller entities are more likely to ``live'' or persist for longer timespans.
Thus the lifeness of a given entity is a measure of how much cumulative information throughout its lifespan is contained by its model of its components.

In summary, the lifeness $L$ is an explicit and quantitative measure that becomes amenable to computation as soon as we specify (1) how and when entity children constitute parents and (2) how to compute the algorithmic complexity of a particular (sub)entity (Section \ref{sec:modCL}).

\subsection{The entities with the best models tend to be longer-lived and reflect more of the world's complexity}

Although closely related, on the one hand synergetics and MEP relate to an entity's model complexity $I_P$, whereas on the other hand the good regulator theorem and FEP relate to the likelihood $I_A$ of existing in the current state given an entity's model\footnote{$I_P$ and $I_A$ are more precisely defined in Section \ref{sec:modCL}}.
Reworded, the former translates as matching the complexity of the entity ('s model of its environment) $I_P$ to the complexity of the environment, whereas the latter translates as minimizing the negative log-evidence $I_A$ (or maximizing the evidence) of the entity's model of its environment, given the current state of the entity or affairs.

An persistent entity's model of its world undergoes an evolutive evidence maximization process, which is nothing else than a sort of automatic model selection or maximization \cite{Friston2007,Friston2011b}, which tends to yield a parsimonious yet accurate account of its surrounding world.
However, a conceptually crucial point (although a digression in the context of this article) is that the complexity of entities dedicated to model their environment $I_P$ typically falls behind by orders of magnitude to the complexity of the actual environment.
So given that entities must ``mirror'' well enough their surroundings to survive, and in a near-critical world with near-infinite correlation lengths there will always be room for finite model improvements, why is this so?
Because the entities only require a just good enough model of the world to survive, and modeling incurs cumulative metabolic energy costs that may not compensate for the increased accuracy or performance $I_A$ \cite{Hasenstaub2010,Hartich2016,
Martinez-Saito2022c}.
Hence, in practice the precepts of maximizing model complexity $I_P$ (derived from MEP and synergetics\footnote{Synergetics prescribes that more complex entities are biologically fitter because they more efficiently consume the available energy (which of course requires an internal model of the world), thus starving other entities.}) and of matching model complexity to its environment's complexity (derived from the good regulator theorem and FEP) are virtually equivalent.

In brief, persistent entities try to both minimize the surprisal (or negative log-evidence) associated with their own and their world's existence $-I_A$, which in turn requires a complex enough model $I_P$.
This conclusion readily follows from FEP, which posits that the behavior of living beings can be cast as minimizing a certain quantity or Lagrangian, which is nothing else than the negative log-evidence of their model integrated over time, which can be construed as a product of complexity and time.
In particular, by seeing the surprisal of the sensory and (internal model's) external states of a system as a Lagrangian, it can be shown that a variational principle applies that minimizing the time integral of the Lagrangian or surprisal is equivalent to minimizing the surprisal at each time point\footnote{Because using generalized coordinates the Lagrangian's direct dependence on coordinate motion disappears \cite{Friston2008}} \cite{Friston2008,Friston2010}.
Then, under some reasonable additional assumptions (being an ergodic random dynamical system) the system's behavior can be explained as if it tried to minimize the entropy of its (ergodic) probability density over sensory and external states (which is accomplished by reconfiguring its internal states, which parameterize its internal model of the external states \cite{Friston2012}).
This is consistent with the tautological tendency of persistent entities to preserve their structure, by minimizing the entropy of their densities (Section \ref{sec:synmepfep}) at all times \cite{Friston2010}.

What can be said about the joint longevity and complexity of persistent entities?
The simplest ``entities'' are atoms. They are eternal, but have zero complexity precisely because they are monolithic, i.e. they lack components.
In contrast, very complex entities are constituted by many children, which in turn have their own children, etc. They have high complexity but are likewise highly susceptible to destruction or death because their existence hinges on the integrity of their many parts.
But what constitutes an entity? One could arbitrarily pick a random domain of the world and declare it to be an entity.
This is precisely why our discussion stresses ``persistent entities'': at any timepoint, the space can be partitioned into arbitrary domains of entities or atoms, most of which, when played out, will shortly break up into their components.
Roughly, short-lived entities, whether simple or complex, are typically just decaying fluctuations, unlike life, whereas long-lived entities can be both simple or complex, but only the latter tend to be ``lifelike'' (Table \ref{tab:entity_examples}).

\begin{table}
\caption{(Over)simplified description of entities by how long they persist (short and long lifespan in rows) and the amount of information needed to describe them (low and high complexity in columns). In parentheses, illustrative examples. \label{tab:entity_examples}}
\small
\centering
\begin{tabular}{c|c|c}
%\diagbox{ lifespan }{complexity}
  & low & high \\
\hline 
short & \vtop{\hbox{\strut small random fluctuations}\hbox{\strut (any ``blips'', thermal motion)}} & \vtop{\hbox{\strut large random fluctuations}\hbox{\strut (wars, Boltzmann brains, Big Bang?)}} \\
\hline
long & \vtop{\hbox{\strut non-autonomous persistent entities}\hbox{\strut (stable molecules, stones, isolated things)}} & \vtop{\hbox{\strut autonomous persistent entities}\hbox{\strut (green algae, ecosystems, life on Earth)}} \\
\end{tabular}
\end{table}

Given the link between criticality and complexity (Section \ref{sec:critworld}), an important question here is how the distribution and the maximum lifeness (maximum longevity and complexity) of the entities that a given world can harbor are related to the configuration parameters of the (IPS) world, and in particular to its proximity from criticality \ref{sec:L_crit}.

\section{Lifelike automata and non-reciprocal forces \label{sec:automata}}

The precursor of interacting particle systems is cellular automata. These are discrete models of computation consisting of nearly isolated cells embedded in a solid-state-like lattice that in general can acquire multiple discrete values.
For example, Conway's Game of Life is a popular two-dimensional cellular automaton that despite its simplicity displays strikingly complex and lifelike behavior \cite{Gardner1970}; and Lenia is a continuous cellular automata that generalizes Conway's Game of Life to continuous states, space and time via evolving scalar fields governed by partial differential equations \cite{Chan2019}.

The popularity of these particular (or rather cellular) systems is not casual: Conway's game rules are such that the system is pulled towards near-critical or peri-critical states ---which seems to be a feature of life (Section \ref{sec:socblobs}). Likewise, the evolution equation of Lenia uses a kernel with a concentric shape that boosts mid-range but suppresses far away and self-excitatory perturbations, thereby achieving a balance between runaway activity and extinction, similarly to solitons.

But in the real world there is no simple ``evolution kernel'' (a rule that specifies the next state given the current state) that enforces the persistence of an entity forward in time:
energy is stored in small degrees of freedom such as molecular vibrations, but at macroscopic scales, energy is leaked into these small scales, and thereby free energy is destroyed.
Here persistent entities must constantly secure sources of free energy, which in general requires a model of the world  \cite{Friston2006} (including oneself).
Lifelike persistent entities adaptively act on the environment via mechanisms such as genetic flows and mutations and epigenetic processes, but simple cellular automata typically only act via a fixed evolution kernel that cannot adapt to changing contexts.
A more fruitful approach is not to hard-code the specific evolution rules of persistent entities and instead simply set interaction forces in a system of many particles and observe whether and what sort of entities emerge and persist.

\subsection{Interacting particle systems}

Most macroscopic objects can be reasonably well described in the classical physics approach of particles or entities that interact through the fields they induce themselves.
This is the sense in which we employ the term interacting particle system or IPS, and not in the sense of continuous-time analogues of cellular automata \cite{Liggett1985}, which typically have static cells.

The behavior of matter at a specific spatial scale can be described by pairwise forces acting between entities composed of the subentities populating the subordinate spatial scale, where the character of the forces is determined by the composition of the entities.
For example, atoms are entities composed of the subentities electrons and nucleons (protons and neutrons), which interact through electromagnetic and residual strong forces. In turn, nucleons are composed of subsubentities (quarks) which interact through strong forces. Moreover, atoms may assemble to form superentities or molecules (clusters of atoms), which interact through intermolecular (residual electromagnetic) forces.
Interaction forces are conservative at the elementary level, but otherwise may not be so.
Some popular simple pairwise (power-law) interaction models are the Lennard-Jones potential, which specifies as a function of the distance $r$ soft short-range repulsive ($\sim r^{-12}$ potential) and mid-range attractive ($\sim -r^{-6}$) forces for realistic intermolecular interactions, and the Hertz potential ($\sim r^{-5/2}$), which describes the elastic interaction of weakly deformable bodies such as soft macromolecules \cite{Pamies2009}.
Even for the simplest entity pairwise interactions, solving the behavior of (even a few) multiple interacting entities is in general an intractable problem.
 
Many sorts of particle systems have been devised and computer simulated to investigate complex behavior, such as molecular dynamics via force fields (interatomic potentials) in chemical physics and biophysics \cite{Rizzuti2022}, schooling \cite{Aoki1982} and flocking \cite{Reynolds1987} in ethology, digital art \cite{Ventrella2017,HackerPoet2017,Abdulrahman2022}, and lifelike behavior \cite{Schmickl2016,Mordvintsev2022}.

\subsection{Interaction force range law \label{sec:ixr}}

The interaction range (more precisely, the dependency of the interaction forces on distance) is perhaps the most important of an IPS's features.
As mentioned in Section \ref{sec:socblobs}, very long-range interactions tend to induce global coupling (subcriticality), whereas very short-range interactions entail independent behavior (supercriticality).
In between, coupling among subsystems mediated by short-range
forces is such that distant subsystems cannot influence each other \cite{Friston2013}.

This is crucial for the sustainability of persistent entities because it affords them a buffer that shields their internal states from \textit{direct} effects of external states: if all nodes of the system were connected to a complex environment, the nodes would be enslaved, at the mercy of the environment. 
In principle this enables an entity to implement of a mechanism (an active model based on perceptual inference) that counteracts life-threatening external stimuli, to preserve itself\footnote{Of course, in subcritical or ``dead'' worlds (cf. Section \ref{sec:critworld}), where all fluctuations decay exponentially to their original state, no models or membranes are needed to ``persist''. On the contrary, in a ``supercritical mess'' where fluctuations are so strong that the whole world is subject to mixing, any subsystem or entity will become jumbled with the environment, i.e. it will not be invariant or persistent, but just a part of an global ergodic system; in other words, it will not exist long enough to ``be''.}.
This model needs to contain enough information (and effectors to bring it to bear) to enhance the survival chances of its host; this is in general feasible because (1) its bulk has capacity to hold more information than its membrane, and (2) a perfect model of the world is not needed, but just a good enough model.
In brief, a NESS with short-range interactions can accommodate persistent entities, which mirror the structure of their habitat.

Distance-dependent power-law interactions are ubiquitous in the physics of our (apparently three-dimensional) world.
The fundamental forces exhibit inverse-square law decay (gravity and electromagnetism: $\sim r^{-2}$ with infinite range), exponential decay (weak interaction: $\sim e^{-r}$ within $10^{-18}$m range), and no decay (strong interaction: $\sim 1$ within $10^{-15}$m range) with distance $r$.
The forces derived from them via aggregates of different constituent particles can display almost any combination thereof, depending on the geometry and types of and amount of particles involved (e.g. $\sim r^{-7}$ for van der Waals forces between atoms or molecules, $\sim e^{-r}r^{-1}$ Yukawa potential for some nuclear forces).

The exact exponent of the power-law of interaction forces is a crucial parameter in determining the dynamics of IPS.
For instance, in a ring network of coupled Lotka-Volterra-like  oscillators modeling ecological population dynamics, the occurrence  and transitions between global synchrony and diverse chimera patterns (a state where localized domains exhibit incongruent activity) is determined by the power-law exponent and the coupling strength \cite{Banerjee2016}.
Another insightful example, despite not being directly related to IPS, is that intermittent foraging trajectories with steps distributed as a Lévy stable distribution (Lévy flights) have been shown to optimize search efficiency under certain conditions \cite{Viswanathan1999,Guinard2021} and observed empirically in albatrosses \cite{Humphries2012} specifically for the inverse-square distribution, which decays as a power-law with exponent 2.

In our two-dimensional ($d=2$) IPS, we focus on interaction force power laws with exponents in the interval $n \in [-2, 1]$, with stress on the subinterval $[-1, 0]$, with infinite range. 
Note that $p = 1-d = -1$ is the exponent of the gravitational or Coulomb force for $d=2$ and that for $n \geq 0$ no distance can prevent particles from interacting with each other.

\subsection{Non-reciprocal interactions \label{sec:nonreciprocal}}

As far as we know, reciprocity always holds for elemental interactions. 
However, starting at the mesoscale, reciprocity may \textit{appear to} break down owing to emergent effects spontaneously arising from the non-equilibrium dynamics of lower scales \cite{Loos2020}.
Newton's Third Law, the \textit{actio et reactio} symmetry for particles, can be ``broken'' when interactions are mediated by a non-equilibrium environment \cite{Hayashi2006,Ivlev2015,Loos2020}.
Newton's Third Law is equivalent to conservation of momentum \cite{Feynman1963}, so in general momentum is not conserved in non-reciprocal systems.
In general, non-reciprocal forces cannot be derived from a classical many-body Hamiltonian, which implies conservation of linear momentum and energy \cite{Ivlev2015,Loos2020}.
If forces are non-reciprocal, then in general (without further constraints) power transfer in interactions is also non-reciprocal, so energy can leave and enter the system in each interaction, similarly to active matter, where the constituent particles at a specific scale can uptake (``create'') and dissipate (``destroy'') energy \cite{Marchetti2013,Ramaswamy2017,Fodor2018}.
Non-reciprocity induces a new contribution to entropy production, in addition to the usual one entailed by thermalization \cite{Dabelow2019} and induces an oscillatory Hamiltonian or total energy that manifests as cyclic behavior \cite{Sompolinsky1986}.
Thus non-reciprocity in networks of asymmetrically coupled elements typically leads to NESS phenomena such as currents \cite{Hong2011,VanZuiden2016}.   
The standard Boltzmann description of classical equilibrium statistical mechanics, which rests on conservation of energy, also breaks down \cite{Ivlev2015} and entropy as a state variable may not be sensibly definable \cite{Gallavotti2004}.

The archetypal examples of reciprocal systems are (attractive) gravitational and (attractive and/or repulsive) electrostatic systems  such as lattices of atoms.
An example of a repelling non-reciprocal interaction system is a dusty plasma \cite{Chaudhuri2011} lattice of microscopic particles, which are typically two closely spaced interacting layers of particles with interactions mediated by plasma wakes \cite{Ivlev2015}.
Conversely, a particle pair moving through a colloidal dispersion exhibits attracting non-reciprocal interactions \cite{Khair2007}.
Non-reciprocal systems with a mixture of attractive and repelling interactions are also possible: the flow solvent around two big fixed particles can induce strong attractive and repulsive domains \cite{Dzubiella2003}; colloids coated with a catalyst can behave like microswimmers \cite{Soto2014}, where the magnitude and presence of the non-reciprocity can be tuned by varying the colloids' surface chemistry and mobility \cite{Ivlev2015}; it is also possible to realize systems with an arbitrary interaction matrix between colloidal particles via laser beam control \cite{Khadka2018}.
There are also models of cognitive agent dynamics that employ non-reciprocal interactions dependent on visibility \cite{Barberis2016,Trilochan2019}, such as unobstructed visible distances to other objects \cite{Moussaid2011}, estimated time to collisions \cite{Karamouzas2014}, and visibility-induced changes of motility \cite{Lavergne2019} that simulate pedestrians and bird flocks \cite{Nagy2010}.

In our IPS, each ``atom'' is a particle of mass $m$ whose dynamics is classically described by a Newtonian equation of motion, the second order differential equation
\begin{equation}
m \frac{d^2\mathbf{x}_i}{dt^2} = \sum_{j=1}^{n_A} \mathbf{f}_{ij} \label{eq:atomdyn1} 
\end{equation}
where $i = 1 \ldots n_A$ is the atom index, $\mathbf{x}_i$ a $d$-dimensional vector indicating its position, $t$ time, and $\mathbf{f}_{ij}$ the force vector exerted by atom $j$ on $i$.
Note that an atom does not interact with itself: $\mathbf{f}_{ii} = 0$.
Atoms pairwise interact through conservative central forces
\begin{equation}
\mathbf{f}_{ij} = - \chi_{ij} \frac{\partial V}{\partial \mathbf{x}_{ij}}, \label{eq:fij}
\end{equation}
where $r_{ij} \equiv |\mathbf{x}_{ij}| \equiv |\mathbf{x}_i - \mathbf{x}_j|$ and $\chi_{ij}$ is the interaction matrix of constant real interaction coefficients ($\chi_{ii}$ = 0) whose sign determines whether forces are attracting (negative) or repelling (positive).
Hence the total force from all other particles on atom $i$ (the r.h.s.\ of Eq.\ \ref{eq:atomdyn1}) is $\mathbf{f}_i = -\sum_{j \neq i}^{n_A} \mathbf{f}_{ij} = - \sum_{j \neq i}^{n_A} \chi_{ij} \frac{\partial V}{\partial \mathbf{x}_{ij}}$.
The forces are derived from a power law function of distance (cf. Section \ref{sec:ixr}) potential
\begin{equation}
V(r) = \frac{1}{1+n}r^{1+n} \quad \text{if} \quad n \neq -1 \label{eq:Vpowlaw}
\end{equation}
so that
\begin{equation}
|\mathbf{f}_{ij}(r)| = \chi_{ij} r^n.  \label{eq:fpowlaw}
\end{equation}
We model atoms as solid (yet overlappable) balls of charge (Appendix \ref{app:solidball}) to avoid the unphysical jagged dynamics typical of classical singularities associated with vanishing volume or infinite density \cite{Feynman1964}.
For two dimensions, as we will see, the exponent range of interest will be around $p \in [-1, 0]$.
Note that $p = 1-d$ for incompressible vector fields such as gravity or electrostatics, so the potential power law would become $-r^{2-d}$.

For the $i = 1 \ldots n_A$ equations of motion to constitute a conservative vector field the factor $\chi_{ij}$ must be symmetric \cite{Ivlev2015,Loos2020}: central forces are conservative if and only if spherically symmetric.
Then we can derive Eq.\ \ref{eq:atomdyn1} as $\frac{d\mathbf{p}_i}{dt} = \frac{\partial H}{\partial \mathbf{x}_{ij}}$, where $\mathbf{r}$ denotes momentum, from the conservative Hamiltonian 
\begin{equation}
H = \sum_{i=1}^{n_A} \left[ \frac{1}{2} m |\mathbf{v}_i|^2 + V_0(\mathbf{x}_i) + \sum_{j \geq i}^{n_A} \chi_{ij} V(\mathbf{x}_{ij}) \right], \label{eq:hamiltonian}
\end{equation}
where in the potential term $\chi_{ij} = \chi_{ji}$, $\mathbf{v} \equiv \frac{d\mathbf{x}}{dt}$ and $V_0$ is an external potential.
The potential energy term is the sum over all pairs of atoms of the mutual energy stored in each pair.
In this case $\frac{\partial H}{\partial \mathbf{x}_{ij}} = \mathbf{f}_{ij} = -\mathbf{f}_{ji} = -\frac{\partial H}{\partial \mathbf{x}_{ji}}$ and all force pairs are reciprocal, which is a manifestation of conservation of momentum.
That all forces are conservative (derivable from a Hamiltonian) entails that the Hessian of the dynamical system equations ---which describe trajectories in phase space--- is symmetric (by the equality of second-order mixed partial derivatives) \cite{Fruchart2021} and is sufficient to determine that the stationary state of the system is in equilibrium \cite{Tome1997}.
However the reverse does not necessarily hold (a symmetric Hessian does not imply that the system is conservative).
Conservative systems satisfy detailed balance lack periodic behavior and flows, and only with exponentially decreasing probability may deviate from equilibrium once reached. 
Exponential decay entails that there exists one characteristic spatio-temporal scale at equilibrium, which precludes hierarchies and complexity.

If we generalize the equations of motion to allow for friction and external forces then Eq.\ \ref{eq:atomdyn1} becomes
\begin{equation}
m \frac{d^2\mathbf{x}_i}{dt^2} + \zeta \frac{d\mathbf{x}_i}{dt} = \sum_{j=1}^{n_A} \mathbf{f}_{ij} + \mathbf{f}_0 \label{eq:atomdyn2} 
\end{equation}
where $\zeta$ is the friction or damping coefficient and $\mathbf{f}_0$ an external conservative force (such as gravity) such that $\mathbf{f}_0 = - \frac{\partial V_0}{\partial \mathbf{x}}$.
The inclusion of friction entails dissipation: Hamilton's equations require that the kinetic (l.h.s.) term be a homogeneous quadratic function in $\mathbf{v}$ for the Hamiltonian to remain constant \cite{Landau1960}.
The friction term entails a power dissipation $\frac{d H}{d t} = - \zeta |\mathbf{v}|^2$ \cite{Feynman1963,Tome1997} (also known as twice the Rayleigh dissipation function).
Thus, reciprocal or symmetric systems with friction eventually reach a motionless state due to a continuous leak of energy.

For $n_A$ atoms, the $n_A \times n_A$ matrix of interaction coefficients $\chi$ contains the interaction coefficients $\chi_{ij}$ between atoms (of the same or different kind) in off-diagonal elements and self-interactions along the main diagonal, which are ruled out here so $\chi_{ii} = 0$.
Here we only consider constant interaction coefficients, but in general they could be functions of e.g.\ the position or motion of atoms or of their pairwise distances.

Effectively, non-reciprocal forces may act as localized sources and sinks of energy, which maintain internal temperature differences.
Some non-reciprocal coupled systems with internal temperature gradients that satisfy detailed balance can be mapped onto reciprocally coupled systems \cite{Ivlev2015,Loos2020}.
In particular, as long as one condition (Eq.\ \ref{eq:DBcond}) is fulfilled, it is always possible for $n_K = 2$ and sometimes (see Appendix \ref{app:symmap}) for $n_K > 2$.
The particles of a system describable via a (pseudo-)Hamiltonian bears an average kinetic energy that can be construed as a temperature \cite{Ivlev2015}.
To explicitly model the effect of stochastically fluctuating forces we can add to the r.h.s.\ of Eq.\ \ref{eq:atomdyn2} a Gaussian white noise force $\boldsymbol{\xi}_i$, with zero mean and correlation function $\langle \boldsymbol{\xi}_i(t) \boldsymbol{\xi}_i(t') \rangle = 2 d \zeta k_B T_i \delta(t-t')$ that models being in contact with a heat reservoir at temperature $T_i$ (the $T_i$ factor in the variance follows from the fluctuation-dissipation relation) \cite{Tome1997,Zwanzig2001,Loos2020}, in $d$ dimensions. This is the Langevin equation
\begin{equation}
m \frac{d^2\mathbf{x}_i}{dt^2} + \zeta \frac{d\mathbf{x}_i}{dt} = \sum_{j=1}^{n_A} \mathbf{f}_{ij} + \mathbf{f}_0 + \boldsymbol{\xi}_i \label{eq:langevin}
\end{equation}
\cite{Loos2020}
 analyzed linear (Hookean; with exponent $n=-1$) diffusive ($m=0$, overdamped, so position jumps discontinuously) motion equations and found that a system may achieve thermal equilibrium even with non-reciprocal couplings: detailed equilibrium is satisfied if\footnote{This is analogous to the Onsager reciprocal relations \cite{Onsager1931}, which establish the symmetry of transport coefficients that more generally relate flows to gradient forces across different properties.} 
\begin{equation}
\chi_{ij} T_j = \chi_{ji} T_i, \label{eq:DBcond}
\end{equation}
which excludes anti-symmetric configurations: $\chi_{ij}\chi_{ji} <  0$ entails that equilibrium is impossible.
Hence, some non-equilibrium systems can be described as equilibrium systems with detailed balance with different temperatures for different subensembles $i = 1 \ldots n_K$ related by $\chi_{ij}$ \cite{Ivlev2015}.
The fluctuation-dissipation theorem couples the strength of force fluctuations to the temperature and the friction coefficient \cite{Zwanzig2001}, so this also entails that the fluctuating forces of different subensembles exhibit different strengths.
However, under overdamped dynamics ($\zeta$ larger on average than the critical damping, which is a function of Eq.\ \ref{eq:fij}) the temperatures $T_i$ of all atom kinds tend to a common background temperature $T_b$, associated with the common friction coefficient $\zeta$ via the fluctuation-dissipation theorem \cite{Ivlev2015}.

It has been shown \cite{Loos2020} via equipartition of energy that equilibrium may be also reached in systems with ballistic (underdamped) transport ($m>0$) and also, under particular conditions, to an arbitrary number of atom species $n_K$.
Following a similar argument (see Appendix \ref{app:DBeq}), it can be shown that the system of Eq.\ \ref{eq:atomdyn2}, where interactions are mediated by power laws of distance central forces, can also achieve thermal equilibrium: the non-reciprocal system of Eq.\ \ref{eq:langevin} satisfies detailed balance if and only if it can be mapped to a reciprocally coupled system that is isothermal \cite{Loos2020} and thus can be described by a pseudo-Hamiltonian (Appendix \ref{app:symmap}).

In summary, non-reciprocal coupling leads to a steady flow of energy through the system, i.e.\ a NESS \cite{Loos2020}, but some (non-conservative) non-reciprocal systems can reach a sort of thermal equilibrium sustained by locally asymmetric interactions: non-reciprocal systems satisfy detailed equilibrium if Eq.\ \ref{eq:DBcond} holds; for $n_K = 2$ this is always feasible, but for $n_K > 2$ the interaction coefficient matrix $\chi_{ij}$ must fulfill some constraints to allow for the possibility \cite{Loos2020}. Such systems may be composed of coexisting subensembles at different temperatures. 
Specifically, a non-reciprocal system can be mapped to a reciprocal system if and only if the interaction matrix of Eq.\ \ref{eq:fij} is \textit{symmetrizable}: $\chi$ is symmetrizable if there exists an invertible diagonal matrix $D$ and symmetric matrix $S$ such that $\chi = DS$.
Symmetrizing is not possible if the number of distinct off-diagonal terms in $\chi_{ij}$ is more than the number of atom kinds $n_K$ (Appendix \ref{app:symmap}): thus the number of distinct $\chi_{ij}$ terms should be no more than $\frac{1}{2}n_K(n_K-1)$ out of an a priori maximum achievable of $(n_K - 1)^2$; this restriction starts being effective for $n_K > 2$ (note that atoms of the same kind always interact reciprocally between themselves).
Note that symmetrizing is not the same as extracting the symmetric component: the interaction matrix $\chi$ can always be broken down into a skew-symmetric $\chi^A = \frac{1}{2}(\chi - \chi^T)$ and a symmetric component $\chi^S = \frac{1}{2}(\chi + \chi^T)$.
If $\chi$ is symmetrizable then a pseudo-Hamiltonian exists and the associated system is conservative and satisfies detailed balance: the interaction forces associated with a symmetrizable $\chi$ may constitute a conservative system if detailed balance (Eq.\ \ref{eq:DBcond}) is satisfied.

\subsection{Non-conservative and conservative forces in non-reciprocal systems \label{sec:nonrecipF}}

Regardless of reciprocity, detailed balance is fulfilled (Eq.\ \ref{eq:DBcond}) if and only if (in the steady state) the total entropy production vanishes. 
Besides, detailed balance implies that the pairwise heat and information flows among all degrees of freedom vanish as well \cite{Loos2020}.
However, as their mesoscopic and macroscopic analogues, some non-reciprocal systems may exhibit NESS (which by definition have positive total entropy production) with zero probability currents, i.e.\ no global particle or energy transport \cite{Roldan2010}: in the absence of external gradients, the source of persistent activity is intrinsic to the system's particular trajectories and interactions and not to an external gradient \cite{Loos2020}, as in active matter \cite{Sanchez2012,Marchetti2013,Fodor2018}.
In non-reciprocal systems who violate detailed balance entropy production occurs but it may or may not be accompanied by heat and/or information flows between degrees of freedom or subsystems: some degrees of freedom may be out of equilibrium without exhibiting heat flows, instead only information flows; conversely, some subsystems out of equilibrium may display heat flows but no information flows \cite{Horowitz2014a,Loos2020}.
A useful fact is that in the steady state a subsystem cannot absorb more information than it can expel heat, and it cannot absorb more heat than it can loose information \cite{Loos2020}.
Discrepancies between information entropy and thermodynamic entropy may occur because (1) thermodynamic entropy only accounts for the degrees of freedom that may store energy and (2) there is not a single way to define information entropy: limitations in accessing information about the system such as spatial resolution and specific criteria will determine what constitutes a degree of freedom (which may or may not store energy).

In thermodynamics or equilibrium statistical mechanics the behavior of a system is assumed to be governed, \textit{for small deviations} $\delta$ from equilibrium, by a conservative field that is the gradient of a thermodynamic free energy function $F$ \cite{Goldenfeld1992,Hohenberg2015,Fruchart2021}.
Then for any parameter $\phi$ representing some thermodynamic variable\footnote{For instance, in Landau theory $\phi$ is an order parameter that signals phase transitions by being non-zero in non-unordered phases and $F$ is the Landau free energy.}
\begin{equation}
\frac{\partial \phi}{\partial t} = \frac{\delta F}{\delta \phi} + w, \label{eq:Landau}
\end{equation}
where $w$ represents thermal fluctuations and $\delta$ represents deviations small enough that equilibrium considerations hold.
If we ignore the noise $w$, the form of Eq.\ \ref{eq:Landau} restricts the sort of bifurcations and hence phase transitions that the system may undergo: any relaxational forces must derive from a potential as $f = \frac{\delta F}{\delta \phi}$.
In regard to describing a system's dynamics the existence of a Hamiltonian function is related to the existence of a thermodynamic free energy function: both rest on energy conservation.
The minimization of free energy additionally ensures that entropy is maximized.

In general, if detailed balance (and its associated fluctuation-dissipation relation) is violated then there is no Boltzmann-like equilibrium density: the noise of NESS defies simple probabilistic description.
But in some systems chaotic and random driving yield identical behavior \cite{Heagy1994,Ding1995} and sometimes a free energy function with its associated steady steady density can be associated with a NESS system. 
Under the assumption that the microscopic dynamics is sufficiently chaotic (Gallavotti-Cohen chaotic hypothesis) it is possible to treat reversible many-particle systems in NESS as thermal systems, where a unique distribution on phase-space asymptotically describes the system \cite{Gallavotti1995a}.
Then the rate of contraction (or minus the divergence) of phase-space volume can be used under some conditions as a metric of internal entropy production \cite{Kurchan1998,Ruelle2003,Sornette2004}).
However very few practical systems satisfy the chaotic hypothesis \cite{Kwon2005}.

Another approach is based on linearization and the Helmholtz decomposition.
If $f$ were an arbitrary function governing the dynamics of the system instead of the gradient of a free energy function, we would have 
\begin{equation}
\frac{\partial \phi}{\partial t} = f(\phi) + w. \label{eq:snlde}
\end{equation}
By the Helmholtz decomposition, under mild conditions any vector field can be decomposed as the sum of a conservative or curl-free field $f^C$ and an incompressible, solenoidal, or divergence-free field $f^S$ as
\begin{equation}
f = f^C + f^S,  \label{eq:HelDec}
\end{equation}
where $f^C = \nabla F$, $\nabla \cdot f^S = 0$ \cite{Stokes1849,Helmholtz1858}, and $f^C$ and $f^S$ are respectively identified with conservative and dissipative forces \cite{Tome1997}
\footnote{The decomposition is not unique because adding to $f$ the gradient of any harmonic function generates another valid decomposition. But it is unique for bounded functions because all non-zero harmonic functions are unbounded (Liouville's theorem in complex analysis, not to confuse with its homonym in physics).}.
The dynamics of systems described by a free energy are characterized by a conservative force $f^C$.
In contrast, the typically ``curly'' dynamics caused by solenoidal forces $f^S$ such as friction (Eq.\ \ref{eq:atomdyn2}) and non-reciprocal forces (non-symmetrizable, which violate Eq.\ \ref{eq:DBcond}) may entail time-dependent phases and bifurcations initiated at exceptional (critical) points, forbidden in equilibrium statistical mechanics, that usher a chiral phase where a rotational continuous symmetry is spontaneously broken but is on average dynamically restored \cite{Fruchart2021} (e.g.\ in non-reciprocal flocking; akin to time crystals \cite{Wilczek2012}).
The most important phase transition enabled in non-reciprocal systems \cite{Fruchart2021} occurs from a static aligned elements phase to a dynamic chiral phase, which can be a sink or a saddle-node becoming a focus (rotating ``center'') roughly corresponding to chiral rotating clusters or chaser-avoider duplets; both ways at the exceptional critical point the eigenvalues vanish, indicating a Bogdanov-Takens bifurcation \cite{Izhikevich2007}.
In steady state, all the dissipated power is exclusively associated with solenoidal forces $\langle f^S \cdot \dot{\phi} \rangle > 0$ whereas $\langle f^C \cdot \dot{\phi} \rangle = 0$, where the chevrons denote time average.
 
If the force field $f$ of a stochastic nonlinear dynamic equation of the form of Eq.\ \ref{eq:snlde} can be decomposed into a sum of divergence-free and curl-free components that are mutually orthogonal as $f = f^S + f^C$ then we can determine the NESS probability density $\rho = \rho(\phi)$ of the state variable $\phi$ around a \textit{stable} fixed point as a solution of its associated Fokker-Planck equation $\frac{\partial \rho}{\partial t} = \nabla \cdot \left[ f \rho - D \nabla \rho \right]$ by simply ignoring the divergence-free component $f^S$ \cite{Tome1997} as 
\begin{equation}
\rho(\phi) = e^{-\frac{1}{D}F(\phi)}, \label{eq:boltzmanneq}
\end{equation}
which is the Boltzmann distribution of systems at thermal equilibrium (canonical ensemble) predicted by statistical mechanics.
Here we have assumed that $f^C = \nabla F$ is the gradient of the potential $F$ and the stochastic noise $w$ has mean zero and correlation function $\langle w(t) w(t') \rangle = 2 D \delta(t-t')$ with $D$ a (transport) diffusion coefficient.
Note that, despite being part of the force field, $f^S$ makes no contribution to the solution of Eq.\ \ref{eq:boltzmanneq} \textit{as long as it is orthogonal to} $f^C$ \cite{Tome1997}.

Crucially, it is always possible to choose the divergence-free component to be orthogonal to the curl-free or conservative component: this is done by subtracting from $f^S$ any conservative subcomponent and passing it to $f^C$ as\footnote{Note that curl-free and divergence-free are neither exclusive nor necessary features: e.g. the gradient of any harmonic function is both divergence-free and curl-free, and spiral fixed points in linear autonomous systems are neither divergence-free nor curl-free.}
\begin{equation}
f^C + f^S = f^D + f^Q, \label{eq:AoDec}
\end{equation}
with $f^D$ a curl-free conservative and $f^Q$ a divergence-free, rotational component that lacks any conservative subcomponent\footnote{Although the term solenoidal usually denotes any divergence-free or incompressible vector field, by its etymology it would be more suitable for divergence-free \textit{and} non-conservative vector fields, exactly as $f^Q$.}.
This is easier to visualize in the neighborhood of fixed points (the system's response cannot be linearized far from fixed points). 
Then the (space-varying) Jacobian of a constructed potential can be linearly approximated as a sum of two linear transformations: a symmetric and a skew-symmetric component real matrix.
A locally linearized system about some fixed point that yields a curl-free or conservative vector field is associated with a symmetric Jacobian matrix (with real eigenvalues); a divergence-free or incompressible field with a zero trace matrix (with zero sum of eigenvalues, not necessarily hollow; cf.\ special linear Lie algebra); and a neutrally stable rotation (a milling \cite{Trilochan2019} ``center'') field with a skew-symmetric matrix (with imaginary eigenvalues in conjugated pairs; cf.\ special orthogonal Lie algebra).
The divergence of the field is just the trace of the Jacobian matrix.
Note that a skew-symmetric Jacobian matrix has zero trace, but the opposite is false: a Jacobian matrix associated with a vector field that is both divergence-free and curl-free (a harmonic function's gradient) induces a squeeze mapping or hyperbolic rotation\footnote{A defective matrix represents a type of squeeze mapping, a shear. It can be combined with any other non-defective matrix by simply adding the appropriate Jordan blocks to its Jordan normal form.}.
Finally, a Jacobian matrix associated with a (linear mapping) vector field that is neither divergence-free nor curl-free induces a spiral (sink or source).

Since any real matrix can be expressed as a sum of a symmetric and a skew-symmetric matrix\footnote{Cf.\ in continuum mechanics the decomposition of the velocity gradient into a symmetric strain-rate tensor (including in turn a shear deformation and an isotropic contraction-expansion) and skew-symmetric spin tensor (rotation) components}, which respectively contain the real and imaginary part of the eigenvalues of the original matrix, it follows \cite{Ao2004,Kwon2005} that Eq.\ \ref{eq:snlde} can be expressed to first-order approximation via the linearization $U \delta\phi \approx -\frac{\delta F}{\delta \phi}$ of the Jacobian of the potential $F$ (with $U$ positive-definite) as  
\begin{equation}
\frac{\partial \phi}{\partial t} = -[D(\phi) + Q(\phi)] U \delta\phi + w, \label{eq:Ao}
\end{equation}
where $Q(\phi)$ is a (zero trace) skew-symmetric and $D(\phi)$ is the semipositive-definite (symmetric) diffusion matrix (cf.\ Eq.\ \ref{eq:boltzmanneq}); both can be assumed constant in a small neighborhood of fixed points $\phi$ \cite{Kwon2005}.\footnote{Equivalently \cite{Ao2004,Kwon2005,Yuan2014}
\begin{equation}
[S(\phi) + A(\phi)]\frac{\partial \phi}{\partial t} = - U \delta\phi + \xi, \label{eq:Ao2}
\end{equation}
with $\langle \xi(t) \xi(t') \rangle = 2 S(\phi) \delta(t-t')$ and $S$ is interpreted as a friction matrix inducing a dissipative force $-S(\phi)\dot{\phi}$ which interestingly is related to the diffusion matrix $D(\phi)$ via a generalized Einstein relation $[S(\phi)+A(\phi)]D(\phi)[S(\phi)-A(\phi)] = S(\phi)$ \cite{Ao2004,Ao2008}.}
The restriction to semipositive-definiteness simply reflects that the diffusive term of the (forward) Fokker-Planck equation induces expansion in the phase-space (positive eigenvalues), i.e. dissipation \cite{Ao2008}.
In the decomposition of Eq.\ \ref{eq:HelDec} rewritten as $f = f^C + f^S = f^D + f^Q$ now we identify $f^D$ and $f^Q$ as the curl-free and divergence-free forces associated with the matrices of their superindices.

A (linearized) conservative process (which satisfies detailed balance), if written as a Hamiltonian system, displays a zero trace evolution matrix, reflecting that the phase-space volume is preserved under time evolution (Liouville's theorem). 
Confusingly, because of the symplectic structure of Hamiltonian systems (its associated symplectic matrix is skew-symmetric), whose evolution equation ``intertwine and swap'' the position and momentum coordinates to compute flow, the conservative nature of Hamiltonian flow is expressed as a \textit{divergence}-free flow.  
Thus, although the forces derived from a Hamiltonian (Eq.\ \ref{eq:hamiltonian}) are conservative, the Hamiltonian flow is a divergence-free vector field (but not necessarily skew-symmetric, since it may display both squeeze and rotation mappings).
Hence, conservative forces are associated with a (non-conservative) solenoidal vector field and vice versa non-conservative forces are associated with a non-solenoidal (non-divergence-free) vector field.
Thus for Hamiltonian systems such as Eqs.\ \ref{eq:atomdyn1}, \ref{eq:hamiltonian} with an added noise term $\boldsymbol{\xi}_i$ (as in Eq.\ \ref{eq:langevin}, with $F=H$ and $\frac{\partial F}{\partial\phi}$ the Hamiltonian vector field with its usual symplectic structure) the semipositive-definite symmetric term $D(\phi)$ of Eq.\ \ref{eq:Ao}, associated with a dissipative friction force $f^D = -\zeta \dot{\phi}$, vanishes leaving only a skew-symmetric $Q(\phi)$ component associated with a conservative force  $f^Q$ which in the phase space is nonetheless a Lorentz force\footnote{A vector cross product force which can be expressed via the product of a skew-symmetric matrix with the velocity vector in phase space.} or solenoidal vector field \cite{Yuan2014}, thereby ensuring that particles run along isoenergy lines.
However, for general (non-conservative) non-reciprocal systems of the form of Eq.\ \ref{eq:langevin} both the symmetric $D(\phi)$ and skew-symmetric $Q(\phi)$ matrices would be non-zero and functions of the location on phase space $\phi$ so they can only be considered constant in a linearly approximable region about a fixed point\footnote{Interestingly, adjusting the $D(\phi)$ and $Q(\phi)$ components while sampling along and jumping between isoenergy lines is roughly how the remarkable Hamiltonian Monte Carlo sampling algorithm for full Bayesian inference of even complex hierarchical models \cite{Betancourt2015,Betancourt2017,Carpenter2017}}.

What can be said about the total energy of our system defined by Eq.\ \ref{eq:atomdyn2} and Eq.\ \ref{eq:fij}?
If our system were purely conservative (reciprocal), we could apply the virial theorem, which relates the time average of the total kinetic $K_{tot}$ and total potential $V_{tot}$ energies as $2\langle K_{tot}\rangle = (n+1) \langle V_{tot}\rangle$ or $m \langle \sum_{i=1}^{n_A} |\mathbf{v}|^2 \rangle = (n+1) \langle \sum_{i=1}^{n_A} \chi^D_{ij}V(\mathbf{x}_{ij}) \rangle$ for $n \neq -1$.
More generally the power or rate of energy non-conserved \cite{Tome1997,Loos2020} has a friction (dissipative) contribution $- \zeta |\mathbf{v}_i|^2$ (Section \ref{sec:nonreciprocal}) and a non-reciprocal contribution $\mathbf{v}_i \cdot \sum_{j=1}^{n_A} \mathbf{f}_{ij}^Q$, where $\mathbf{f}^Q_{ij} = - \chi_{ij}^Q \frac{\partial V}{\partial \mathbf{x}_{ij}}$ \footnote{A similar expression could be obtained for entropy production \cite{Tome2006}.}. Hence the total non-conservative power is 
\begin{equation}
\frac{d H}{d t} = - \sum_{i=1}^{n_A} \left[ \zeta |\mathbf{v}_i|^2 - \mathbf{v}_i \cdot \sum_{j=1}^{n_A} \chi_{ij}^Q \frac{\partial V}{\partial \mathbf{x}_{ij}} \right], \label{eq:disspow}
\end{equation}
where the first term of the r.h.s.\ is always negative, and the second may be positive (energy inflow) or negative (energy outflow). 
In the steady state its average must be zero: $\langle \sum_{i=1}^{n_A} \left[ \zeta \mathbf{v}_i + \sum_{j=1}^{n_A} \chi_{ij}^Q \frac{\partial V}{\partial \mathbf{x}_{ij}} \right] \rangle = 0$.
At any rate, the conservative forces $\mathbf{f}^D_{ij} = - \chi_{ij}^D \frac{\partial V}{\partial \mathbf{x}_{ij}}$ applied to atom $i$ can be derived from a Hamiltonian (cf. Eq.\ \ref{eq:hamiltonian}) as $\frac{d\mathbf{p}_i}{dt} = \sum_{j=1}^{n_A} \mathbf{f}^D_{ij} = -\sum_{j=1}^{n_A} \chi_{ij}^D \frac{\partial V}{\partial \mathbf{x}_{ij}} = \frac{\partial H}{\partial \mathbf{x}_{ij}}$ because the forces are reciprocal and conservative so their contribution to the Hamiltonian is zero at all times\footnote{The analysis \cite{Tome1997} is carried out in overdamped regime ($m=0$) but the same applies in underdamped regime \cite{Loos2020}.} \cite{Tome1997}.

By definition $\mathbf{f}_{ij}^Q$ lies along the straight line incident with atoms $i$ and $j$, so $\mathbf{f}_{ij}^Q \cdot \mathbf{v}_i = 0$ if and only if $i$ moves orthogonally to said line.
Thus for each atom $i$ the non-reciprocal power $\mathbf{v}_i \cdot \sum_{j=1}^{n_A} \mathbf{f}_{ij}^Q$ depends on the velocity and relative position with respect to $i$ of every other atom $j$.
In the steady state, Eq.\ \ref{eq:disspow} becomes 
\begin{equation}
\langle \sum_{i=1}^{n_A} \mathbf{v}_i \rangle = - \frac{1}{\zeta}\langle \sum_{i=1}^{n_A}\sum_{j=1}^{n_A} \chi_{ij}^Q \frac{\partial V}{\partial \mathbf{x}_{ij}} \rangle, \label{eq:Vness}  
\end{equation}
which shows that the total momentum is a function of the friction, the skew-symmetric component of the  interaction matrix and variables such as pairwise distances and power exponent (Eq.\ \ref{eq:Vpowlaw}).

This equation also holds for any isolated subsystem in steady state, and is more useful in such form because we can apply it to any subensemble of atoms that at least temporally does not interact with the rest of the system, perhaps due to being spatially segregated.
How often this occurs depends essentially on the density of atoms.
Taking for example just one duplet of atoms $i=1,2$, typically one of them acts as a chaser and the other as a avoider, thus reflecting the circularity inherent to skew-symmetry, where either of them may slow down or speed up depending on a subtle interplay among their current mutual distance and velocities, their (opposite sign) interaction coefficients, and the power law shape.
This asymmetric interaction yields a steady state average ``cluster velocity'', which is the chase velocity at which the pair of atoms will keep moving together.
With zero friction the non-reciprocal forces will keep injecting energy into the system and the total energy grows unbounded \cite{Ivlev2015} (Appendix \ref{app:symmap}).

As we increase the size of a subsystem, its analysis becomes exponentially more complex.
We have just mentioned that a non-reciprocal duplet is a single entity endowed with an ``intrinsic motor'' (energy source) that permanently propels it with some average non-zero velocity.
On the other hand, a reciprocal duplet would be static (conservative) and can only exist as long as it is attractive: otherwise its two constituent atoms would part ways and cease being an entity.
For entities composed of three or more atoms, the combinatorial possibilities quickly become intractable.
Nonetheless, the sole condition of being an entity places a strong constraint: all entity members must be bound by predominantly (slightly) attractive interaction forces (e.g. stable fixed points), with perhaps some non-reciprocal interactions creating ``intrinsic motors'' (sink spirals and cycling centers), so that they stay in a coalesced state that can be recognized as a single persistent entity.
Conversely, repellors such as unstable fixed points or sources (or source spirals) would simply disintegrate and dissolve in the surrounding medium as soon as they are formed, thus never becoming persistent entities.
Section \ref{sec:lifefinder} expounds an approximate method to discover and quantify the number and type of such entities or groups of atoms that tend to stick together via a dynamic filtering algorithm.

\subsection{Our particular world: atoms of multiple kinds interacting via power law central forces \label{sec:IPS}}

Our IPS builds on an implementation \cite{Abdulrahman2022} of Clusters or Particle Life \cite{Ventrella2017}, an algorithm contrived to imbue a soup of moving particles with lifelike behavior. 
Unlike in other implementations where particles' pairwise forces dependency on distance is a linear piecewise function with short-range repulsion and long-range attraction or repulsion \cite{HackerPoet2017}, here we use power law central forces (Section \ref{sec:ixr}).

For simplicity, for all simulations we set the following parameters thus: rectangular domain edge lengths $L_1 = 800$ and $L_2 = 600$; variables data information content $I_0 = \log{N} = 32\log{2}$; atom mass $m = 1$; BUNCH algorithm forward time step $\delta t = 1$; atom radius $r_a = 3$.

The most important parameters are: space dimensionality $d$; number of atom kinds $n_K$; total number of atoms of each kind $n_i : i=1, \ldots, n_K$ and of any kind $n_A \equiv \sum_{i=1}^{n_K} n_i$; interaction force (power law) exponent $n$ and coefficients $\chi_{ij} : \forall i,j = 1 \ldots n_K$; friction or damping coefficient $\zeta$; time step $\delta t$ between simulation frames; walled (with boundary) or wrap around (compact without boundary) edges domain topology.
The larger $\delta t$ is, the faster the simulation runs, but the more inaccurate it becomes. A too large $\delta t$ introduces spurious effects such as reciprocal systems with friction never reaching a still state (cf.\ Section \ref{sec:nonreciprocal})---ostensibly, physical time may theoretically also be discrete on a scale of less than $10^{-33}$s \cite{Feynman1982,Wendel2020}.
We ran the BUNCH algorithm in $d = 2$ dimensions with $n_K = 4$ atom kinds for interaction force power law exponents in the interval $[-2,0]$. 

Damping is the loss of energy associated to a finite velocity by dissipation.
It determines the difference between dry (damping due to friction with an embedding medium) and wet (no damping) active matter systems \cite{Marchetti2013}.
A non-zero damping coefficient prevents the total energy of the system from growing unbounded \cite{Ivlev2015} (Section \ref{sec:nonrecipF}).
The coefficient $\zeta$ was set to be constant across atoms kinds; in general it would be associated with some background temperature $T_{bg}$ via the Einstein relation between dissipation and energy and diffusion, although we do not explicitly use it here.
 
To scout the $d^2$-parameter space of the between-species interaction matrix, each element is independently initialized to a real number in the interval $[-1,1]$. 
To enhance exploration, the provided \texttt{C++} version graphical user interface (see Appendix \ref{app:roving_params}) also affords two sliders that specify the step size and the stepping probability per time step in the parameter space of matrix interactions.

The type of world wall or spatial boundary affects the evolution of the system: boundary effects are crucial to stabilizing ordered patterns in NESS \cite{Reinken2020}.
First, a wall may absorb or release energy into the system by being either perfectly hard (reflecting velocities with no transfer of energy) or elastic (non-conservative relaxational Hookean force).
Second, the boundary defines the space where entities may exist. The wall defines the size of the world and thus its density, which is a key variable determining the frequency of interaction between clusters. 
For a finite space without boundary (closed manifold), the choice of topology restricts the sort of entities that may abide in the system. This is accomplished via the orientation in which the boundaries of a e.g.\ plane are sewn (fundamental polygon): here we only implement a toroidal topology by simply identifying the opposite sides of a 2-dimensional domain while preserving their orientation. 
However, we will mainly run simulations on a hard walled world because it is computationally cheaper (cf.\ Appendix \ref{app:closedmanifold} for BUNCH's trouble in running on a closed manifold) and because the scale of the typical entities we found were smaller than the scale of their encompassing world.

\section{A viable lifefinder algorithm \label{sec:lifefinder}}

We have set out the theory behind the emergence of lifelike behavior in interacting particle systems.
Now our goal is to describe an algorithm to identify, tag, and store the location and lineage or relationships between the persistent entities emerging in a system of interacting particles, and to quantify the lifeness of these entities.
In particular, in line with Section \ref{sec:lifeness}, each node of a dynamically evolving tree (hierarchy) of nested nodes (clusters), starting from a single entity (the omnicluster or archicluster), can be regarded as an entity.
The nodes of the tree or dendron are identified with entities of a hierarchical clustering structure. 
As nodes, persistent entities are by definition discrete; each entity is \textit{one} entity; and clearly, clusters of them are also discrete.
Each entity or cluster of entities inhabits a particular domain in space and time, with their characteristic space and time scale.
Since entities can only cluster in discrete units, inevitably their attending spatio-temporal scales must also be distributed as discrete lumps.
The ensuing structure is characterized by a set of discrete scales \cite{Sornette2004,Martinez-Saito2022c} that match the levels of a hierarchy of persistent entities (cf. Section \ref{sec:critworld}).

Just as the particles of a random attractor (cf. Section \ref{sec:socblobs}) or cluster define the cluster itself (via its members and location) by virtue of their sticking together through time, so we could say as well that the children of a particular cluster are constrained by belonging to their parent cluster.
This duality is related to the circular or bidirectional interaction between entities and environment, and lies at the core of the autopoietic \cite{Varela1974} and FEP \cite{Friston2013} views on biotic self-organization.
However, our algorithm proceeds \textit{mostly} on a bottom-up fashion, from atoms upwards.
This is pertinent because the IPS lacks higher-order parameters.
But note that teleonomic or seemingly purposeful behaviors may spontaneously emerge in large enough systems with only pairwise interactions between particles, as in nature.

Given a distribution of particles, how can we find what is the most likely (hierarchy of) clusters that generated them?
For this, we need (1) a characterization of how particles are (spatially) distributed over clusters (Section \ref{sec:HMGC}) and (2) an algorithm that adheres to this characterization to tractably find and identify likely clusters-entities (Section \ref{sec:bunch}).

\subsection{Modeling a multiscale world with a hierarchical mixture of Gaussian clusters \label{sec:HMGC}}

For the sake of generality, we do not assume that the generative process of the atom soup is known (although here we did design our own IPS).
As mentioned in Section \ref{sec:socblobs}, a straightforward yet costly way to identify persistent entities is rolling autocorrelation kernels of different scales over space and time and selecting peaks from the output.
Here the scale of the kernel corresponds to the characteristic space-time scales of persisting entities, where bigger scales are typically associated with larger, slower, and more complex entities.
However, this requires either choosing the scale of the kernels, which in general is unknown, or computing all, which is intractable.
A lighter and more elegant approach is assuming that persistent entities, just as NESS and natural complexity (Sections \ref{sec:critworld}, \ref{sec:synmepfep}, \ref{sec:socblobs}), are organized in a hierarchical fashion, as cluster or entity trees.

A more efficient (at the cost of some flexibility) approach is modeling the particle soup world as a hierarchical mixture of (elliptical) Gaussian clusters (HMGC)\footnote{Not to confuse with hierarchical clustering of mixtures \cite{Goldberger2004}, which is an information compression technique for reducing mixtures within a single level.} that has \textit{no free parameters}.
The HMGC is mixture model where, at each level, the weights (the probabilities of children belonging to each possible parent cluster) are determined by the lineage probability, i.e. the product of recursively compounding the nested probabilities of children belonging to their parents, of parents belonging to their grandparents, etc.
Thus it explicitly models a genealogy of entities via the probability of nested memberships through a lineage of cluster-entities.
At each level, clusters are (1) parents that aggregate children elements of the subordinate level, and (2) their centroids are the children of the supraordinate level.

Therefore at the lowest (atomic particle) level the aggregates of particles are modeled as a mixture of scattered bivariate Gaussian functions (clusters or entities), whose iso-density loci are in general ellipses. In two dimensions, a more flexible, yet overly sophisticated, approach to model ring-like structures would be replacing Gaussian blobs with (squared) Bessel functions, which yield the radial part of circular membrane oscillations, such as the sombrero potential or jinc function.
We use Gaussian blobs because they are the most parsimonious (in the entropy maximizing sense) densities that carry information about location, size, anisotropy, and orientation, in virtue of being defined by only the first two moments or cumulants \cite{DeGroot2001}.
This is similar to atomic nuclei are modeled as balls that can be stretched and squeezed into ellipsoids \cite{Andersen1972}.
Thus, clusters and entities are defined by (1) the location of the centroid, their (2) Gaussian blob or ellipse $d$ semi-axes and orientation, and finally by (3) their children.
At the second level, the cluster centroids of the first level are in turn modeled as clusters themselves, whose centroids are in turn clustered at the third level, etc.
In a mixture, each elliptic blob carries a weight, which allows representing not only ovals and circles, but also oblong shapes, sticks (near-degenerate ellipses), and annuli (``donuts'' made of a wide hill concentric with a narrow pit, akin to a high-pass frequency filter).
All these in turn can be combined to form arbitrarily complex shapes such as polygons, whorls, tadpoles or chains.
The top (omnicluster or universal cluster) and bottom (atoms) entities are the only eternal structures.

Till now, we have simply referred to both clusters and entities as sets defined by their member elements (if we consider their spatial location and size and orientation non-essential contingent attributes).
Although both are sets defined by their constituting members, for the sake of bookkeeping here we establish an important distinction between them, that arises only when considering multiple kinds of atoms: entities are invariant under permutations of atoms or entities of the same kind, but clusters are not.
To wit, for clusters all atoms and clusters are distinguishable, whereas entities only distinguish between atoms or entities of different kinds \footnote{Entities are closer to the physical definition of object, because in physics e.g. elementary particles or atoms of the same kind are indistinguishable.}.
It follows that all clusters are unique and distinguishable, but some entities (at the same level) can be indistinguishable (and hence interchangeable).
Thus, while each cluster is a set (of atoms at level 1, otherwise of subordinate sets), entities are multisets (of atoms at level 1, otherwise of subordinate multisets).
This distinction is crucial for computing the evolution of the system state: the indistinguishability of atoms and entities, together with a limit on the minimal spatio-temporal resolution with which particles can be tracked (see Footnote \ref{fn:gibbsHt}) immediately entails a loss of information\footnote{Because microscopic variables evolve towards high entropy asymptotic distributions (become mixed or averaged out) at the macroscopic scale.} and hence an increase in entropy. This is the mechanism that underpins the Second Law. 
  
In a $d$-dimensional IPS, atoms have two properties: atom kind or species and location coordinates.
Cluster and entity (non-atom) children or members have the following properties:
\begin{itemize}
\item Kind: cluster or entity identity, defined by the list of its children.
\item Location coordinates ($d$ variables).
\item The ellipse's semi-axes lengths ($d$).
\item Ellipse orientation ($d-1$).
\end{itemize}

How do we know that a particular child entities belongs to a particular parent entity?
Via inference: a procedure to fit the HMGC model's parameters to the data (atoms) determines entity memberships.
There are many feasible ways to perform inference. 
One of the main contributions of this article is precisely the BUNCH algorithm for fitting HMGC models.

The virtues of HMGC are simplicity, versatility and adaptability, requiring zero externally adjusted parameters (except perhaps the initial conditions), and perhaps a near-optimal trade-off between simplicity and accuracy.
Crucially, not only the location, size, and orientation of blobs require no parameter adjusting, but neither do the number of blobs at each level \textit{and even} the number of hierarchy levels (the minimum number of levels is one).
This is important because by penalizing ``idle'' parameters that make little contribution to model accuracy, inference automatically selects the number of levels and blobs at each level, which is a sort of model selection \cite{Friston2007,Friston2011b,Hobson2012}.
Intuitively, the complexity of a model is proportional to the amount of information (e.g. number of levels and clusters in them) needed to describe it \cite{Kolmogorov1963,Wallace1968}, where more parsimonious models are deemed to be better representations of the modeled object.

Notice that we have neglected time.
Here we use a static HMGC model, i.e. in it no time derivatives are explicitly modeled.
Instead, time-dependent changes are simply accommodated by iteratively refitting the model to the IPS configuration at each time step. 
Thus cluster members can belong in a static or dynamic manner by either staying still or moving e.g. cyclically or chaotically, as long as they remain in the vicinity of the cluster centroid.
A more sophisticated model could comprise, besides centroid locations and ellipse orientations, their time derivatives \cite{Friston2008}, e.g.\ centroid velocities (and accelerations) and ellipse axes angular velocities.

\subsection{Model complexity and likelihood of the world \label{sec:modCL}}

Both the HMGC and the cluster dendron model $D$ are specified by $i = 1 \ldots L$ levels, each with clusters $j \in {C_{i,j}}$, each defined by a Gaussian blob with parameters $\mu_{i,j}$ and $\Sigma_{i,j}$.

How much information is needed to describe a particular hierarchy of clusters with a cluster dendron or HMGC model? 
What is the surprisal or information associated with describing a particular world (configuration of atoms) with a cluster dendron or HMGC model?
The former is called model complexity; the latter, likelihood.

Let us first discuss model complexity. It is important to remark that although the HMGC (and cluster dendron) model is hierarchical in the sense that clusters are connected in a tree graph and are recursively dependent on their children (the subordinate level clusters), their parameters are not connected as a Bayesian hierarchical model: any given cluster's parameters are independent from all the other parameters of the model, and in fact in our particular implementation all model parameters have the same fixed prior.

The HMGC model has no arbitrary adjustable parameters, but it is parameterized by the covariance $\Sigma$ and location $\mu$ of every cluster in every level.
Each cluster is defined by a Gaussian blob density that yields the conditional probability of having a child at a location $x$ as $\mathcal{N} \left(x \mid \mu, \Sigma \right) = \frac{1}{\sqrt{\det{(2\pi \Sigma)}}} e^{-\frac{1}{2}x^T\Sigma^{-1}x}$.
This blob is defined in $d$ dimensions by $3d-1$, of which $d$ correspond to spatial coordinates, $d$ to the sizes of semi-axes, and $d-1$ are required to specify spatial orientation.  

Regardless of the spatial dimension $d$ of the world, the total probability density of any Gaussian blob, being a probability measure, must amount to 1 and is related to its covariance by 
$1 = \sqrt{\det{(2\pi \Sigma)}}$, which for isotropic covariance of variance $\sigma^2$ becomes $(2\pi)^{d/2} \sigma^d$.
Hence the conditional probability of finding a child within one standard deviation of the blob decreases exponentially with the number of dimensions. 
More precisely, the expected density of a spherical Gaussian blob (not to confuse with the average location) is $ E[\mathcal{N}(0,\sigma^2)] = \int \mathcal{N} \left(x \mid 0, \sigma \right)^2 dx = \frac{\sqrt{\det{(\frac{1}{2} \cdot 2\pi \Sigma)}}}{\det{(2\pi \Sigma)}} = \frac{\pi^{d/2} \sigma^d}{(2\pi)^d \sigma^{2d}} = (2 \pi^2 \sigma)^{-d}$, which also decreases exponentially with the number of dimensions.

Note that the expected densities so calculated depend on the choice of the length unit, via $\mu$ and $\sigma$.
This is because we are implicitly measuring surprisal through Shannon's differential entropy, which is a continuous version of (discrete) information entropy, with units of probability per geometrical measure.
The differential entropy of a $d$-dimensional normal distribution is $\int \mathcal{N} \left(x \mid \mu, \Sigma \right) \log{\mathcal{N} \left(x \mid \mu, \Sigma \right)} dx = \frac{1}{2}\log{\det{(2\pi e \Sigma)}} = \frac{1}{2}\log{\det{\Sigma}} + \frac{d}{2}\log{2\pi e}$.
But unlike entropy, differential entropy lacks a simple interpretation because it can be negative and is not invariant under changes of variables (or variable units).
Instead, Jaynes suggested to use the limiting density of discrete points \cite{Jaynes1968}, which is an invariant measure of entropy of a density $f(x)$ defined as $\lim_{N \to \infty} H_N(f) = \log{N} + \int f(x) \log{\frac{m(x)}{f(x)}} dx = \log{N} - D_{\text{KL}}(f \parallel m)$, where $m(x)$ is a uniform probability density over the quantized domain of $f(x)$ with $N$ discrete points, and $D_{\text{KL}}(f \parallel m)$ denotes the Kullback-Leibler divergence from $m$ to $f$. 
Here, the continuous entropy is obtained in the limit $N \to \infty$,
This suggests that in general the entropy of a continuous variable is no less than infinity\footnote{This is intimately related to the problem of reconciling the irreversibility of thermodynamics typically described via a system with countable many states (Second Law), and the reversibility of classical statistical mechanics as described with a density in continuous phase space (Liouville's theorem). Gibbs' H-theorem \cite{Jaynes1965} provides an enlightening connection: an entropy increase follows from a ``blurring'' of the phase space trajectories (which corresponds to a loss of spatial resolution) traced by Liouville equation trajectories.\label{fn:gibbsHt}} and that actually information entropy is unambiguously well defined for discrete variables only, with which it acquires the extensive character of thermodynamic entropy. Alternatively, differential entropy may be simply restricted to spatial scales large enough to accommodate a phenomenological probabilistic description of matter: e.g. in classical mechanics this would correspond to scales larger than the size of the system's particles.
We will see that $\log{N}$ is associated with the complexity of the priors $I_P$, and that $D_{\text{KL}}$ can be used to compare models (cf.\ Section \ref{sec:omniclus}).

A subtle but common problem for describing model complexity is that there is no single legitimate manner to specify priors.
As we just discussed, any continuous variable contains infinite information unless one measures it with (the problematic) differential entropy. 
Hence, we used uniform discrete probability distributions, whose entropy over $N$ values is simply $\log{N}$ (also called Boltzmann's entropy).
Arbitrarily, we can set $N$ to the typical single-precision floating-point number type, which has a size of 4 bytes or 32 bits and an information content of $\log{N} = 32 \log{2} \approx 22.18$.
A $d$-dimensional cluster dendron with $i = 1 \ldots L$ levels and $n_{C_i}$ clusters at each level $i$ has $(3d-1) \sum_{i = 1 \ldots L} n_{C_i}$ parameters and the complexity of its priors is 
\begin{equation}
I_C =  \log{N} \left[ (3d-1) \sum_{i = 1}^L n_{C_i} - (2d-1)\right], \label{eq:I_C}
\end{equation}
where the $-(2d-1)\log{N}$ term takes away the information of the top cluster corresponding to its spatial location and orientation to render $I_C$ invariant to translations and rotations.

However our particular implementation of the HMGC and cluster dendron models\footnote{That is, if we ignore the complexity associated with the random generator of the BUNCH algorithm.} has no parameters except for the initial conditions, which are simply the $3d-1$ parameters of the starting cluster ellipse blob. And since we consider any pair of entities equal up to rotations and translations, this number reduces to $d$. Hence, from Eq. \ref{eq:I_C} we get
\begin{equation}
I_P =  d\log{N}. \label{eq:I_P}
\end{equation}

Although not pertaining to the complexity of the cluster dendron, 
for the purposes of estimating its parameters the E-step of the BUNCH algorithm (Section \ref{sec:bunch}) requires storing the information $I_B$ needed to specify the distribution of children over clusters at each level. The computation of $I_B$ is described in Appendix \ref{app:bellnumber}.

Computing the log-evidence of a HMGC model given a configuration of atoms is in general intractable because at every level each child belongs (proportionally to the weight probability vector) to all possible cluster parents, where the contribution of each cluster parent to the likelihood in turn depends on their weight vector and all their own cluster parents, recursively up to the top cluster.
In contrast, the log-evidence of a cluster dendron given all atoms is simply given by the sum over all levels of the log-probability of the single ascendant of the atom\footnote{Because the cluster dendron model lacks independent parameters, the log-likelihood becomes identified with the log-evidence}.
Concretely, for each atom, this can be calculated recursively in top-bottom order by adding the likelihood of each cluster in level $i$ having its children at level $i-1$ as $- \sum_{i = 1}^L \sum_{j \in C_{i,j}} \log{P \left( x_j \mid D \right)}$, where $P \left( x_j \mid D \right)$ is the probability of the element $j$ at level $i$ being located at $x_j$, given by its parent cluster at level $i+1$, defined by a Gaussian density $P \left( x_j \mid D \right) = \mathcal{N} \left(x_j \mid \mu, \Sigma \right)$ with its own parameters $\mu$ and $\Sigma$. Then the surprisal of a countable number of atoms at any time point given a particular cluster dendron $D$ is the model accuracy or negative log-likelihood $I_{A|D}$ (the sum of the negative log-likelihood of every parent cluster given its children atoms; $l=1$), the dendron model complexity $I_D$ is the sum  of the negative log-likelihood of every parent cluster given its children clusters ($l=2 \ldots L-1$) and thus the information content $I_{A,D}$ of the tree model or dendron complexity $I_D$ with its atom leaves or likelihood $I_{A|D}$ is the sum
\begin{equation}
I_{A,D} = I_{A|D} + I_D = - \sum_{a \in A} \sum_{l = 1}^{L-1} \log{P \left( x_{l,a} \mid D \right)},
\end{equation} 
where $A = C_0$ is the set of all atoms and $x_{l,a}$ is the location of the unique ascendant of atom $a$ located at level $l$. 
The cluster dendron (and HMGC) model describes the probabilistic distribution of atom locations as a categorical (or mixture) density that is recursively defined at each level by the probability assigned to each cluster by its parent cluster. 
The cluster dendron model ignores any other atom attributes such as color or spatial or temporal correlations of any order beyond first.
Under the assumption that the dendron model is a reasonably good representation, $I_C$ (Eq.\ \ref{eq:I_C}) can be considered, in the spirit of the Schwarz and Akaike Information Criteria \cite{Bishop2007}, a rough approximate value $I_C \approx I_D$ that is proportional to the number of free parameters.

Collecting summands, for $d=2$ and discrete variables of cardinality $N=2^{32}$, the log-evidence or total information or surprisal required to describe an IPS world given a particular cluster dendron model becomes
\begin{equation}
I_W \equiv I_P + I_D + I_{A|D} = 64 \log{2}  - \sum_{a \in A} \sum_{l = 1}^{L-1} \log{\mathcal{N} \left(x_{l,a} \mid \mu_{l+1,a}, \Sigma_{l+1,a} \right)},
\end{equation}
where $\mu_{l+1,a}$ and $\Sigma_{l+1,a}$ are the mean and covariance  matrix of the unique cluster ascendant of atom $a$ located at level $l+1$. 
Note that here the model complexity of the cluster dendron $I_D$ corresponds to the (Kolmogorov \cite{Kolmogorov1963}) complexity of the algorithm that enables the cluster dendron to dynamically adapt its hierarchical structure to the state of the world. 
An equally sensible definition for the negative log-evidence would have been $I_W \equiv I_{A,D} = I_D + I_{A|D}$.

\subsubsection{Comparisons and a benchmark model: the omnicluster \label{sec:omniclus}}

Although the surprisal or negative log-evidence of the cluster dendron $D$ given a configuration of the atoms $I_W$ evaluates how well the model describes the world, without a yardstick of what constitutes a mediocre fit its significance is obscure.

Assuming the world is isolated, the simplest HMGC would be the omnicluster $O$ (cf. Section \ref{sec:socblobs}), defined as the single cluster ($n_C=1$) best fitting all the atoms. It affords a yardstick through the surprisal of all atoms conditioned on the omnicluster $I_{A|O} = - \sum_{j \in A} \log{P \left( x_j \mid O \right)} = -\sum_{j \in A} \mathcal{N} \left(x_j \mid \mu_O, \Sigma_O \right)$, 
where $\mu_O$ and $\Sigma_O$ contain the only $3d-1$ parameters of the model.
This enables to assess the negative log-evidence of a model $D$ given the world with reference to its homologous value for the omnicluster: 
\[
I_{A|D} - I_{A|O}.
\]

As in the previous section, the complexity of the omnicluster is $I_P(O) = 64 \log{2}$.
For the omnicluster, the contribution associated with the hierarchical structure of the model vanishes $I_O=0$.
The distance between the likelihood of the models $D$ and $O$ can be measured as the Kullback-Leibler divergence from the density $P_{A|D}$ to $P_{A|O}$ over the atoms $a \in A$:
\[
D_{\text{KL}}(P_{A|O} \parallel P_{A|D}) = - \sum_{a \in \mathcal{A}} P_{A|O}(x) \frac{\log{P_{A|D}(x)}}{\log{P_{A|O}(x)}}.
\]

\subsection{Lifeness \label{sec:lifeness_eq}}

Given an entity dendron $D$ (an entity dendron is just a class of indistinguishable cluster trees, cf. Section \ref{sec:HMGC}), an entity subdendron $D_{i,j}$ is defined by the subset of $D$'s nodes that comprises entity $C_j$ at level $i$ \textit{and} all its descendants (recursively nested children). The largest $D_{i,j}$ is $D$ itself, the second smallest are any of the first level entities and the smallest are the atoms (elementary or zeroth level ``entities'').

The lifeness $L_{i,j}$ of an entity subdendron $D_{i,j}$ is defined as
\begin{equation}
L_{i,j} \equiv I_{D_{i,j}} \cdot \tau_{i,j}^{cum}, \label{eq:L_cum}
\end{equation}
where $I_{D_{i,j}}$ is the subdendron complexity\footnote{Instead of $I_{D_{i,j}}$ we could alternatively use the difference of the subdendron and omnicluster complexities $I_{D_{i,j}} - I_O$.} and $\tau_{i,j}^{cum}$ the cumulative lifespan of the entity subdendron $D_{i,j}$, which is defined as the sum of all time intervals during which \textit{at least one} instance (cluster) of the entity subdendron $D_{i,j}$ was alive\footnote{Another sensible definition is: the sum of all time intervals during which instances of the entity subdendron $D_{i,j}$ were alive. Here all the multiple time intervals associated with the simultaneous existence of multiple instances of $D_{i,j}$ are included in the sum.}.
Here only the subdendrons associated with currently alive clusters $D_{i,j}$ are considered.
An alternative measure of lifeness exclusively associated to the currently alive entities is
\begin{equation}
L_{i,j}^{pres} \equiv I_{D_{i,j}} \cdot \tau_{i,j}^{pres}, \label{eq:L_now}
\end{equation}
where $\tau_{i,j}^{pres}$ is simply the lifespan of $D_{i,j}$ exclusively associated with the present life.
Finally, another definition of lifeness can be made by including not only the currently alive entities but all the entities that were ever alive during the simulation $D_{i,j}^{all}$ as $L_{i,j}^{all} \equiv I_{D_{i,j}^{all}} \cdot \tau_{i,j}^{cum}$. 

The definition of entity (as opposed to cluster) provided in Section \ref{sec:HMGC} suggests that another important clarification is in order: do multiple (cluster) instances of an entity contribute one or multiple lifespan units per simulation frame?
Both are sensible ways to define lifeness, bearing different interpretations.
For a given entity, the former definition (one-instance) adds a lifeness unit if at least one entity instance is currently alive, whereas the latter (multiple-instances) adds as many lifeness units as entity instances are currently alive.
Here we chose the latter, which has a spatial dimension to it that better captures the notion of biological species.
For example, glycine is a common molecule, both the simplest (proteinogenic) amino acid and a spinal cord neurotransmitter, that can be found in many living systems. 
Hence its lifeness (assuming that the BUNCH algorithm detects it correctly) using the multiple-instances definition would be many orders of magnitude larger than using the one-instance definition.
However in this article this choice is inconsequential because our simulations consist of only a few hundred atoms so occurrences of multiple instances are only sporadic.

Note $L_{i,j}$ and its variations comprise the log-likelihood $I_A$ because $I_A$ describes the complexity linking particular subentities $D_{i,j}$ to the extra information needed to describe the atoms of $D_{i,j}$ given $D_{i,j}$.

An approximate and simpler way to define lifeness $\Lambda_{i,j} \approx L_{i,j}$, in the spirit of Eq.\ \ref{eq:I_C}, is: an entity subdendron $D_{i,j}$ in a $d$-dimensional space within a dendron of $i = 1 \ldots L$ levels has an approximate lifeness 
\begin{equation}
\Lambda_{i,j} = n_{i,j} \cdot \tau_{i,j},  \label{eq:approx_lifeness1}
\end{equation}
where $n_{i,j}$ is the cumulative complexity of subdendrons, recursively defined as
\begin{equation}
n_{i,j} = 
\left\{ \begin{array}{lll}
160 \log{2} + \sum_{h \in H_{i,j}} n_{i-1,h}  & \text{if} \quad i > 0  & \text{(nodes)} \\
64 \log{2}  & \text{if} \quad i = 0 & \text{(leaves),}
\end{array} \right. \label{eq:approx_lifeness2}
\end{equation}
where $H_{i,j}$ is the set of entities that are children of the subdendron $D_{i,j}$ and $n_{i,j}$ is just the total number of dendron nodes comprised within the subdendron $D_{i,j}$, which can be calculated recursively by simply counting the total number of descendants, multiplied by the complexity of each node. 
Hence the r.h.s.\ second term for dendron nodes ($i>0$) is a sum only over the children of $n_{i,j}$.
The r.h.s.\ first term for dendron nodes ($i>0$) comes from $160 \log{2} = 5 \log{(2^{32})} = (3d-1) \log{(N)}$ parameters (cf.\ Eq.\ \ref{eq:I_C} in Section \ref{sec:modCL}), and analogously for dendron leaves (i.e.\ atoms) except that they are determined by 2 parameters (spatial coordinates for $d=2$) instead of 5.
Note that the reason the approximation $\Lambda_{i,j} \approx L_{i,j}$ holds reasonably well is related to the reason that small fast fluctuations are not part of persistent entities per se (as mentioned in Section \ref{sec:critworld}) and so their algorithmic complexity or information content can be safely discounted.

\subsection{The BUNCH algorithm: Bottom-up bisect-unite nodes clustered hierarchically \label{sec:bunch}} 

The BUNCH algorithm is a dynamical filter that estimates on the fly the parameters of a hierarchy of nested Gaussian blobs (HMGC, cf. Section \ref{sec:HMGC}) that fits a set of moving particles $A$, by maximizing its log-evidence $\log{P_{\text{HMGC}}(A)}$.
Because this inference problem quickly becomes intractable as the number of atoms (and hence model complexity; Section \ref{sec:modCL} ) increases, even more so for a dynamically evolving system, the BUNCH algorithm accomplishes approximate inference via heuristics such as point estimation, stochastic sampling, and ``lazy'' or greedy estimation.

Briefly, starting from the first (atom) level, BUNCH performs a bottom-up sweep, where cluster features and the identity of their children are estimated in a hybrid Bayesian and maximum likelihood fashion.
In addition, the number of clusters and levels is adjusted via a lazy scheme that performs checks to decided whether to split or merge pairs of randomly picked clusters.
The code is available on the web hosting service GitHub repository (\url{https://github.com/mmartinezsaito/racemi}) under the MIT license (see also Appendix \ref{app:software}).

\subsubsection{Rendering inference tractable \label{sec:varbay}}

The BUNCH algorithm can be construed as a recognition model that approximates and renders tractable estimating the parameters of a  generative model that enacts the true generative process underlying the world state \cite{Hinton1993a,Friston2003}, where the generative process is our IPS (Section \ref{sec:IPS}) and the generative model is the HMGC model (Section \ref{sec:HMGC}).

All levels are modeled as Gaussian mixtures \cite{Bishop2007} estimated via an elliptical K-means algorithm \cite{Sung1998}, which is the Expectation-Maximization (EM) algorithm \cite{Dempster1977} for Gaussian mixtures with an E-step where atoms are categorically (hard) assigned to clusters \cite{Bishop2007}.

Similarly to the EM algorithm, BUNCH is a combination of Bayesian and maximum likelihood approaches in a iterative scheme that yields a HMGC density together with maximum a posteriori (MAP) estimates of their parameters (e.g.\ centroids) and latent variables (e.g.\ cluster membership), such that the MAP estimates at one level become the data for the the supraordinate level. 
Variational Bayesian approaches are a generalization of the EM algorithm, where all parameters are treated as latent variables, endowed with a (typically approximate) density, that are sequentially optimized within each iteration step \cite{Friston2006,Bishop2007}.
Variational Bayes renders estimation tractable by assuming a factorized (independent factors) and simplified (usually Gaussian) form of the posterior, which enables optimizing one parameter at a time. In general, this is done by minimizing the variational free energy \cite{Feynman1972,Friston2003,Friston2006,Bishop2007} function
\begin{equation}
F = D_{\text{KL}}(Q_{\text{BUNCH}}(\theta) \parallel P_{\text{HMGC}}(\theta|A)) - \log{P_{\text{HMGC}}(A)}, \label{eq:FE}
\end{equation}
where $A$ is the data or the state of the atom soup that we want to explain by fitting to it the HMGC model with generative density $P_{\text{HMGC}}(A, \theta)$ that we assume generated it; $\theta$ is the set of parameters that defines a particular configuration of the HMGC model; and $Q_{\text{BUNCH}}$ is the recognition density that estimates the parameters $\theta$ of $P_{\text{HMGC}}$.

However, in our case $Q_{\text{BUNCH}}$ degenerates, from a probability density, to a Dirac delta because the BUNCH algorithm simply selects a MAP point estimate of $\theta$ in a tractable manner.
Although the first term of Eq.\ \ref{eq:FE}'s r.h.s. enters the mechanics of BUNCH in general it is intractable to compute explicitly due to its stochastic and combinatorial complexity. 
Thus we simply ignore it and instead of $F$ we use the log-evidence $\log{P_{\text{HMGC}}(A)}$ as a measure of goodness of fit.

Overall, BUNCH inherits the flexible sequential updating scheme from both EM and variational Bayes, and the simpler hybrid maximum likelihood-Bayes estimation approach from EM.
BUNCH enables tractable estimation of a HMGC, a tree of nested Gaussian mixture densities.
This is achieved via dynamical updates, which combine (1) cycling through the cluster tree parameters at each hierarchy level during a bottom-up sweep, similarly to dynamical filtering approaches resting on variational inference, especially DEM \cite{Friston2008} and similar derivations (e.g.\ Hierarchical Gaussian filtering \cite{Mathys2011,Mathys2014}), (2) random sampling of log-evidence in the neighborhood, and (3) local exploration of the parameter space with stepsize determined by a learning rate $\alpha$, as in gradient descent or temporal regularization \cite{Friston2007}.

\subsubsection{Clustering: separately at each level from bottom to top \label{sec:bu_clustering}}

How to group particles into clusters and determine the location and number of clusters? 
A metric or notion of distance is needed to associated a measure of proximity to pairs of points; here we employ the default choice of Euclidean distance.
Now the first idea that comes to mind is to allocate atoms to the nearest cluster center or centroid; this yields a polygonal partition of the space into Voronoi cells. 
This is precisely what the k-means clustering algorithm \cite{Lloyd1982} does, after randomly initializing the cluster centroids.
k-means is a form of hard clustering, as opposed to fuzzy clustering, which preserves the probabilistic membership of children inherent to Gaussian mixtures.

However, a major limitation of k-means is only accommodating isotropic blobs, i.e.\ disks or spheres for 2- or 3-dimensional environments.
In k-means, only cluster centroids are estimated; this precludes modeling both the size and anisotropy of clusters.
This limitation is overcome by elliptical k-means \cite{Sung1998,Bishop2007}, an algorithm that also estimates cluster covariances.
This affords ellipses or ellipsoids, which can roughly and efficiently represent any shape via combinations of elongated (rod-like) or disk-like ellipses (prolate or stick-like and oblate or disk-like ellipsoids for $d=3$).
This is important for modeling membrane-shaped structures such as micelles and lipid bilayers.
Concretely, elliptical k-means \cite{Sung1998} is a hard-assignment version of the Gaussian mixtures with general covariance matrices.
By using the Mahalanobis distance, it generalizes k-means to enable modeling of non-spherical or elliptical clusters \cite{Sung1998,Raykov2016}, which effectively is the same as computing the covariance matrix of elliptical blobs.
  
In our case, each EM iteration consists of two steps: the E-step for children element reallocation among the parent clusters, which reuses both children and parents from the previous iteration, and the M-step, which updates the clusters' sizes and positions \cite{Bishop2007}.
Due to the tree topology of the cluster hierarchy, at every E-step iteration a cluster may adopt new children only among the children of its sibling clusters (i.e. grandchildren of its parent cluster).
Just as variational inference, EM is a heuristic not guaranteed to converge to a global optimum \cite{Neal1998}.
At each level, the estimated clustering is simply a Gaussian mixture where each mixture's weight vector degenerates from a probability vector of length $K$ (number of clusters) into a categorical variable that can take $1$ of $K$ possible categories.
This means that each child is hard-assigned to (``locks onto'') a single cluster parent, instead of belonging to each of the $K$ cluster parents with a probability specified by a weight vector.
This makes the cluster tree or dendron likelihood function much faster to compute than for the full HMGC model.

The so far described EM and gradient-based approaches to Gaussian mixture and naive elliptical k-means have still three major limitations: (1) they require setting in advance the location and number of centroids; (2) when a cluster has only one (or no) child assigned its covariance matrix collapses so the algorithm gets stuck in singularity of the likelihood function \cite{Bishop2007} (the log-likelihood contains from the normalization coefficient of Gaussian blobs terms such as $-\log{|2\pi\Sigma|}$ that approach infinity as the covariance matrix $\Sigma$ becomes degenerate); and (3) it does not accommodate nesting of clusters (hierarchies).
Instead, we want a scheme that averts singularities and automatically determines all cluster parameters for every level of the hierarchy \textit{and} the number of levels of the hierarchy.

\subsubsection{Bisect-unite and birth-death \label{sec:BUBD}}

This section describes lazy candidate selection for splits and merges, and the role of birth and death of clusters in BUNCH.

Determining the number of clusters without enough constraints is an ambiguous problem.
By increasing the number of clusters, one can arbitrarily increase the accuracy or negative log-likelihood $I_A$ of the model (by having as many clusters as atoms and vanishing covariances).
However this entails increasing also the complexity $I_D$ of the model, which contributes to the total log-evidence $I_W = I_A + I_D$. 
Accuracy and complexity are traded off when attempting to maximize model evidence \cite{Bishop2007,Friston2010}.

Some approaches, such as k-means \cite{Lloyd1982} and its variants (e.g.\ k-medoids) and mean shift \cite{Comaniciu2002}, require a priori specifying the number $K$ of clusters.
Other methods such as DBSCAN \cite{Ester1996} and its OPTICS \cite{Ankerst1999}, allow non-convex shaped clusters by relying on sample density, but nonetheless still require beforehand one or more parameters specifying a density threshold to form clusters.
Yet other schemes, such as hierarchical clustering and bisecting k-means \cite{Steinbach2000}, altogether may do away with prior settings.

In hierarchical clustering, a tree of nested clusters is built by successively merging clusters in a bottom-up sweep (agglomerative clustering) or splitting clusters in a top-down sweep (divisive clustering).
In the resulting dendron or tree, the root node is the omnicluster, the boughs are the next-to-root nodes, the twigs are the smallest non-terminal nodes or first level clusters, and finally the leaves or terminal nodes are the atoms. 
BUNCH is a dynamical filter that blends features of both agglomerative and divisive clustering approaches:
\begin{itemize}
\item Initial condition: The cluster dendron is initialized with a single cluster (an omnicluster) that fits all atoms of the world.
\item Bottom-up sweep: All clusters are nested and, at any level, solely determined by the configuration of all potential children (in the subordinate level), so each iteration of the algorithm must proceed bottom-up.
\item Random (lazy) selection of cluster candidates for merging and splitting: At each level and iteration step, one cluster pair is randomly selected for potential merging and then likelihood ratio test between the log-likelihood of the merged parent cluster given the combined children and the separate two parent clusters given their respective children is virtually calculated to decide whether to merge (if positive) or not (if negative).
Then and similarly, one cluster pair is randomly selected for potential splitting and a decision to split or not based on an analogous likelihood ratio test is made. Further details in Appendix \ref{app:BUBD}.
\end{itemize}

Note that randomly selecting candidates for splits (one) and merges (two) is analogous to applying stochastic perturbations to the cluster tree.
After the perturbation the resulting log-evidence is computed to decide whether to keep or discard the changes.
This is conceptually the same as the forward gradient used to optimize objective functions in machine learning \cite{Baydin2022,Ren2022}.

At every time step, clusters (more precisely, their index in memory pointing to a list of their attributes) may (1) be born, (2) be forwarded to the next time step (i.e. stay ``alive''), or (3) die.
Clusters are born when (1) one cluster comes out from a merge, (2) two clusters come out from a split, and (3) the world is initialized with a single cluster (the omnicluster) and starts up.
Any change to the cluster tree is guided by the direction in which the log-evidence of the atoms given the cluster tree is on average increased (since lazy rules may sometimes reduce log-evidence). 
A cluster is carried over to the next time step as long as it better predicts or models the configuration of its children relatively to other (extant or not) clusters.
Otherwise, it dies (its index in memory stops being used) as a result of being split, merged, or ``killed''.
A childless cluster (that has lost all its children) is killed (its index is excluded from future usage) because its existence does not explain anything (see Appendix \ref{app:BUBD}).

\subsubsection{Bottom-up cluster tree updating}

As in agglomerative clustering the BUNCH algorithm uses a level-wise bottom up approach, to evaluate the likelihood of the cluster parents of the current level children and to lazily evaluate merges, splits, and deaths. 
As in divisive clustering \cite{Kaufman1990}, BUNCH recursively may split the initial omnicluster into descendent clusters; in this respect, it is akin to Bisecting k-means \cite{Steinbach2000}, X-means \cite{Pelleg2000} (which repeatedly attempts subdivision until a criterion such as the Bayesian information criterion is reached) or G-means  \cite{Hamerly2003} (which iteratively selects clusters and bisects them if they do not pass a test of Gaussianity
dependent on a significance level parameter).
However, BUNCH may also merge pairs of selected clusters at every step.
Divisive clustering with exhaustive search is $\mathcal{O}(2^L)$, but BUNCH employs lazy selection of candidates for splits and merges.
Note however that because cluster likelihood is evaluated bottom-up, a given cluster tree can only grow or collapse by at most one level each time step: each top-down recursive split can occur at most once per time step.   
Just as cluster reshaping and centroid shifting, in general splits, merges and deaths occur whenever model evidence is \textit{locally} higher at newly sampled points in parameter space.
See Appendix \ref{app:BUBD} for details.

But among all hierarchical clustering, perhaps BIRCH (Balanced Iterative Reducing and Clustering using Hierarchies) \cite{Zhang1996} is conceptually closest to BUNCH.
The nodes or branches of the tree created by BUNCH are analogous to BIRCH's Clustering Feature Nodes.
Like BIRCH \cite{Zhang1996}, BUNCH clusters only keep statistics (e.g. number of children, centroid location, distance from centroid to children, cf. Appendix \ref{app:svd}) of subclusters, and it is bottom-up hierarchically built.
However BIRCH uses two parameters, viz. branching and threshold factor, to respectively limit the number of children in a cluster parent and the distance or eligibility of a child to become adopted  by the candidate clusters parents, whereas BUNCH is parameter-free.

The output of BUNCH is a dynamically evolving cluster tree or dendron whose leaves are the atoms and nodes are the hierarchically nested clusters or entities (with higher degree or thicker branches typically representing more complex and persisting entities) that define the current best model or explanation of the IPS configuration.

\section{Numerical simulations: hierarchically clustering particles with BUNCH \label{sec:simul}}

In this section we illustrate the computations of the BUNCH algorithm on multiple IPS instances. 
For simplicity, we fix the number of dimensions to $d=2$. 
Hence, by the divergence theorem, the flux of a particular atom's interaction force field through any surface enclosing it is constant if it decays with exponent $p = 1-d = -1$ (cf. Section \ref{sec:ixr}).
We fixed the damping coefficient to $\zeta = .7$, the number of atoms to $n_A = 256$ with $n_K=4$, $n_i = 64 : \forall i=1 \ldots n_K$; and the bounding universe box to 800 pixels (arbitrary unit) of width and 600 pixels of height. 

The interaction matrix can be decomposed into symmetric and skew-symmetric components $\chi = \chi^S + \chi^A$ (Section \ref{sec:nonreciprocal}), which enables defining a few useful ---if somewhat arbitrary--- descriptive statistics\footnote{Remember that skew-symmetry is incompatible with equilibrium (Eq. \ref{eq:DBcond}).}: 
\begin{enumerate}
\item The average balance of repulsion and attraction among atoms
\[ r_{\chi} = \frac{1}{n_A(n_A-1)}\sum_{i,j=1}^{n_K} n_i (\chi_{ij} n_j - \chi_{ii}) \in [-1,1], \]
where a positive (negative) sign indicates repulsion (attraction).
\item The ratio of the sum of skew-symmetric ($\chi^A$) to symmetric ($\chi^S$) matrix component element absolute values
\[ s_{\chi} = \left( 1+\frac{n_i|\chi_{ij}^A|n_j}{n_i(|\chi_{ij}^S|n_j - |\chi_{ii}^S|)} \right)^{-1} \in [0,1], \]
with $s_{\chi}=0$ only for purely symmetric and $s_{\chi}=1$ only for purely skew-symmetric interaction matrices.
\item Another sensible index of skew-symmetry would be $s'_\chi$, a measure of the distance between the symmetrizable and non-symmetrizable components of $\chi = \chi^{A'} + \chi^{S'}$, where $S'$ denotes the symmetrizable matrix closest to $\chi$ under some suitable metric (such as an ''entry-wise'' matrix norm) and $A' \equiv \chi - S'$.
The motivation for $s'_\chi$ is that only symmetrizable systems can always be mapped to a reciprocal system described by the Gibbs-Boltzmann distribution (underpinned by a pseudo-Hamiltonian).
\item The averaged difference of the absolute value of symmetric and skew-symmetric matrix elements
\[ d_{\chi} = \frac{1}{n_A(n_A-1)} n_i((|\chi_{ij}^S| - |\chi_{ij}^A|)n_j - |\chi_{ii}^S|) \in [-1,1], \]
where a positive (negative) value denotes a larger contribution from the symmetric (skew-symmetric) component, vertical bars denote absolute value and Einstein summation notation was used in defining $s_{\chi}$ and $d_{\chi}$.
\end{enumerate}

We will denote each simulation instance with interaction matrix $\chi$ by its parent family of IPS defined by the triad $(p, r_{\chi}, s_{\chi})$.

\subsection{Illustrative example: World $(-1,.09,.30)$ \label{sec:p09_30}}

\begin{figure}
\centering
\includegraphics[width=.87\textwidth]{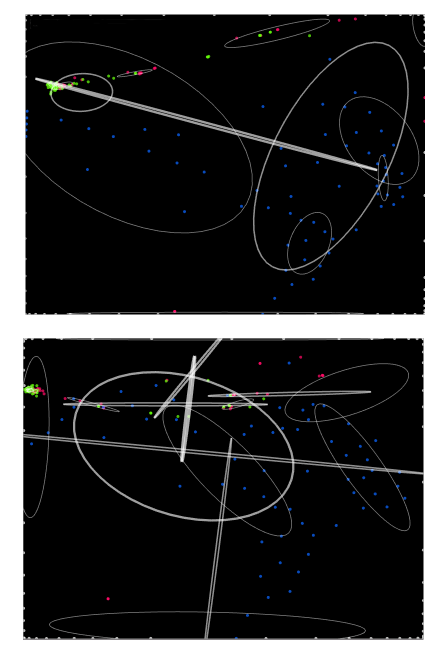}
\caption{Two arbitrarily chosen frames (1316 and 1541) of World $(-1,.09,.30)$. Atom colors (red, green, blue, white) are assigned from one of four species. The BUNCH algorithm was started up at frame 711. Each gray ellipse represents one live cluster; its major and minor axes are twice the standard deviation of the average distance of its children with respect to is centroid (Appendix \ref{app:svd}) and its thickness is proportional to its height in the cluster hierarchy.  \label{fig:p09_30_f1406+1541}}
\end{figure}

First we show a simulation run of an IPS (Fig. \ref{fig:p09_30_f1406+1541}) with interaction force power law exponent $p=-1$ and matrix $\chi_{(-1,.09,.30)} \approx \begin{bmatrix} 
-.19 & -.55 & .21 & -.05 \\
-.97 & -.94 & -.52 & .34 \\
.96  & .20 & .07 & .31 \\
.11  & .99 & .79 & .73 \end{bmatrix}$, which is representative of dissipative NESS with long-range interactions exhibiting chaser-avoider duplets and complex behavior characteristic of skew-symmetric interaction matrices (Section \ref{sec:nonrecipF}).
$\chi_{ij}$ is the coefficient of the force on any atom of kind $i$ exerted by any atom of kind $j$, with the sole exception that atoms do not exert force on themselves. A negative (positive) sign indicates attraction (repulsion).
Atom kinds are arbitrarily colored: 1=red, 2=green, 3=blue, 4=white. 
Calculating yields $r_{\chi} = .094$, $s_{\chi} = .300$, and $d_{\chi} = .265$; thus interaction forces are on average slightly more repulsive than attractive and comprise both symmetric and skew-symmetric components, with more of the former.
We will denote this IPS instance as World $(-1,.09,.30)$, a notation that lists some of the most characteristic parameters and statistics of the system, viz. power law exponent $p$, repulsion-attraction balance $r_{\chi}$ and interaction symmetry $s_{\chi}$.

Examining $\chi$ reveals that blue ($i=3$) and especially white ($i=4$) atoms are repelled by all atoms, including their kind; red ($i=1$) is attracted by all except blue, and green ($i=2$) is attracted by all except white.
This results in jittery white atoms ditched along the edges of the world box and jittery blue atoms strewn all over the world trying to stay away from all surrounding atoms, while a bubbling clump of red and green atoms scurries along convoluted trajectories while bouncing off the box edges (Fig. \ref{fig:p09_30_f1406+1541} shows two snapshots).

\begin{figure*}
\includegraphics[width=1\textwidth]{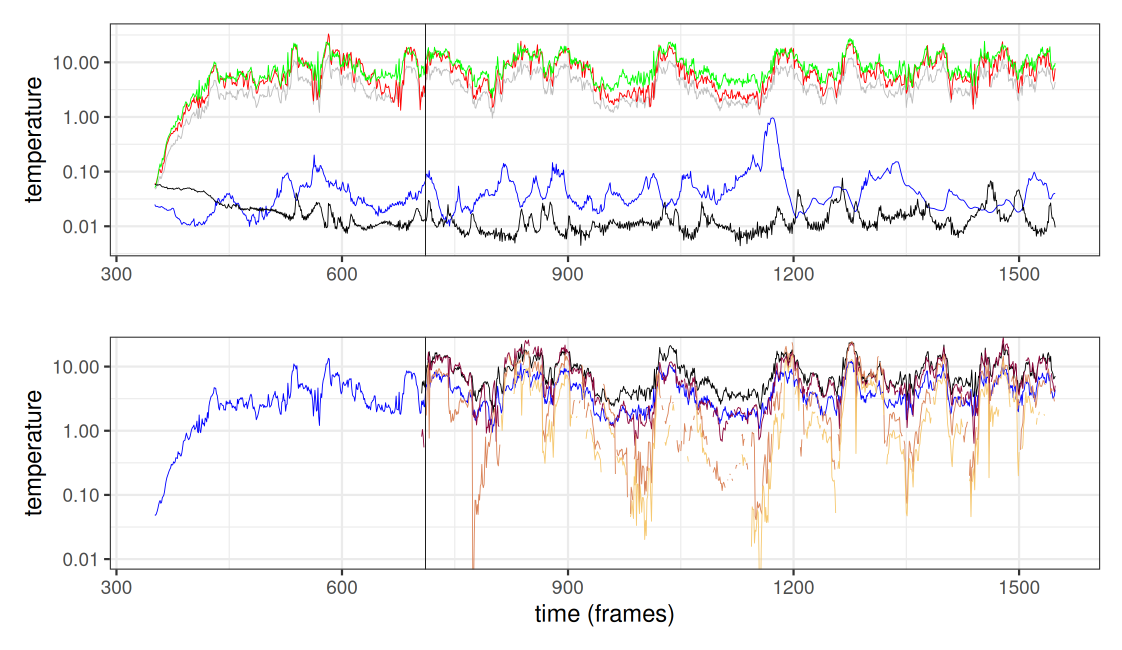}
\caption{The temperature (squared velocity in px/s) of atoms at each stimulation time step (frame) for World $(-1,.09,.30)$, in logarithmic scale. Top: By atom kind, for $n_K=4$. Red, green, blue, and black denote the atom kinds $k=1\ldots 4$; grey denotes aggregate mean. Bottom: By the dendron level ($n_L=4$) encompassing the entity. A heat colormap is used where darker (lighter) colors denote lower (higher) levels; blue denotes aggregate mean. The vertical line denotes the start of the BUNCH algorithm run at frame 711. \label{fig:p09_30_T_K_L}}
\end{figure*}

Each of the 4 atom kinds in World $(-1,,.09,.30)$ has an associated temperature or squared velocity \cite{Feynman1963}. Consistent with their associated $\chi_{ij}$ signed coefficients, red and green have higher temperature than blue and white atoms (Fig.\ \ref{fig:p09_30_T_K_L}, top).
The mean temperatures for each atom kind are approximately $T_1 = 7.6, T_2 = 9.5, T_3 = .06,T_4 = .01$.
We are not aware of any simple way to deduce the temperatures given an arbitrary $\chi$, as is the case for symmetrizable systems \cite{Ivlev2015,Loos2020}, which satisfy Eq.\ \ref{eq:DBcond}.

By the fundamental (equal a priori probability) postulate of statistical mechanics, it follows that the average energy (squared velocity) of akin atoms is the same.
Hence if a cluster as an object that weights the aggregate weight of all its children, then we expect its average energy to be the average energy of its children divided by the number of children, which is analogous to the calculation of the center of mass velocity of composite bodies in kinetic theory of gases \cite{Feynman1963}.
A similar argument explains that entities are slower roughly in proportion to their number of children (Fig.\ \ref{fig:p09_30_T_K_L}, bottom).

\begin{figure*}
\includegraphics[width=1\textwidth]{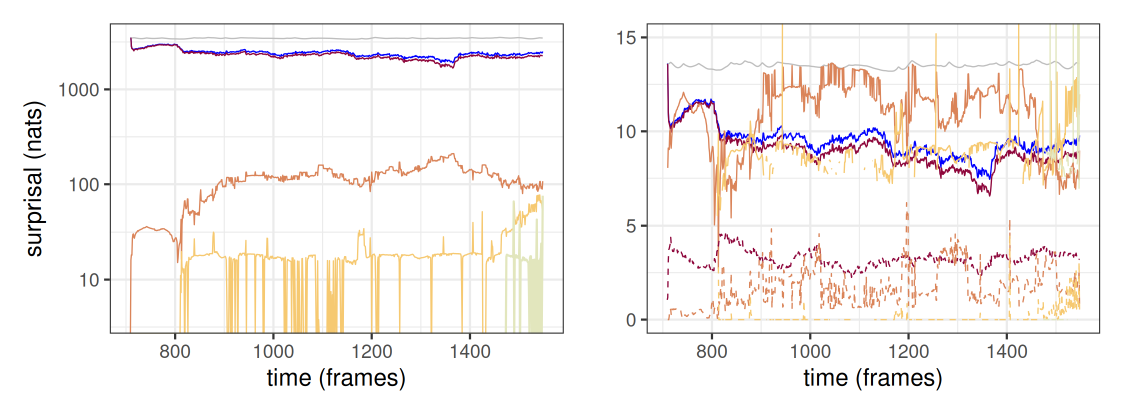}
\caption{The surprisal of entities at each of the $n_L=4$ dendron levels at each simulation time step (frame) for World $(-1,.09,.30)$. The heat colormap associates darker (lighter) colors with lower (higher) levels; gray denotes the reference model or omnicluster. Left: Sum of surprisals over entities within each level; blue denotes aggregate sum. Right: Within-level mean (solid) and standard deviation (dashed) surprisal of entities. Blue denotes aggregate mean. Second level clusters have on average the largest surprisal until approximately frame 1450, when they are exceeded by the third and fourth level clusters. \label{fig:p09_30_su_L_sum_mean}}
\end{figure*}

\begin{figure}
\centering
\includegraphics[width=.5\textwidth]{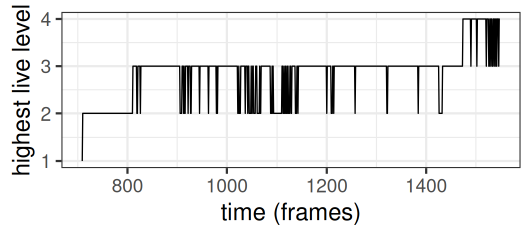}
\caption{Evolution of the current maximum dendron level for World $(-1,.09,.30)$. \label{fig:p09_30_hystlev}}
\end{figure}

The surprisal of the dendron and its constituting clusters (cf.\ Section \ref{sec:lifeness_eq}) dynamically evolving under the BUNCH algorithm is shown in Fig.\ \ref{fig:p09_30_su_L_sum_mean}, with the evolution of the highest live or top cluster displayed in Fig.\ \ref{fig:p09_30_hystlev}\footnote{Note that BUNCH or in general hierarchical clustering algorithms would have trouble fitting a static solid-state-like world: fitting a lattice with a hierarchical model is an ill-posed problem, and the resulting cluster dendron is likely to have only one or two levels while struggling to minimize surprisal.}.
Although not fully shown, in the first few frames of the BUNCH algorithm the total suprisal plummets to a value of around 2400 nats or 9.4 nats per atom, and thence oscillates in its neighborhood.
This is a pattern typical of coevolving species in Wright fitness landscapes \cite{Kauffman1991,Ao2008}.
The rugged plateaus indicate that a NESS has been attained. 
The hierarchical structure of the tree entails that most of the clusters belong to the first (atomic) level, so the atomic level aggregate suprisal $I_{A|D}$ accounts for most of the total suprisal $I_{A,D}$ (Fig.\ \ref{fig:p09_30_su_L_sum_mean}, left).
The dendron surprisal $I_D$ is the sum of surprisals for all levels except the atomic, and the total surprisal $I_{A,D}$ is the sum of surprisals across all levels (Section \ref{sec:modCL}).
The surprisal associated to the dendron components at each level as estimated by the BUNCH algorithm is almost always smaller than that of the reference model of the omnicluster (Fig.\ \ref{fig:p09_30_su_L_sum_mean} in grey; cf. Section \ref{sec:omniclus}).

The mean surprisal per children (cluster or atom) at each level is typically larger at supra-atomic levels (Fig.\ \ref{fig:p09_30_su_L_sum_mean}, right).
Because the current version of BUNCH does not model velocities and hence fast atoms cannot be well predicted, one would expect that within-level averaged surprisal is also higher for lower levels.
However here this is not the case, likely because continuous smooth models do not explain well sparse data, and cluster blobs have far fewer children at e.g.\ the second level than the first.

\begin{figure}
\centering %width=.5
\includegraphics[width=.5\textwidth]{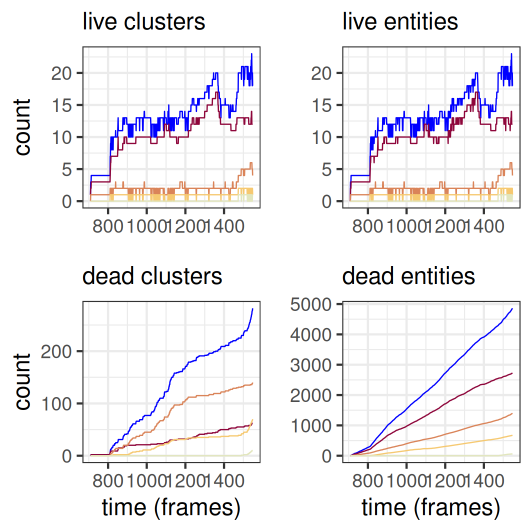}
\caption{Number of live (top) and dead (bottom) clusters (left) and entities (right) at each level. Darker (lighter) colors denote lower (higher) levels; blue denotes whole dendron sum. \label{fig:p09_30_LivDed_C_E}}
\end{figure}

The evolution of the amount of live and dead clusters and entities separately for each level is shown in Fig.\ \ref{fig:p09_30_LivDed_C_E}.
Note that there are more dead entities than dead clusters because during its lifespan any given cluster typically takes the form of multiple entities as it adopts and loses children (Section \ref{sec:BUBD}).

\begin{figure}
\centering
\includegraphics[width=.5\textwidth]{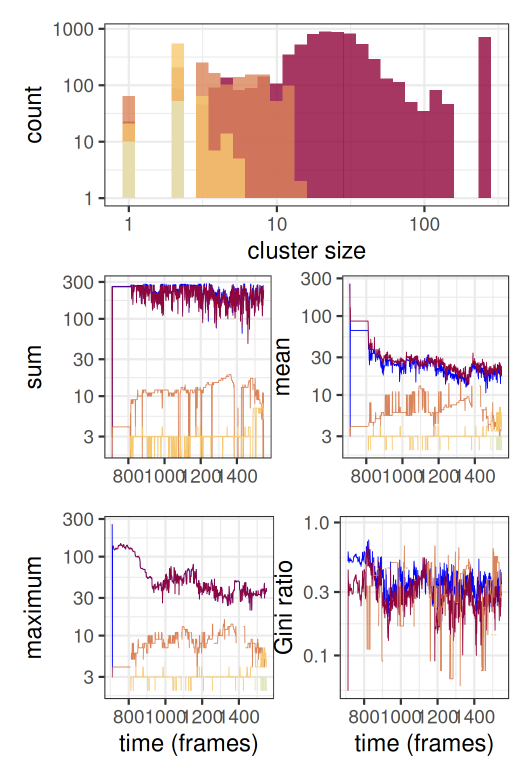}
\caption{Cluster size histogram (top row) and time-series of statistics (middle and bottom rows), for each level. The data consists of the cumulative count of cluster sizes over all frames for each level. Colors as in Fig.\ \ref{fig:p09_30_su_L_sum_mean}.  \label{fig:p09_30_CluSiz_hist_statime}}
\end{figure}

More precisely, the distribution of cluster sizes (i.e.\ their number of children, whether atoms or subclusters) is displayed in Fig.\ \ref{fig:p09_30_CluSiz_hist_statime}.
Ignoring the bar at size 256 (corresponding to the initial configuration of a single cluster of size $n_a=256$) and its possible bipartitions around 128, the first level cluster size histogram's right side is consistent with a power law of exponent roughly -2 in the approximate range $[20, 200]$ for first-level clusters, akin to the distribution of species within genera proposed by Willis (although he measured -1.5) \cite{Willis1922,Newman1997}, and of roughly -1 for second-level clusters (Fig.\ \ref{fig:p09_30_CluSiz_hist_statime}).
The statistics sum, mean, maximum and Gini ratio of cluster sizes at each frame and for each level are also shown (Fig.\ \ref{fig:p09_30_CluSiz_hist_statime}, middle and bottom).

\begin{figure*}
\includegraphics[width=1\textwidth]{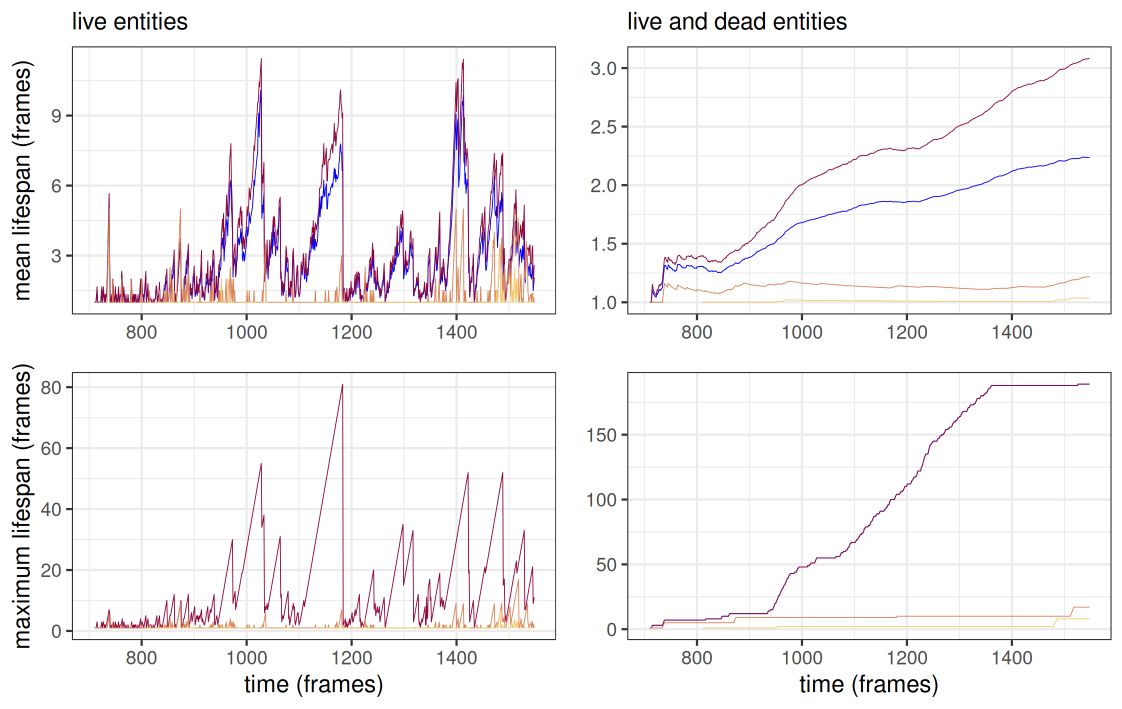}
\caption{Current mean (top) and maximum (bottom) lifespans for World $(-1,.09,.30)$ for only live (left) and both live and dead (right) entities. Colors as in Fig.\ \ref{fig:p09_30_su_L_sum_mean}. \label{fig:p09_30_L_cur}}
\end{figure*}

\begin{figure}
\centering
\includegraphics[width=.5\textwidth]{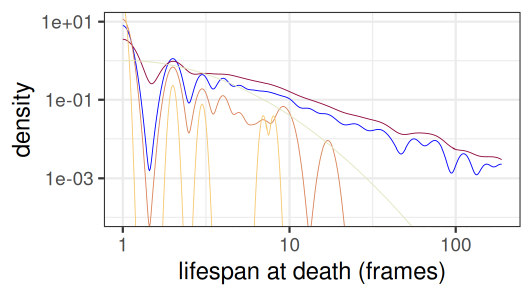}
\caption{Distribution of lifespans across all entities at the end of the simulation (frame 1549) on log-log scale for World $(-1,.09,.30)$. Colors as in Fig.\ \ref{fig:p09_30_su_L_sum_mean}. The curves are kernel density estimates, as implemented by default in ggplot2 package's \texttt{geom\_density} function \cite{Wickham2009}. \label{fig:p09_30_LTcum_dens} }
\end{figure}

The mean and maximum of current lifespans for each level are shown in Fig.\ \ref{fig:p09_30_L_cur} for live (left) and all (live and dead; right) entities.
Most top level entities have a lifespan of just 1 frame.
This is a common phenomenon also in many other simulations (e.g.\ Appendix \ref{app:moresims}).
The top entities are, at least at the beginning of the world, ephemeral because by the definition of entity (Section \ref{sec:HMGC}) if any of its children changes the entity ``dies'', in the sense that it becomes a different entity.

The temporal distribution of lifespans for live entities hints at a scale-free distribution of entity lifespans \cite{Sneppen1995a,Newman1997}.
Plotting confirms that a power law of exponent close to -2 seems to govern the distribution of lifespans\footnote{The contribution of live entities to the distribution of lifespans can be safely neglected, as evinced by Fig.\ \ref{fig:p09_30_LivDed_C_E}'s right column.} for all (having ever existed) entities at the end of the simulation (frame 1549) at least across two orders of magnitude for first-level entities and one order or magnitude for second- and perhaps third-level entities (Fig.\ \ref{fig:p09_30_LTcum_dens}). 

\begin{figure}
\includegraphics[width=1\textwidth]{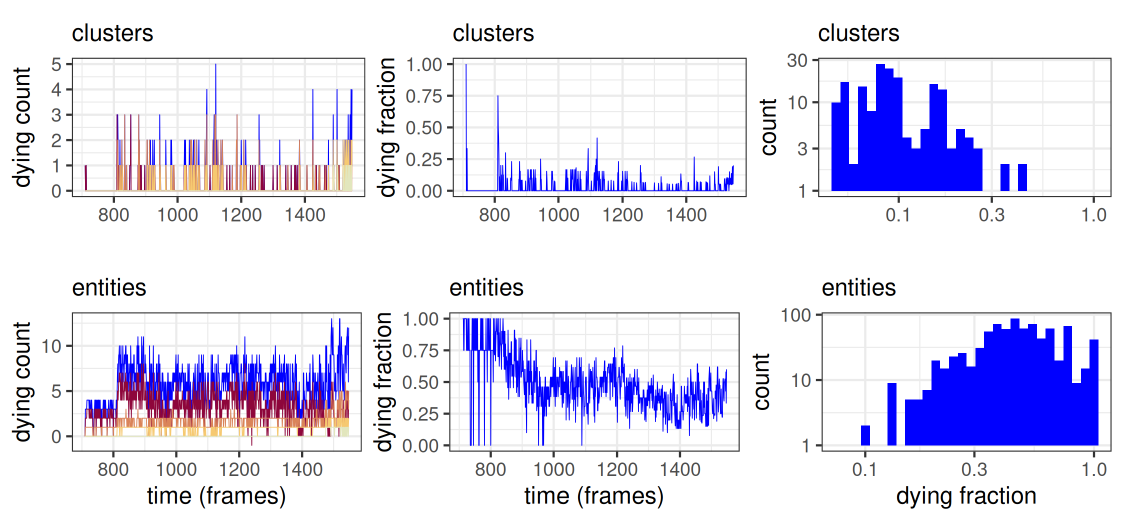}
\caption{Count (left column; for each level) and fraction (middle and right columns; dendron aggregate) of dying clusters (top) and entities (bottom) for World $(-1,.09,.30)$. Colors as in Fig.\ \ref{fig:p09_30_su_L_sum_mean}. \label{fig:p09_30_Dying_C_E}}
\end{figure}

The count of dying clusters and entities for each level at each time point is shown in Fig.\ \ref{fig:p09_30_Dying_C_E}'s left column.
Recall that a cluster dies when it loses all its children (Section \ref{sec:BUBD}).
Rescaling the count of dying clusters as the fraction of the total yields an intermittent pattern (Fig.\ \ref{fig:p09_30_Dying_C_E} top middle) analogous to extinction records \cite{Newman1997} with a comparable power law distribution \cite{Raup1991,Sole1996a}.
But it is the entities, as opposed to clusters, that are closest to the notion of a persistent being.
Nonetheless the interpretation of the dying fraction of entities for World (-1,.09,.30) is less straightforward (Fig.\  \ref{fig:p09_30_Dying_C_E} bottom), perhaps due to the relative paucity of data in this specific run of World (-1,.09,.30) precluding the expression of recognizable patterns.

\begin{figure}
\includegraphics[width=1\textwidth]{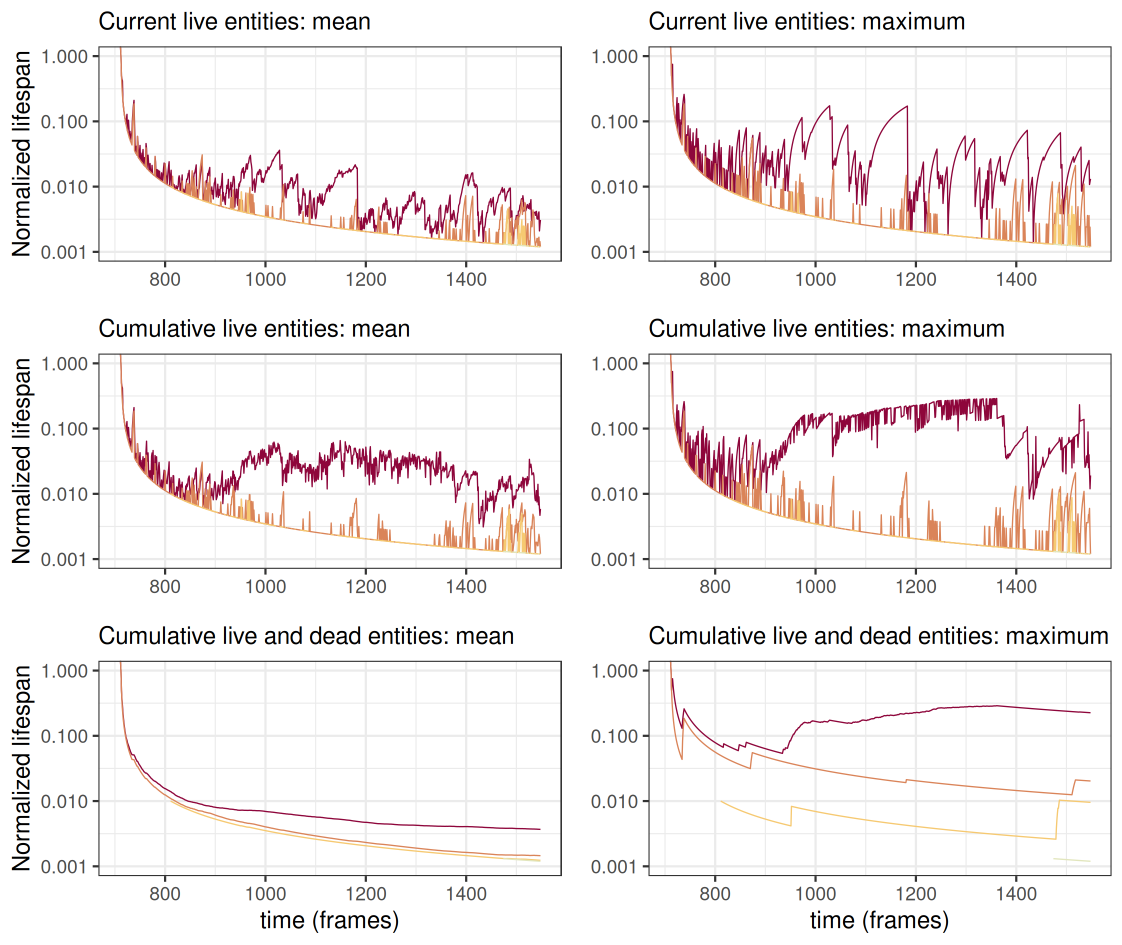}
\caption{Normalized live current (top), live cumulative (middle) and all cumulative (bottom) lifespans statistics (mean and maximum on the left and right) for World $(-1,.09,.30)$. Colors denote levels as in Fig.\ \ref{fig:p09_30_su_L_sum_mean}. \label{fig:p09_30_LN_cur}}
\end{figure}

Normalized lifespans (the lifespan divided by the time passed since the BUNCH algorithm start) afford yet another view of the precarious existence of entities.
Non-decreasing trends indicate persistence of entities.
In the present short run of World (-1,.09,.30), only first level entities are relatively long-lived judging by the level-wise mean and maximum as measured by both current and cumulative lifespans (Fig.\ \ref{fig:p09_30_LN_cur}). 

\begin{figure}
\centering
\includegraphics[width=.6\textwidth]{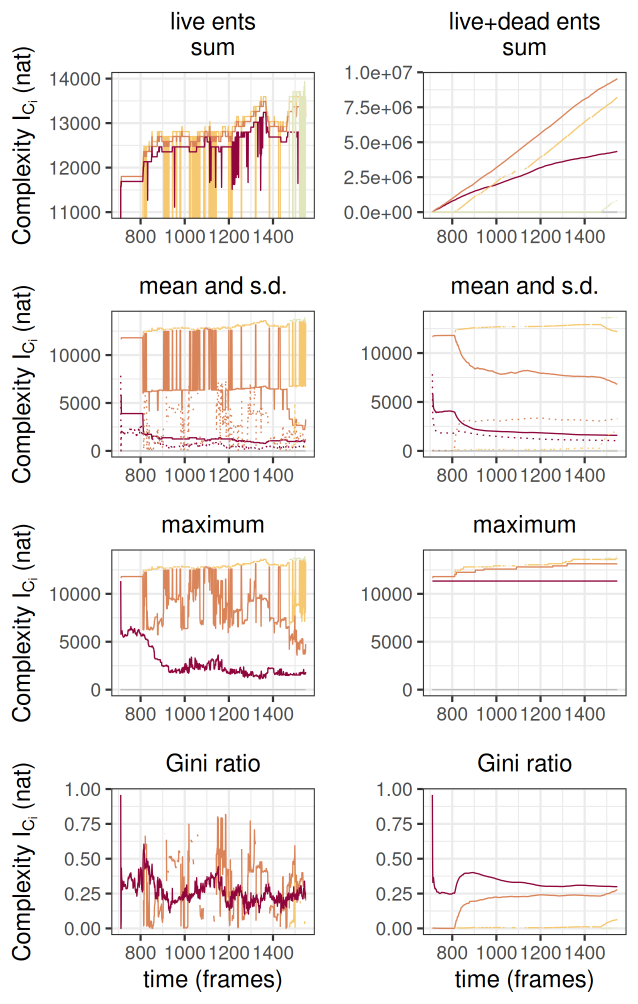}
\caption{Live (left) and all (right) entities complexity level-wise statistics. From top to bottom: sum, mean and standard deviation, maximum, and Gini ratio. The complexity of a given subdendron entity encompasses the information content of all its descendants. Colors denote levels as in Fig.\ \ref{fig:p09_30_su_L_sum_mean}. \label{fig:p09_30_LivDed_Ic}}
\end{figure}

Calculating the complexity of entities is in general computationally expensive (e.g.\ via autocorrelation functions, cf.\ Section \ref{sec:socblobs}), but BUNCH naturally affords multiple measures of complexity by simply counting the number of entities at each level.
The complexity or information content of the structure of any given subdendron $I_{D_{i,j}}$, in the spirit of the Schwarz information criterion, can be roughly approximated proportionally to the count of the free parameters of all its (proximal and distal) descendants via $I_{C_{i,j}}$ (cf.\ Section \ref{sec:modCL}).
Level-wise subdendron complexity $I_{D_i}$ statistics for currently live and all live and dead subentities are shown in Fig.\ \ref{fig:p09_30_LivDed_Ic}.

\begin{figure}
\includegraphics[width=1\textwidth]{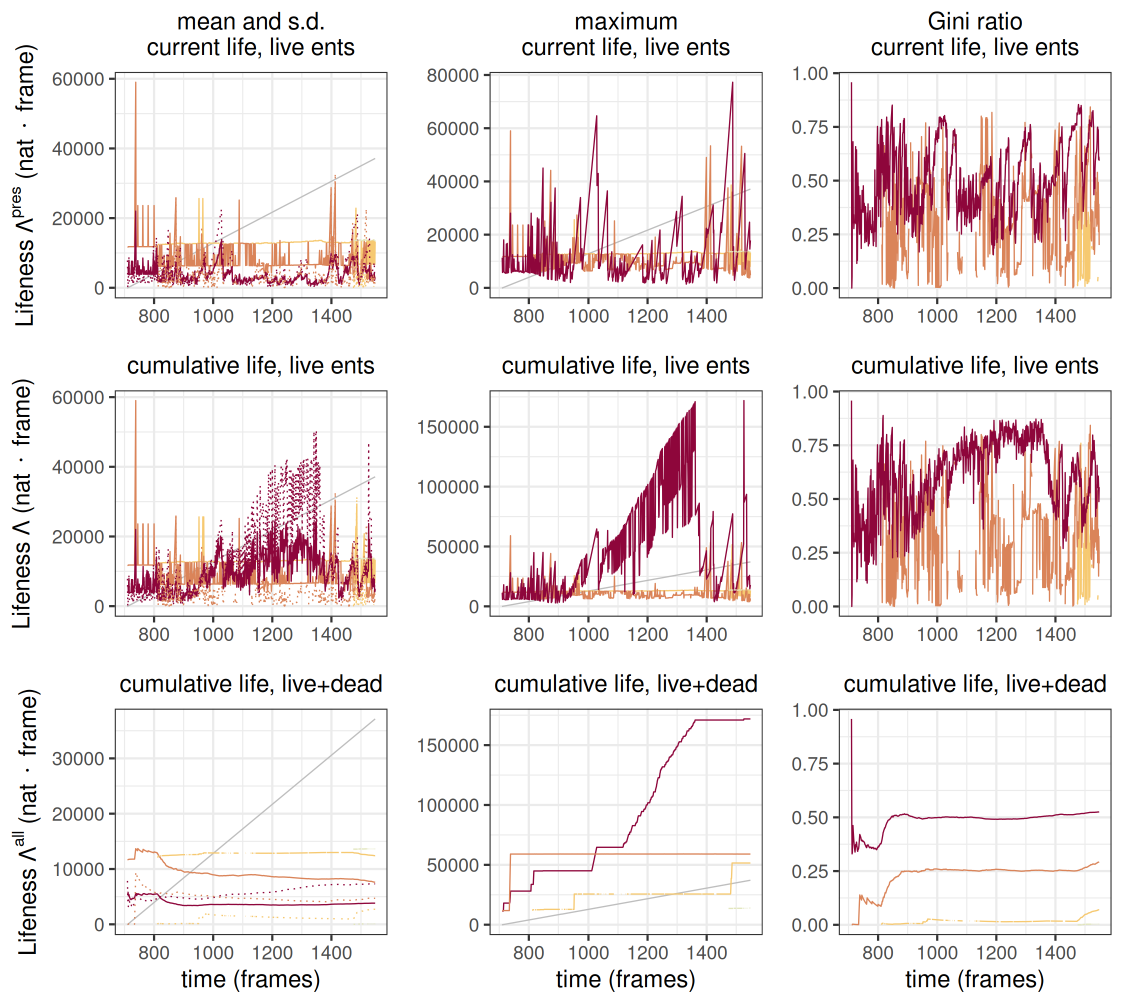}
\caption{Lifeness. Gray lines indicate the (reference) lifeness of any individual atom of $64 \log{2}$ times the frame count since BUNCH is started at 711. The other colors denote levels as in Fig.\ \ref{fig:p09_30_su_L_sum_mean}. \label{fig:p09_30_Lifeness}}
\end{figure}

Finally, the evolution of lifeness (Sections \ref{sec:lifeness} and \ref{sec:lifeness_eq}) over simulation frames is shown in Fig.\ \ref{fig:p09_30_Lifeness} for current live entities (top), cumulative live entities (middle) and cumulative all entities (bottom)\footnote{Although it has virtually no effect on the results, recall that here we use the multiple-instances definition of lifeness (Section \ref{sec:lifeness_eq}).}.

\begin{figure}
\includegraphics[width=1\textwidth]{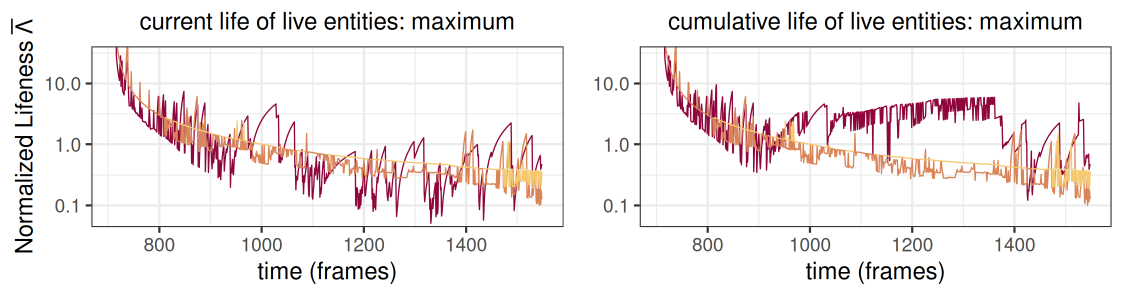}
\caption{Normalized maximum lifeness for the current life of live entities (top and middle rows of the central column in Fig.\ \ref{fig:p09_30_Lifeness}). \label{fig:p09_30_LifenessN}}
\end{figure}

The lifeness definition most relevant to lifelike behavior is likely $\Lambda$, which rests on cumulative lifespans of live entities (middle row in Fig.\ \ref{fig:p09_30_Lifeness}), because it accounts for an entity's lived time regardless of the lifespan being continuous or on-off.
Recall that here we consider that each entity instance contributes to the lifeness of that entity (multiple-instances definition, cf.\ Section \ref{sec:lifeness_eq}). 
Likewise, another statistic relevant to lifelike behavior is the maximum because the within-level distribution of lifeness is typically non-uniform or even power law for cluster sizes (cf. Fig.\ \ref{fig:p09_30_CluSiz_hist_statime}) and lifespans (cf. Fig.\ \ref{fig:p09_30_Dying_C_E}), especially when there is a large number of entities.
A measure of non-uniformity is the Gini ratio (Fig.\ \ref{fig:p09_30_Lifeness}, right column).
The normalized lifeness (lifeness divided by the lifeness of any individual atom, which is $64 \log{2}$ times the frame count since BUNCH started running) for the within-level maximum of current and cumulative live entities (corresponding to the top and middle plots of the center column in Fig.\ \ref{fig:p09_30_Lifeness}) is shown in Fig.\ \ref{fig:p09_30_LifenessN}, where non-negative trends indicate non-trivial lifeness increases.

\subsection{A stroll on the landscape of interacting particle systems: Looking for lifeness \label{sec:stroll}}

The vast parameter space and lack of analytical solutions preclude, at this stage, achieving a comprehensive understanding of which regions of the parameter landscape harbor lifelike behavior.

To illustrate the effect of the force range, sign, and symmetry, we plotted statistics for two representative IPS with $p=-1$, one with $p=-.333$ and one with $p=-2$ in Appendix \ref{app:moresims}.
World (-1,0,1) has purely skew-symmetric interactions, which yields a persistently chaotic motion of chaser-avoider pairs of atom kinds and no possibility of coalescence that precludes the amalgamation of persistent atom groups and hence the formation multi-level clusters (cf.\ \ref{fig:n17_0_and_0_1_hystlev}, right).
In contrast, World (-1,-.17,.00) has predominantly attractive and purely symmetric interactions which enabled the consolidation of cluster hierarchies of three and even fleetingly attained four levels (cf.\ \ref{fig:n17_0_and_0_1_hystlev}, left), similar to World (-1,.09,.30) (Appendix \ref{app:n1n1}).
By virtue of its strong short-range interactions World (-2,.24,.45) is a few rapidly scurrying particles within a grid of mostly static particles with punctuated velocity spikes (\ref{fig:n11_46_n.33_and_p24_45_n2_T_K_L}); contrastingly World (-.333,-.11,.46) has long-range interactions that result in a chaotic behavior of clusters of various sizes displaying complex and rapidly varying patterns (Appendix \ref{app:n.33n2}).
Other instances of IPS that we ran but do not show are World (-1,-.04,0), with purely symmetric and predominantly attractive forces, where a single star-like clump coalesces at the center with a few scattered slow atoms at the boundary.
In World (-1,.15,.63), a triad of loosely persistent clusters roams the box.
Predominantly static, wobbly frozen or solid-state-like structures arise in low-energy and predominantly repulsive systems.

After selecting a fixed representative set of 41 parameters $p \in [-3,0]$, with emphasis on -1 and its vicinity, we randomly sampled corresponding 41 configurations of IPS worlds on the $n_K^2 = 16$-dimensional space of interaction matrices $\chi$ and ran simulations.
The loose criterion for stopping the simulation was attaining and staying at NESS for a long enough timespan to collect useful statistics. The average of the number of frames simulated since the start of BUNCH $t_{B_0}$ was 1608. The average of the total number of frames simulated $t_{run}(\text{f})$ was 8235 because the frames running BUNCH were much slower to compute than those not running it (Table \ref{tab:simstats2} in Appendix \ref{app:statble}).
We collected relevant statistics such as $r_{\chi} \in [-1,1]$, $s_{\chi} \in [0,1]$, $d_{\chi} \in [-1,1]$, maximum attained level $L$, lifeness $\Lambda$, normalized lifeness $\bar{\Lambda}$, and normalized lifeness averaged over a simulation interval within NESS $\langle \bar{\Lambda} \rangle_l$ over the entities within level $l$ (Table \ref{tab:simstats1} in Appendix \ref{app:statble}).

At NESS, we expect most of the collected statistics to become stationary, such as temperature and surprisal, the number of live clusters, but also the number and complexity of distinct dead and live entities, the maximum attained cluster dendron level, and the  currently live clusters' (or entities') lifespan (which yields the lifeness $\Lambda^{pres}$).
However some statistics will not plateau or become stationary for a forbiddingly long timespan.
For example, maximum statistics in general are notoriously difficult to come by. 
More importantly, the number of distinct dead and live entities increases exponentially with the number of atoms and atom kinds, so in populous systems the probability that all possible entity types arise within a feasible simulation interval is extremely small.
Similarly, the mean or other aggregate statistics of currently or cumulative live \textit{and dead} entity lifespans will not start plateauing until all entity types arise.

On the other hand, the cumulative lifespan (associated with the lifeness $\Lambda$) should linearly increase with time \textit{for subentities that are lifelike} (i.e./ persisting), but it is likely to increase sublinearly for non-lifelike entities (i.e.\ entities that pop up and fade out intermittently by thermal chance).
This is reflected in the typical power law distribution of lifespans (e.g.\ Fig.\ \ref{fig:p09_30_LTcum_dens}), upon which any statistic of the lifespan distribution such as the mean will depend.
More trivially, the number of dead clusters will also increase linearly without bound.
By definition, the lifeness statistic is likely to bear similar behavior to the lifespan statistic.

What is the effect of the interaction force law range on lifeness?
For $d=2$ dimensions, $p=-1$ is a special power law because then the flux of a particular atom's interaction force field through any surface enclosing the atom is constant (Gauss's Law; cf. Section \ref{sec:simul}); in contrast for $p < -1$ it decreases with distance and for $p > -1$ it increases with distance. 
For $p > -2$, the tail of the interaction law has a characteristic length scale defined by its mean, and for $p < -3$ it also has a variance.
Hence large absolute values of $p$ tend to yield a multiplicity of isolated subsystems ---e.g.\ solid-state-like if repulsive as World (-2,.24,.45)--- where a mostly static lattice is occasionally strewn with local spurts of single atoms slung by the deep and narrow potentials (Appendix \ref{app:solidball}) in the neighborhood of other atoms' centers (Fig.\ \ref{fig:n11_46_n.33_and_p24_45_n2_T_K_L}).
In contrast, smaller absolute values ($-1 < p < 0$) increase the range and strength of interactions, until at $p=0$ the system becomes fully connected regardless of distance.

With regard to lifeness as defined in this paper, an important special type of IPS are frozen or solid-state-like worlds, wobbly lattices that may arise in predominantly repulsive symmetric systems.
This is because their nearly-static structures clearly yield a high lifeness that disagrees with its lack of lifelikeness.
This discrepancy is due to neglecting the time derivatives in the computation of lifeness: the BUNCH algorithm implementation described in Section \ref{sec:bunch} does not account for entity velocities, let alone accelerations, thus leaving out the dynamic aspect of lifelikeness.
This could be achieved e.g.\ by calculating velocity vector fields in the neighborhood of centroids.
Although extending the current implementation of the BUNCH algorithm to include the temporal derivatives of entity properties (e.g.\ location and blob size) is conceptually straightforward \cite{Friston2008,Friston2008b}, its computational burden would not be warranted by the scope of this paper.
If temporal derivatives were included in the computation of surprisals, lifeness would readily distinguish between solid-state-like systems and dynamically rich systems because the complexity associated with temporal derivative variables (e.g.\ velocity and acceleration) would outweigh the atemporal variables (e.g.\ blob position).
This readily would enabled distinguishing (subcritical) static complexity from (critical) dynamic complexity via the hierarchy of cluster velocities, along with locations.
This is akin to the difference between spatial and spatio-temporal autocorrelation functions.

Because of the vastness of the IPS configuration parameters landscape, finding meaningful patterns in it is troublesome.
Thus we focused on simple regression analyses for a few selected variables.
To test the effect of interaction force power law exponent and other IPS parameters on lifeness, we fitted one generalized linear model with a logarithmic link function for the range $p \in [-1, 0]$ and another for $p \in [-3, -1]$ (Tables \ref{tab:glm1}, \ref{tab:glm2}).
Since here we cannot distinguish static from dynamic lifeness we simply excluded from the analysis the solid-state-like systems occurring in repulsive symmetric IPS. These were the 3 systems World (-1,-.04,0), World (-1,-.17,0), and World (-1,.23,0) among the simulated 41 systems (Appendix \ref{app:statble}).

For IPS with $p \in [-1,0]$, which a priori is the long-range correlation domain favorable for life(ness), only $p$ for $l=3,4$ (the highest level) was predictive of higher lifeness (Table \ref{tab:glm1} in Appendix \ref{app:statble}): as $p$ approaches zero in $[-1,0]$ lifeness becomes higher.
We performed the same analysis but using a slightly different averaging of lifeness with $\mathring{\Lambda_l}$, which is as $\bar{\Lambda_l}$ except that $l$ indicates the ordinal rank of the level starting to count from the top level, which has rank 1 (as opposed to starting to count from the atomic level).
This allows to assess the lifeness at different levels referenced from the top, and its necessary because IPS attain different top levels, which in our case is either 3 or 4 (cf.\ Table \ref{tab:simstats1}). 
The results were analogous: only $p$ for $l=1$ was predictive of higher lifeness (the highest level) (Table \ref{tab:glm2}).
For IPS with $p \in [-3,1)$, $p$ was not predictive of higher lifeness (Appendix \ref{app:statble}).

We similarly analyzed how  IPS parameters affect surprisal $I_W$ relates.
More repulsion ($r_\chi$) was associated with higher surprisal for IPS with $p \in [-1,0]$, but not with $p \in [-3,-1]$.
Higher ratio of skew-symmetric to symmetric interactions ($s_\chi$) was associated with higher surprisal regardless of $p$, but oddly a higher absolute value of symmetric than skew-symmetric components was also associated with higher surprisal regardless of $p$ (Appendix \ref{app:statble}, Table \ref{tab:glm3}).

In short, these preliminary analyses suggest that in the range $p \in [-1,0]$ IPS worlds with interaction force exponent $p$ values closer to zero wer associated with higher lifeness at higher cluster dendron levels, which agrees with the notion that lifeness is related to long-range correlations.
Higher surprisal was associated with more repulsive interactions and with skew-symmetry but in a non-straightforward manner: while some skew-symmetry seems to be propitious to lifeness, too much is likely to be counterproductive.

\section{Discussion}

We have presented a quantitative definition of life called lifeness that rests on physical and biological arguments and suggests that lifelike structures may arise spontaneously in complex enough environments.
Next we surveyed the generative complexity of interactive particle systems, focusing especially on the role on non-reciprocal or asymmetric interactions.
Then we specified a particular implementation of BUNCH, an approximate algorithm that computes lifeness on the fly.
Finally, we ran numerical simulations to illustrate the viability of the BUNCH algorithm and the feasibility of computing lifeness for moderately large systems.

\subsection{Which worlds are (the most) livable? \label{sec:livability}}

The multiple configuration parameters of IPS, viz.\ atom kinds and quantity, reciprocity, interaction range, space geometry and topology, lead to a vast diversity of NESS systems that is difficult to scour for inhabitable or life-compatible regions.
However the following parameter statistics serve as beacons that illuminate the picture of the NESS that will emerge in the IPS simulation.

First, the interaction force range, parameterized by the power law exponent $p$, together with the particle density ---which we kept fixed here--- largely determine the resulting NESS type, between two relatively uninteresting extremes: strong interconnectivity leads to mean-field-like states where sometimes a generalized central limit theorem-like \cite{Levy1955} result yields a n asymptotic solution but only probabilistically, and contrarily for no interconnectivity we just get independent nodes.
By adjusting the interaction strength and frequency, a spectrum of phases from ideal-gas-like states ($p<-1$) passing through
diverging correlation length states ($p \approx 0$) and onto mean-field-like states emerge which displays a vast diversity also in analogous systems \cite{Perez1996}.
The exponent $p$ is one measure of distance to criticality (and thus of algorithmic complexity, cf.\ Section \ref{sec:critworld}) because it is associated with correlation strength. 
Another possible measure is susceptibility, typically defined as the derivative of an order parameter with respect to some external force field; however it is difficult to calculate near criticality \cite{Goldenfeld1992}.

Second, the ratio of repulsion to attraction $r_\chi$ largely determines whether the system will gravitate towards a predominantly attractive plasma-star-like clump of convecting particles or contrarily be prone to spread e.g.\ as a solid-state-like lattice layout of vibrating particles.
Beyond a certain value of $r_\chi$, the system will sunder out as isolated subsystems\footnote{The system becomes reducible, in Markov chain terminology.}.

Third, interaction reciprocity in many-body systems can be coarsely assessed via the ratio of skew-symmetric to symmetric interactions $s_\chi$.
Non-reciprocal or skew-symmetric interactions, which break the action-reaction symmetry, govern the steady-state dynamic properties of entities, as demonstrated e.g.\ by colloid coating symmetry-induced motion \cite{Soto2014} and diffusing mixtures, and can induce self-organized traveling patterns such as itinerant clusters of particles (cf.\ chaser-avoider duplets; Sections \ref{sec:nonrecipF}, \ref{sec:p09_30}) and chiral oscillatory patterns \cite{You2020} (Section \ref{sec:nonrecipF}).
Such effective interactions often are mediated by a non-equilibrium environment and are pervasive in biotic systems.
More generally, non-reciprocity may usher phase transitions to  time-dependent phases where spontaneously broken continuous symmetries are dynamically restored (Section \ref{sec:nonrecipF}); namely, NESS phases that self-sustain as oscillatory structures such as synchronization, flocking and pattern formation \cite{Fruchart2021}, which are organic to biotic dynamics.
Some degree of interaction non-reciprocity is likely to be essential to life (Sections \ref{sec:nonrecipF}, \ref{sec:stroll}, \ref{sec:livability}).
This is not because non-reciprocal systems display a complex NESS that eludes analytic description and may at best be treated probabilistically, since most chaotic Hamiltonian or reciprocal systems are ergodic and could be similarly treated \cite{Gallavotti1995a}, but because biotic systems need to replicate the inherent skew-symmetry of their environment.
Biotic environments typically evince nonlinear non-conservative feedback functions that in general preclude reduction to fluctuations around a single favored state or energy minimum \cite{Laughlin2000}.
In short, non-reciprocity is an essential feature of NESS associated with matter and energy flows.
 
Biotic skew-symmetry is for example reflected in the stress gradient hypothesis, which is an evolutionary theory stating that slow symbiosis or mutualism \cite{Bergstrom2003} is more common in stressful or resource-scarce environments, whereas resource-abundant environments induce rapid competition or parasitism \cite{Holmgren2010}.
This is illustrated by the snowflake yeast, a yeast mutant that develops clusters of clones \cite{Ratcliff2015}, developing a macroscopic multicellular life in a resource-scarce (anaerobic) environment \cite{Bozdag2023}.
Resource scarcity encourages thriftiness and zero-sum game playing (growth constrained by a divergence-free or also a contracting field of resources), which can be modeled by systems displaying higher clustering via specialization and symbiosis that engender super-entities . 
Contrarily, resource abundance (unconstrained growth) in general allows for some squandering and lends itself to less higher-order clustering and asymmetric conditions that grow from the initial surfeit of resources that encourage segregation, asunder goals and thereby competition and parasitism.

We found some evidence that longer interaction range is associated with more lifeness in a subcritical interval (while 
$-1 < p < 0$).
How can we reconcile this with the thesis (cf.\ Section \ref{sec:socblobs}) that short-range interactions are necessary for the integrity of a membrane or skin (Markov blanket) \cite{Friston2015c} and hence life?
This may be explained via a conceptual renormalization.
Life (on Earth) abides in a discrete spatial hierarchy \cite{Martinez-Saito2022c} across several orders of magnitude, from nanometers to kilometers.
Each discrete scale is characterized by a interaction range and strength that can be considered short-range without loss of generality; this explains the persistence of membranes or blankets at each scale.
Then the entities thus shrouded in their respective membranes may assemble to form larger superentities, and it is these larger superentities that interact between themselves but with a correspondingly larger characteristic range and strength, rescaled as a function of their new larger size.
This rescaling process occurs sequentially from the atomic level until there is not more free energy left to drive life (cf.\ Section \ref{sec:symbrkcasc}).
This explains that although characteristic short-range interactions occur within each level, the hierarchy keeps growing until it ``fills'' the NESS system that encloses it.
Further work will be needed to elucidate this issue.

Although all three factors are likely to by and large determine inhabitability, there was no evidence that asymmetry and the attraction-repulsion ratio were associated with lifeness.
We speculate that this is because we neglected modeling temporal derivatives, especially velocity, thereby losing the dynamical contribution to lifeness (Section \ref{sec:stroll}).

\subsection{The sweet spot is slight subcriticality \label{sec:L_crit}}

Much evidence suggests that living organisms typically thrive in pericritical (near critical) regimes \cite{Kauffman1991,Chialvo2006}, mostly in subcritical phases with intermittent excursions into supercriticality \cite{Friston2012a,Martinez-Saito2022c} (cf.\ Section \ref{sec:critworld} and \ref{sec:socblobs}). 
In statistical terms, the susceptibility of the entity to parameter variations or Fisher information is near a maximum at pericriticality, which affords sensitivity and fast adaptability \cite{Hidalgo2014}.

Our model of the world, the brain, may be typically poised in a slightly subcritical state, with occasional incursions into a supercritical state via self-induced instabilities ---analogous to SOC and chaotic itineracy--- that enable efficient and sensitive perceptual transitions \cite{Bonachela2010,Friston2012a,Priesemann2014}.

In general an adaptive mechanism is required for a system to steadily remain in a pericritical regime.
In networks of neurons or pulse-coupled nodes, this can be balancing the ratio of excitatory to inhibitory neurons to stay at the ``edge of chaos'' \cite{VanVreeswijk1998,Bressloff1999}; here coupling strength and connectivity degree would roughly match the $p$ exponent of IPS.

An example of a supercritical biotic model is Peskin's sinoatrial node (the heart's pacemaker) model of a population of pulse-coupled oscillators, which is arranged as a complete graph, i.e.\ all-to-all connected, and typically evolves toward synchronous firing \cite{Mirollo1990}. Similarly, some configurations of supercritical chains of oscillators may exhibit traveling waves \cite{Kopell1986}.
Restricting connectivity to e.g.\ local topologies displays a phase transition between incoherent and synchronous firing on lattices depending on dimensionality \cite{Daido1988} and on chains \cite{Kopell1990} and yields configurations closely related to slider-block models of earthquakes, which display self-organized criticality \cite{Hopfield1994,Strogatz2001} (cf.\ Section \ref{sec:critworld}).
Similarly, at the ecosystem level the links or relationships between species entities may be shaped by and for coevolution to optimize sustained fitness \cite{Kauffman1991}.
In Kauffman's (NK) ecosystem model, this optimal is found ``poised at the edge of chaos, i.e.\ at pericriticality \cite{Kauffman1991}.

Biotic systems can be likened to our particle simulations by matching entities to living organisms and interaction force power laws to simplified models of the more sophisticated long range interactions that pervade nature at macroscopic scales.
In both views, spontaneous coevolution of entities may induce dynamical instability and power law correlations \cite{Kauffman1991}.

\subsection{Metaselection: Stacking pericritical systems \label{sec:stacking}}

Interactions among entities occur not only within level, but also between adjacent levels in an embedding hierarchy.
Wright visualized evolution as the motion of a species (entity) on an \textit{adaptive} fitness landscape \cite{Wright1932}. 
Because species can mutually deform their fitness landscapes \cite{Kauffman1991}, the ensuing Darwinian dynamics \cite{Ao2008} cannot in general be described as a conservative system, but an interspecies potential function can be found that represents Wright's landscape \footnote{This is analogous to Type II complexity, which is used to describe neural variables considered as part of the whole brain, and Type I complexity, which ignores interactions with other parts of the (eco)system or brain \cite{Friston2000g}.} when defined together with  a skew-symmetric matrix \cite{Ao2008} (cf.\ Section \ref{sec:nonrecipF}).

This corresponds to a metadynamics of coevolution of multiple species that mutually deform each other's fitness landscapes (e.g.\ ruggedness and peak locations) such that Nash equilibria just tenuously form across the ecosystem \cite{Kauffman1991}; this metadynamics and all ecosystem parameters are also attracted to a peri-critical state \cite{Kauffman1991}.
Analogously, BUNCH finds or selects live entities, which are those that persist because their parameters enable them to.

At this poised state avalanches of coevolutionary changes unleashed by perturbations propagate with their size following a power law \cite{Kauffman1991}.
In BUNCH simulations power laws turn up for example in the distribution of (sub)entity sizes, with exponents between -1 and -3 for first- and second-level clusters (Fig.\  \ref{fig:p09_30_CluSiz_hist_statime}).
This reflects the distribution of the number of living biota genera within some species follows a power law, which evince an exponent typically around 1.5 \cite{Willis1922,Willis1922a,Newman1997}.
This phenomenon extends to higher taxa (ranks of the Linnean hierarchy) \cite{Newman1997} at least across protists, fungi, plants and animals in the form of limited fractal self-similarity with higher differentiation in marine than continental groups \cite{Burlando1993}. 

Metaselective mechanisms may manifest explicitly as external dynamic control variables that drive a subordinate system to hover about the limit between laminar (subcritical) and bursting (supercritical) phases in an on-off intermittent manner \cite{Platt1993} \footnote{This is distinct from Pomeau-Manneville scenarios, where intermittency occur for fixed bifurcation parameter values.}.
The driving variables may tether the system to pericriticality either by construction \cite{Dickman2000,Martinez-Saito2022c} or via self-organization \cite{Friston2012a}.

That pericritical systems may naturally stack hierarchically suggests that fine-tuning for abiogenesis may not be necessary: the frameworks of synergetics and MEP predict that driven NESS systems may spontaneously evolve toward complex self-organized entities by naturally filling all empty biotic niches at each discrete spatio-temporal hierarchy level \cite{Sharma2007} until pericriticality is reached \cite{Bonachela2009,Mastromatteo2011} (Section \ref{sec:synmepfep}), whereas highly optimized tolerance systems and FEP illuminate how such entities accomplish prebiotic lifelike self-organization and behavior \cite{Friston2013} (Section \ref{sec:socblobs}).

\subsection{Biotic complexity as a bottom-up cascade of spatio-temporal scales separated through symmetry breaking \label{sec:symbrkcasc}}

Biotic systems may spontaneously arise, abide in, and support pericritical environments (cf.\ Sections \ref{sec:critworld}, \ref{sec:synmepfep}, \ref{sec:socblobs}), but how do they accrue complexity across multiple spatio-temporal scales? 
When there is a scale dependence of the interaction forces, scale dependent structures or frustration appear: droplets (entities) made of particles (subentities) with a characteristic scale strew the system, but no drops or aggregates of droplets (superentities) coalesce \cite{Laughlin2000}.
However, most systems relevant to life are governed by laws that are symmetric or scale-free across several orders of magnitude.

The key mechanism might be an upward (bottom-up) emergence of information as opposed to laterally (within-level) diffusion of information.
The structure peculiar to each level of a complex hierarchical system in general may not be easily reconstructed from its sublevels ---supraordinate levels are \textit{different} from subordinate levels \cite{Andersen1972}.

Symmetry is broken by a combination of small scale fluctuations and exogenous driving.
No dynamically sustained pericritical state could be attained without an exogenous source of free energy: the constant presence of some external driving force is what maintains a cascade of free energy flows within and between weakly and hierarchically coupled discrete entities \cite{Andersen1972,Martinez-Saito2022c}.
Matter in supercritical state is ``fuzzy'', i.e.\ it behaves like a continuous variable in a ``stochastically predictable'' manner: a central limit theorem \cite{Levy1955} may yield a predictable Gaussian or stable distribution governing the behavior of the (ergodic) supraordinate level from e.g.\ atoms (a thermally noisy atomic level), or probabilistic laws may prescribe that (quantum) mechanics itself be non-deterministic. 
Contrarily, matter in subcritical state is locked into a mostly static frame (along with its information) which was not fully symmetric from the beginning.
But in general in pericriticality each new level may be attended by new complexity non-trivially arising from its subordinate level.
A large system typically has less symmetry than its governing laws \cite{Andersen1972}.
The information that determines these new structures typically is relayed ``upwards'' via a symmetry breaking mechanism initiated by random fluctuations from the subordinate level.
``At some point we have to stop talking about decreasing symmetry and start calling it increasing complication'' \cite{Andersen1972}.

The specific mechanism may be different for each level.
Barring intrinsic stochastic processes that produce information, no new information is actually created.
Instead some of the information contained in lower levels, such as fluctuations originating from initial conditions percolates upwards, akin to how critical phases preserve fluctuations (information) across spatio-temporal scales, whereas bulk phases erase them \cite{Goldenfeld1999}.
The origin of this information lies in the localized or discrete nature of matter or elemental particles (and hence of entities built upon them) and in the initial conditions of the system or cosmogony of information (the ``bit birth'')\footnote{Barring the intrinsic probabilistic nature of the most popular interpretations of quantum mechanics.}.
In some systems this information finds its way into and emerges as hierarchical structures \cite{Whitesides2002}---for example biotic systems store the (symmetry-breaking) information needed to model their environment so as to offset destructive stimuli.
Once a hierarchy is assembled, the interactions between successive levels can be construed as mutual \cite{Haken1977,Auletta2008} even if they evolved in a purely bottom-up manner (as in biotic morphogenesis \cite{Friston2015c}), or can be explained with a single unifying principle (descent on a variational free energy landscape \cite{Friston2012,Palacios2020}). 

Biological entities seem to flourish at the transitional interface between a continuous source of energy and a domain nearly depleted of available energy, such as submarine volcanic vents or the ``warm little ponds'' \cite{Darwin1858} of coastal intertidal shallow lagoons of primordial soup.
The processes that enable the bottom-up emergence of (biotic) information in sequential order are likely to require a continuous slow input of free energy \cite{Bak1987} constituting the gradient or non-dissipated excess energy that both induces the genesis of higher levels of complexity (due to the input energy enabling the exploration of energetically expensive configurations \cite{Sharma2007}) that incorporate the new entities taking form and  dissipates (relaxes) the free or excess energy at each discrete level or scale.
At each level the new emergent entities stabilize when they cannot find ways to dissipate energy more effectively \cite{Prigogine1971,Haken1977,Martinez-Saito2022c}.
The result is a hierarchy of NESS (eco)systems.
Finally, the top level is defined by it dissipating all remaining free or available energy. 

Lifeness is the product of the complexity ultimately originated in symmetry breaking across hierarchy levels \cite{Andersen1972,Martinez-Saito2023} and its ability to persist.
The highest biotic complexity on Earth is observed in multicellular organisms.
Cells are the key subentities constituting multicellular entities.
But cells are also clusters of their subentities.
Cellular functions are carried out by subentities made up of many species of interacting molecules (clusters of atoms) \cite{Hartwell1999} in pericritical networks.
For example, cellular signaling pathways are arranged in complex networks exhibiting emergent integration of signals across multiple time scales and self-sustaining feedback loops \cite{Bhalla1999} that generate mass, energy, information transfer and cell-fate specification \cite{Jeong2000}.
The topological scaling properties of metabolic interactions among cellular constituents and reactions may emerge from the same blueprint for the organization of complex non-biological systems \cite{Kuchling2020,Palacios2020}.
Thus life may have evolved from the non-biological scale-free complexity widespread in nature, with the addition of some mechanism or engine driving it. For instance, an active matter version of liquid crystals (a state of matter between solid crystal and liquid) that spontaneously assembles hierarchically can be created by putting together biological molecules that are capable of converting chemical energy into motion \cite{Sanchez2012}. 
Turbulence is a striking example of exogenously driven non-biological hierarchical complexity with multiple spatio-temporally separated scales, from a top level of the largest scale from where energy is input, through several levels of (Taylor microscales) of energy cascading down to increasingly smaller eddies till the smallest, last, viscous scale where energy dissipates or free energy vanishes \cite{Davidson2015}.
Other examples are diffusion limited aggregation \cite{Charbonneau2017}, earthquakes \cite{Olami1992} and solar flares \cite{Lu1993}.

Thus biotic and non-biotic entities in NESS may accrue complexity through similar fundamental mechanisms: a bottom-up process of entropy proliferation (Section \ref{sec:synmepfep}) that is whittled or hacked at by a ``top-down'' process of natural selection or culling (Section \ref{sec:socblobs}) resulting in persistent complexity. 
The same idea can be paraphrased in terms of synergetics \cite{Haken1977}: the entities that are best at consuming free energy gradients (fittest) are the ones that survive and displace the rest.
The dissipative nature of driven NESS biotic systems works not only at the lowest scale ---as in the viscous scale of turbulence--- but at all levels.
For example, at the level of multicellular organisms the cooperation among their cells for development, maintenance and reproduction is subject to destructive fluctuations or forces ---not unlike the thermal fluctuations dissipate the kinetic free energy of macroscopic entities such as vortices in turbulence--- of which cancer is a conspicuous phenomenon \cite{Green2024} of cheating within this cooperative multicellular entity \cite{Aktipis2015,Kuchling2020}, a ``reversion to an ancestral quasi-unicellular phenotype'' \cite{Lineweaver2021}.

The exogenous force driving a NESS may build hierarchies not only in space but also in time.
For example biological species or more generally taxa's structure and evolution.
``The transition from unicellular to multicellular life is the paradigm case of the integration of lower-level individuals (cells) into a new higher-level individual---the multicellular organism'' \cite{Herron2009}, and it has occurred independently multiple times, e.g.\ in red algae, brown algae, land plants, animals, and fungi \cite{Bonner1999} likely when individuals from the same and/or different species specialize to accomplish different tasks in building an econiche (Black Queen hypothesis \cite{Morris2012}).
A compelling example of individual cells coalescing into a higher multicellular organism through evolution is the volvocine algae, which resemble snapshots of an evolution toward multicellularity starting 200 million years ago from a creature like the unicellular flagellate green algae Chlamydonomas, clustering into colonies like Gonium (4-16 members), Pandorina (8-32), Eudorina (16-64), Pleodorina (32-128), and finally Volvox (up to 50000) \cite{Herron2009,Jimenez-Marin2022,Green2024}.

The tree of life was a metaphor of Darwinian evolution originally envisioned as a temporal branching dendron of phylogenies \cite{Darwin1859}, but it may well be generalized to the spatio-temporal domain by simply considering all subentities of entities as bearing lifeness.
This multiscale dynamics of taxonomic hierarchies not only  may be an expression of pericritical evolutive processes \cite{Sole1996a} but more generally of the dynamics of coupled systems constituted by organisms and their ecological niches, which coordinate patterns across large spatio-temporal scales \cite{Ramstead2018,Ramstead2019}.

\subsection{Estimating the lifeness of biological entities \label{sec:bioL}}
 
By regarding ecosystems and biotic organisms as cluster dendrons of finite scale invariance ---once the scale of living organisms is reached, subclustering proceeds all the way down till the microscopic level \cite{West1999}--- BUNCH furnishes an approximate yet feasible method to quantify lifeness or the product of algorithmic complexity and persistence.
In our simulations we noted that the within-level lifeness statistics of mean, maximum, and Gini coefficient unveil complementary aspects of each of the subdendrons' lifeness (Section \ref{sec:p09_30}). 
Most of the simulations lacked the complexity and length necessary to evolve entities that evinced maximum lifeness at levels close to the top (e.g.\ Figs.\ \ref{fig:p09_30_Lifeness}, \ref{fig:n11_46_n.33_and_p24_45_n2_Lifeness}), but in general we expect that maximum lifeness is to be found somewhere in the middle (way \cite{Laughlin2000}) but closer to the top than the bottom (cf.\ Section \ref{sec:p09_30}).
We expect that if we applied the BUNCH algorithm to some region of Earth's surface the cluster or entity holding the highest lifeness would belong to a multicellular living organism, which most likely will not occupy the highest level of the dendron, but a level not too far from it.
The top entity would enclose the whole universe, thus aggregating a jumble of children subentities and likewise down the hierarchy, where most of the subentities would display intermittent lifespans to various extents.

For example, consider the lifeness of a specific RNA segment entity.
We would need an estimate of its cumulative lifespan and its algorithmic complexity.
We can assume the RNA world hypothesis that RNA was the key self-replicating molecule that enabled abiogenesis some 4 billion years ago \cite{Neveu2013}.
The algorithmic complexity of the entity can be derived from our knowledge of molecular biology that it is a chain defined by some specific sequence of four nucleotide bases.
In turn each of the nucleotide bases is itself a persistent subentity constituted by multiple atoms in a specific spatial configuration.
Although physical atoms are not atomic in the sense of elemental particles, they behave as such in biotic systems.
Hence the three levels of the dendron entity would be constituted by atoms, nucleotide bases, and RNA respectively, and we can calculate the dendron complexity of our RNA segment using the formulas of Section \ref{sec:modCL}.
Finally we obtain our RNA segment's lifeness by simply multiplying the estimated lifespan and dendron complexity.
Note that here we are assuming that this specific segment did not become extinct and then reappeared, i.e.\ its lifespan is uninterrupted, and we are using the one-instance definition of lifeness (see \ref{sec:lifeness_eq}).
Finding multiple-instance lifeness would require further multiplying by the estimated number of copies of our RNA segment entity.

\subsection{Are ``aliveness'' and purposefulness (teleonomy) intuitive expressions of persistent complexity (lifeness)?  \label{sec:alivelife}}

The \textit{subjective} quality of looking alive or lifelikeness and its dichotomy is an expression of our intuitive assessment of whether something is alive or not.
Far from just being an amusing ability, the skill of telling apart live from dead holds clearly fitness advantage.
But how do we do it?

Our chain of thought is typically poised on a fickle interface between randomness and constancy, and we tend to look on neither uniform randomness nor determinism as very interesting.
This predisposition reflects both the unstable dynamics of natural coevolution \cite{Kauffman1991} and the intrinsic volatility of the world we live in \cite{Chialvo2006} that engendered it in the first place.
One such interesting object of attention are lifelike creatures, which are associated with e.g.\ long-range correlations, purposeful (teleonomic) and adaptive behavior or motion, and persistence.
Teleonomy is the apparent purposefulness of lifelike entities, that are brought about by natural selection---while it manifests on the level of organisms, evolution itself is not teleonomic \cite{Mayr1961}.
Persistent entities are teleonomic: they seem to be trying to survive but in actuality they simply exist because their behavior enabled the persistence of their ancestors in the same environment that they inhabit, so their behavior can be construed as top-down effects in hierarchies of biotic entities \cite{Auletta2008,Palacios2020}.
The first-order perspective of teleonomy corresponds to free will as the apparent willed purposefulness of our own behavior \cite{Martinez-Saito2023}.

Thus \textit{ceteris paribus} we expect higher lifeness to strongly correlate with lifelikeness via the conjunction of complexity and lifespan.
The complexity of living organisms may seem a ``source of causal chains which cannot be traced beyond a terminal point because they are lost in the unfathomable complexity of the organism'' \cite{Elsasser1987}.
However, human perception is highly sensitive to the lifelikeness of random objects and groups thereof, through summary statistics \cite{Leib2016}.
Mid-level features such as second-order statistics, shape or texture, as opposed to high-level features, are the basis of lifelikeness (or animacy) perception for both objects and clusters of objects \cite{Markov2021}. 
Hence human perception is likely to use a simple model of lifelikeness perception that enhances fitness just enough to be useful \cite{Martinez-Saito2023}.

As a subjective validity check, here I judged the entities having more lifeness to be correlated with lifelikeness.
But arguably Particle Life's implementation \cite{Abdulrahman2022} of IPS is perhaps more lifelike, which is likely to be due to: (1)
removing altogether collision (no solid balls, Appendix \ref{app:solidball}); (2) fixing $p=-1$; (3) ignoring viscosity, i.e.\ $\zeta=0$; (4) setting a maximum  interaction range, typically much smaller than the bounding box; (5) assuming the time step $\delta t$ between frames is the unit (Section \ref{sec:IPS}); (6) not storing simulation statistics, such as those required for plotting their temporal evolution; (7) not plotting on the fly (for the JavaScript app, cf.\ Appendix \ref{app:software}); this configuration improves performance and enables simulating thousands of particles.
For reference, our settings of $\zeta = .7$ and $p=-.1$ was roughly similar to Particle Life's range of 80 and $p=-1$.

\subsection{Living entities are what persists in non-equilibrium: good enough mirrors of their environment}

We asked in Section \ref{sec:critworld} why our world is complex and critical.
Although somewhat unsatisfactory, the most straightforward answer is that it is the sort of (and perhaps the only) world where you (or someone) could ask such question (anthropic principle) \cite{Carter1983,Martinez-Saito2021}.
Roughly, this is because to ask that, a subject needs a complex model of the environment \cite{Conant1970}, which, crucially, is necessary only to the extent that the world it inhabits is complex and hence non-trivial to predict or calculate and act upon.
That is, the typical inhabitants of a pericritical world are both imperfect mirrors of their world and lifelike (as we perceive them) entities.

A pericritical NESS system is the foundation of life as persistent complexity, but it is not enough to ``ferment the prebiotic broth''---complexity \textit{per se} \cite{Edlund2011} is expedient to fitness insofar as it reflects the complexity of the world.
Fortunately ---similarly to how sauerkraut or kimchi are fermented by lactic acid bacteria found already on cabbages--- the missing ingredient may be already found in the broth: the property of subentity of revising its characteristic phase space domain or attractor \cite{Winfree2001,Friston2013}.
This is accomplished through an approximate model or mirror of the world, which is necessary because a NESS world is dissipative, so clusters or transient structures tend to wander away to never return back to themselves ---a ``thermal death reign'' sucks in and kills persistent entities.
The ability to avoid this fatal region, via the ``burning of biotic fuel'' or consumption of free energy, defines life.

Not all persistent entities need to possess an explicit model of every component of their action-perception chain: e.g.\  for some organisms just stochastic switching can replace perception when the environment changes infrequently \cite{Kussell2005}, and some microbes adopt probabilistic diversification strategies upon entering poorly identifiable (sensed) new environments \cite{Wolf2005}.
However a persistent entity must typically comprise a good enough model or ``mirror'' of its habitat \text{and} itself in order to identify noxious exogenous forces and thereby act to avoid death by reconfiguring its internal milieu and boundary \cite{Varela1974}. 
The entity's model of itself will in general be much simpler yet good enough to persist than the entity itself \cite{Martinez-Saito2023}.
In the process of trying to  avoid death by predicting and acting on their environment (minimize surprisal through variational free energy \cite{Friston2006}), persistent entities not only offset the dispersive forces outside their bodies but also inside them \cite{Friston2012}, where the difference between them is just a matter of external action (on the space outside the entity) versus internal signaling (on its subentities).
Thus it ensures furthering its own existence as an entity distinct from the environment, i.e.\ a self-sustaining perturbation or ``biotic soliton'', that has built-in mechanisms to cancel out the dissipative forces \cite{Friston2006}, from the microscopic scale of xenobiotic metabolism \cite{Becker2011} via a mesoscopic scale of the immune system to the macroscopic scale of behavior \cite{Friston2010} and perception.
Similarly, the BUNCH algorithm implicitly (lazily) selects with highest probability the trees (cluster dendrons) that bear the highest (HMGC) model evidence.
This information or algorithmic complexity integrated over time yields an averaged log-evidence (lifeness).
BUNCH emulates evolution because it culls out low evidence clusters in an all-or-none fashion (since cluster membership is a binary variable; cf.\ Appendix \ref{app:bellnumber}). 

All subentities of a persistent entity must have been dealt a role in the fitness of their parent entity. For instance, vertebrate eyes are high precision optical devices that adaptively tune their axial length in response to cornea and lens refractive power \cite{Grosvenor1987}, image blur and focal plane location \cite{Meng2011}, during all the lifespan of the device.
Hence that myopia can be caused by near-work (hyperopic defocus) \cite{Kepler1611,Goldschmidt1968,Goss2000} and worsened by wearing glasses \cite{Wildsoet1997,Sun2017} but can be similarly reversed through natural lifestyle such as outdoor activity \cite{Wu2013} and ecologically congruent eye accommodation \cite{Brown1989,DeAngelis2008,Becker2014,Steiner2022,Delshad2020}.
Overall living entities and their subentities can be construed as a physical embodiment of automatic model selection \cite{Friston2011b}: only the best models remain.
The BUNCH algorithm mimics this selective mechanism in a simplified fashion.

But unlike typical biotic systems, the prebiotic entities identified by BUNCH are unlikely to be ergodic due to their simplicity, in turn a consequence of the relatively small number of atoms and duration of the simulations.
A persistent entity in a complex enough world would not only need (and has \cite{Friston2013}) a built-in model of its habitat \cite{Conant1970} but also exhibits adaptive behavior ---adaptive active inference \cite{Kirchhoff2018}--- embedded in a closed dynamic action-perception loop of circular causality that inextricably and bidirectionally interlocks the entity and its environment's structures \cite{Friston2012}.
These coupled systems are the ones that were represented at the inception of evolutionary systems theory as an own ever-warping Wright fitness landscape for each entity or via Fisher's Fundamental Theorem \cite{Frank1992,Ao2005}.
Such (living) entities are able to make inferences about distant future states in order to preserve themselves via homeostasis and more generally action, which endows them with autonomy in a complex world of limited predictability, all the way from the top level to the atomic level entities. 
But here we find non-autonomous entities, which are unable to anticipate distant future states and instead reflexively (shallowly) react to current stimuli because they lack a (deep temporal) model of the world \cite{Friston2006}.
This is also denoted as mere active inference \cite{Kirchhoff2018} and such entities may be rather thought of as subentities dependent on an enclosing entity upon which their persistence depends.

Likewise the entities found here lack a well defined boundary, which entails that their Markov blankets \cite{Friston2013} cannot be readily identified as specific subentities.
In contrast, in nature biotic systems exhibit Markov blankets than can readily be identified as e.g.\ membranes, epithelial tissue or skin.
However, we could have devised a more sophisticated system mimicking biotic mesoscopic membranes.
Surfactants are amphiphilic molecules that create emulsions (typically oil in water) where micelles of oil droplets are suspended in water.
Lipid bilayers can be modeled analogously, hence liposomes and cell membranes too.
We could have replicated these arrangements e.g.\ with three elemental particles corresponding to water, oil, and surfactant, with an interaction matrix where water and oil are repelled between species but attracted to each other and the surfactant is mutually attracted to both.
This system may display its own idiosyncratic (critical) phase transitions between regular periodic and aperiodic random fluid-like phases \cite{DeGennes1982}.

Moreover, here the size of the simulated IPS entailed a typical maximum dendron height of 4 levels.
Typically many thousands of frames passed before the dendron grew to the top fourth level (Figs.\ \ref{fig:p09_30_hystlev} and \ref{fig:n17_0_and_0_1_hystlev}).
In general the organic emergence of higher order structures may take a long time.
This is likely to have been the case on Earth, assuming abiogenesis occurred on it: from its coalescence it took $\approx 5 \cdot 10^8$ years for the earliest self-replicating and self-assembling entities to come about; from them to the advent of Ediacaran biota (the earliest known complex multicellular organisms) and the subsequent Cambrian explosion of life diversification another $\approx 2 \cdot 10^9$ years.

\subsection{Death as a way to prolong life}

The origin, proliferation and extinction of species in biological ecosystems is analogous to the slow driving input, spreading and avalanches, and fast relaxation of energy in self-organized critical systems (Section \ref{sec:critworld}).
The BUNCH algorithm, at each level and time step, lazily and stochastically spares the lives of fit clusters or entities, gives birth to new fitter entities by fusing subentities that ``work'' better together than apart, or vice versa splits them, kills the less fit entities, and caps the dendron hierarchy with a new top entity when the top level is not simple enough (Section \ref{sec:bunch}).
Although an expression of model fitting to evolving estimated clusters rather than an exactly defined evolution of entities, the resulting fitting dynamics are similar to those of other self-organized critical models of sequential coevolution, such as Kauffman and Johnsen's NK model, where avalanches of fitness changes or mutations \cite{Kauffman1991} may trigger extinction and speciation events due to low fitness---this is exactly analogous to how biological perceptual transitions are enabled by self-induced instability \cite{Friston2012a}.
This is also a lazy stochastic form of extremal dynamics, an evolution rule that removes unfit entities by at each time step culling the least fit \cite{Bak1993,Sneppen1995a}, which is one of the mechanisms resulting in self-organized criticality \cite{Dickman2000}.
Similar culling schemes, that ignore coevolution, are at each iteration stochastically perturbing the interspecies interactions and culling those with negative fitness \cite{Sole1996c}; or establishing a stochastically fluctuating environmental stress that sets a ``coherent'' threshold, the entities whose fitness is below it become extinct\cite{Newman1996a,Newman1997}.

This culling scheme is replicated at multiple levels of biological organization whenever there exists a well-defined unit of selection or evolutionary information: from species or genera (taxa) via the phylogeny of subpopulations (group) \cite{Ramstead2018}, asexually reproducing cells, and genes \cite{Dawkins1976} to self-replicating molecules.
Note this scheme does not translate well to a sexually or post-mutation reproducing organism because it does not constitute a unit of selection \cite{Lewontin1970}.

Power law distributions turning up in multiple statistics in a particular system may be an indication of pericriticality. 
For instance mass extinctions may be caused by avalanches of  coevolutionary fitness that manifest as power law distributions of extinction events \cite{Newman1997}.
Bak-Sneppen's \cite{Bak1993,Sneppen1995a} model of punctuated equilibrium (periods of stasis alternate with avalanches of evolutionary change) in the macroevolution of an ecology of interacting and interlocked \cite{Kauffman1991} species uses a self-organized critical interpretation of power laws for taxon lifetimes and extinction sizes.
In our simulations of pericritical IPS worlds, the cumulative lifespan of subentities typically evinces a power law distribution with exponent between -1 and -2, especially for the lower dendron levels (e.g.\ Fig.\ \ref{fig:p09_30_LTcum_dens}, \ref{fig:n11_46_n.33_and_p24_45_n2_LTcum_dens}).
This range is consistent with both the exponent of $-1.9 \pm 0.1$ 
estimated over the Phanerozoic eon for genera lifetimes \cite{Raup1986,Sneppen1995a}, and the fluctuating stress model exponent of $\approx -1$ \cite{Newman1997} and our simulations.
The distribution of entities lifetimes in a pericritical system could also be related to the distribution of laminar phases in systems exhibiting on-off intermittency, which have a nearly universal asymptotic exponent of -3/2 \cite{Heagy1994}. 

Concerning (cluster or entity) membership, fossil data indicates that for some clades the number of genera of living biota follows a power law with an exponent $-1.7 /pm 0.3$ \cite{Willis1922,Newman1997,Newman2005}.
We also found inklings of power law distributed (sub)cluster sizes in a limited range.
Relatedly, there is a correlation between speciation (as number of competing genera) and age-independent extinction rates in most higher taxa, which may be explained with the Red Queen hypothesis that coevolving taxa engage in an arms race where they must constantly adapt to equally constantly adapting competitors \cite{VanValen1973}.
 
Fossil data from the extinction of Paleozoic and Mesozoic
marine invertebrate families \cite{Sepkoski1993,Raup1986} has been fit in agreement with a power law, distribution of exponent $\approx -1.95$ \cite{Sole1996a,Newman1997}. 
A similar analysis on the fraction of species killed yielded an
exponent of $-1.9 \pm 0.4$ \cite{Raup1991,Newman1996}.
Our simulation results are also compatible with this interpretation (cf.\ Fig.\ \ref{fig:p09_30_Dying_C_E}, top right) and with the fluctuating stress model's exponent of $\approx -2$ \cite{Newman1997}.

The time series of extinction sizes over the Phanerozoic \cite{Sepkoski1993} is qualitatively consistent with the time series entailed by Bak-Sneppen's self-organized critical model of evolution \cite{Sneppen1995a} and our simulations (cf.\ Figs.\ \ref{fig:p09_30_Dying_C_E}, top middle; and Figs.\ \ref{fig:n11_46_n.33_and_p24_45_n2_Dying_C_E}, \ref{fig:n17_0_and_0_1_Dying_C_E}), and displays a power spectrum that approximately follows an inverse power law \cite{Sole1997}.
Other death-related phenomena showing similar statistics of frequency-sizes are forest fires and wars \cite{Roberts1998}.

However many other mechanisms, such as preferential attachment and multiplicative stochastic processes, can explain power laws \cite{Sornette2000,Mitzenmacher2004,Newman2005}.
For example, if a taxon gains and loses species over time like a random walk and it becomes extinct when the number of species reaches zero (the gambler's ruin; one type of birth-death process) then the time for which taxa live should have the same -1.5 power law distribution of the first return times of random walks. The similar yet less realistic Yule's birth process yields power exponents below -2 \cite{Yule1925,Newman2005}.

So what is the role of death in this scheme?
From the perspective of BUNCH, (cluster) death is just an expression of (lazy stochastic) model selection---a transition in the space of cluster dendrons fitness or model evidence landscape.
Here selected entities persist only if they conform well to a HMGC explanation of the particle configuration.
Although somewhat misleading, here there is an analogy on the one hand between entity HMGC models and a biotic species, and on the other hand between the IPS configuration of particles and the ecosystem embedding a biotic species. 

In phylogenetics death takes a similar form: taxa undergo speciation and (with or without leaving descendants) extinction processes that mimic the creation and elimination of particles characteristic of mutation-selection genetic algorithms or particle filter methods \footnote{Unlike in Monte Carlo and Markov chain Monte Carlo methods, these (mean-field interacting) particles interact with empirical (simulation) measures of the population of particles \cite{DelMoral1997}.}, where sequential interacting samples (e.g.\ species, organisms, clusters) simultaneously explore and deform the fitness space they inhabit \cite{DelMoral1997}, just as in coevolutionary ecosystems (cf.\ Section \ref{sec:stacking}).
Thus here death is similarly a way to cut short unpromising search paths in the fitness landscape.

Death in ontogenetics corresponds to the common sense of death: programmed obsolescence (senescence) of mortal old biotic organisms, with or without progeny.
Senescence is an expression of (epigenetic and genetic) information loss: epigenetic drift (faulty gene expression modulation) and genetic mutations correlate with aging or the biological clock across mammals \cite{Mendelsohn2017,Yang2023,Bertucci-Richter2023}.  
This sort of death is different from phylogenetic death (extinction) if the organism reproduces sexually or asexually with mutations because then the organism is not anymore the unit of selection; then genes are. 
Anyhow death here also reignites exploratory dynamics by pulling out old (past) organisms and putting in new (future) organisms with some perturbed (randomized) phenotypes.
In other words, mortality affords evolution with a stochastic component that enables it not to become stuck in suboptimal minima, thereby increasing the survival odds and prolonging life, at the cost of losing some information \cite{Hinton2022}.

Death is (a particular form of) information loss.
An important factor that may modulate this information (and hence complexity) loss is horizontal information transfer, i.e.\ information shared among organisms other than through reproduction.
Unicellular organisms achieve this by transferring genetic material such as plasmids. 
Humans and other animals can transfer information in the form of traditions, culture and science, which instead of copying bits involves learning to reproduce some probability distribution (distillation) \cite{Hinton2015}.
Death is also risky and metabolically costly because it is attended by reproduction, whether sexually due to the loss of most gametes or asexually, e.g.\ via self-replication such as fission or parthenogenesis. 
Self-replication requires a minimum rate of heat production (determined by the growth rate, internal entropy, and durability of the replicator) \cite{England2013} that is entailed by Landauer's principle that performing irreversible logical computations entails producing a minimum of thermodynamic entropy \cite{Landauer1961}. 
However the downsides of mortality are offset by its essential role in enabling exploration on the fitness landscape.

\subsection{Caveats and prospective research}

The current implementation of the BUNCH algorithm is a proof of concept that can be further expanded in many directions.
First, time and computational power constrained the total number of atoms, number of IPS configurations and length of the simulations, which resulted in too little data to make strong statistical claims and too shallow cluster dendrons (too few levels) to give rise to lifelike entities exhibiting adaptive inference.
Second, by fixing the world boundary to a perfectly reflective hard wall we forwent examining the effect of varying the world topology and boundary conditions (Section \ref{sec:IPS}).
We also fixed the damping coefficient, the relative number of atoms and atom kinds (Section \ref{sec:simul}).
Third, we constrained BUNCH to only fit strictly tree-like graphs---a child can only have at most one parent. Loosening the graph structure constraints would lead to the evolution of a wider range of entities.
Fourth, similarly to other random sampling fitting schemes such as Markov Chain Monte Carlo and particle filtering \cite{DelMoral1997}, and unimodal posterior density schemes such as variational Laplace and in particular Dynamic Expectation Maximization \cite{Friston2008}, BUNCH is a lazy random sampling approximate inference algorithm.
But this drawback is justified by the complexity and intractability of the problem it comes to grips with, because approximate inference simplifications tend to perform much better than what their simplifying assumptions (mean-field) warrant \cite{Friston2007}, and because intriguingly some biological minimization problems have one dominant basin of attraction instead of a rugged energy landscape, like the folding of the proteins that are biologically relevant \cite{Laughlin2000}.
Fifth, as mentioned in Section \ref{sec:HMGC}, the current BUNCH implementation neglects time derivatives such as velocity that would substantially increase goodness of fit.
Sixth, given the vast IPS interaction force and range parameter space, 41 different IPS configurations were just enough to only work out an exploratory statistical analysis.
Prospective lines of research may address these shortcomings.

Another plausible way to analyze spatial configurations of particles, aside from hierarchical clustering schemes such as BUNCH, is using spectral graph theory.
Given the adjacency matrix $A$ of a graph where directed edges indicate interactions, its largest eigenvectors are the most interconnected sets of vertices, so it can be used to delimit the clusters encompassing the most densely connected subclusters.
The edges could be defined via e.g.\ some measure of distance or force strength, and a Markov blanket (parents, children, and spouses) matrix ---which roughly corresponds to the boundary of a persistent entity--- can be calculated as $M = A + A^T + A^TA$ \cite{Friston2013}.
The Laplacian matrix can be similarly used to find clusterings of multiple scales through its eigenvectors, which express spatial frequency modes \cite{Belkin2003}. 
Spectral graph theory could also be adapted and incorporated into an iterative scheme to yield a hierarchical model, perhaps in a hybrid scheme combining hierarchical clustering, akin to the hierarchical trophic links of food webs \cite{Williams2000}.
The properties of the graphs thus obtained could be collated with the those of actual biotic systems \cite{Sporns2000,Strogatz2001}.
For instance, it is plausible that complex enough persistent entities would display enhanced internal connectivity that enabled quick propagation of signals from e.g.\ destructive perturbations, and be laid out like small-world graphs \cite{Watts1998} (e.g.\ primate brains \cite{Stephan2000}), which can be efficiently navigated with lazy (local) heuristics \cite{Kleinberg2000}, or like the solutions to Steiner tree problems \cite{Courant1996}.
Scale-free networks \cite{Barabasi1999}, which have been observed in biological metabolic networks \cite{Jeong2000}, are also likely to be relevant because of their robustness and scalability \cite{Sporns2004}.

\subsection{Conclusion}

We have proposed a quantitative definition of life, lifeness, a scalar equal to algorithmic complexity (information) integrated over lifespan (time) of persistent structures (entities).
Lifeness thus defined conforms with our intuitive notion of lifelikeness.
We also devised an approximate filtering algorithm without adjustable parameters that computes the lifeness of arbitrary dynamical particle configurations under the assumption that the searched entities are generated by a hierarchical mixture of Gaussian clusters model.
We showed preliminary evidence suggesting that the lifeness of entities is associated with the distance to criticality of the worlds they inhabit. 

An implication for theoretical research is that discarding the notion that life is a binary category could avoid unnecessary debates around biological category mistakes.
On the practical side, measuring the lifeness of biotic systems and their subsystems could afford a useful marker for computer simulations and simple physical or biological systems to assess properties such as number of hierarchical levels, metabolic regime, and foreseeable longevity in computational biology, biochemistry and cell biology.
It also may enable hierarchical classification of artificial or biological entities such as taxa, organs, organelles, and molecules (e.g.\ the RNA example in Section \ref{sec:bioL}) in fields such as physiology, taxonomy, cytology, and molecular biology.
This will require, both in artificial and in physical systems,  identifying the persistent subcomponents of the target entity (Section \ref{sec:critworld}).
Since structures that can be modeled as hierarchical mixtures of clusters are fairly common, BUNCH is likely to be applicable in other areas of the natural sciences, in particular where  hierarchical clustering algorithms are already being employed.
Finally, the link between lifeness and criticality could be built on to inform the birth (abiogenesis) and persistence (autopoiesis) of biotic systems with the physics of criticality and in reverse to support the notion that we could not exist in a world far from criticality.

\section{Acknowledgments}

This article was supported by the HSE University Basic Research Program.

%\bibliography{bunch}  % for bibtex
\printbibliography[heading=bibintoc] 
%\printbibliography[type=article,heading=bibintoc] 
% for biblatex, and display in TOC, ,title={References}

\appendix

\section{Software notes \label{app:software}}

Our interacting particle system was developed starting from Particle Life \cite{HackerPoet2017,Abdulrahman2022}, a loose replica of Jeffrey Ventrella's Clusters algorithm \cite{Ventrella2017}.
We provide JavaScript and \texttt{C++} implementations that build on Particle Life code \cite{Abdulrahman2022}.
The JavaScript and \texttt{C++} source code are available on the web hosting service GitHub repository (\url{https://github.com/mmartinezsaito/racemi}) under the MIT license.
Both versions include a simple graphical user interface for tweaking the physics and visualization of the simulation.

The JavaScript version additionally includes online plots for monitoring statistics by means of the open source library Plotly (v2.20.0) \cite{Plotly2015}.

The C++ version stands on the open source image rendering library openFrameworks (v0.12.0) \cite{openFrameworksCommunity2004} and  enables simulating a larger number of particles and visualizations of the matrix of interaction force coefficients.
To compile the C++ code \cite{Abdulrahman2022}, download the GitHub repository (\url{https://github.com/mmartinezsaito/racemi}) and openFrameworks (\url{https://openframeworks.cc/}). Then generate a new openFrameworks project, add the addon ofxGui; after the project files are generated replace the \texttt{/src} folder with the one provided on GitHub.
At the time of publication, the GUI slider that sets the number of atoms does not allow decreasing it.

Data analysis was performed on R on text data files written by the \texttt{C++} program.

\section{Interacting particle system implementation details \label{app:IPS}}

\subsection{Solid finite atoms \label{app:solidball}}

We assume solid atoms of a small (but arbitrary) radius of $r_0=3$ pixels, within which mass is uniformly distributed.
Hence the power law potential of Eq.\ \ref{eq:Vpowlaw} is actually
\begin{equation*}
|\mathbf{f}(r)| = 
\left\{ \begin{array}{lll}
\chi_{ij} r^n   & \text{if} \quad r > r_0 & \text{(outside)} \\
\chi_{ij} r_0^n \frac{r}{r_0}  & \text{else} & \text{(inside)}
\end{array} \right.
\end{equation*} 
This assumption of continuously distributed charge avoids the singularity of discrete charged point-particles, effectively capping interactions forces between overlapping atoms.

\subsection{Detailed balance \label{app:DBeq}}

Eq. \ref{eq:langevin} has the associated Fokker-Planck equation \cite{Zwanzig2001,Tome2006} for the evolution of the density   $\rho = \rho(\bar{\mathbf{x}}, \bar{\mathbf{p}}; t)$ of the position $\bar{\mathbf{x}} = [\mathbf{x}_1, \ldots,\mathbf{x}_{n_A}]$ and momentum $\bar{\mathbf{p}} = [\mathbf{x}_1, \ldots,\mathbf{p}_{n_A}]$ of all atoms in phase space (or Klein-Kramers equation)
\begin{eqnarray*}
\frac{\partial \rho}{\partial t} &=& -\frac{\partial}{\partial \bar{\mathbf{x}}} \cdot \left[ \bar{\mathbf{v}} \rho \right] - \frac{\partial}{\partial \bar{\mathbf{p}}} \cdot \left[(\sum_{j=1}^{n_A} \bar{\mathbf{f}}_j - \zeta \bar{\mathbf{v}})\rho \right]
+ \zeta k_B \bar{T} \frac{\partial^2 \rho}{\partial \bar{\mathbf{p}}^2} \\
&=& -\frac{\partial }{\partial \bar{\mathbf{x}}} \cdot \left[\bar{\mathbf{v}} \rho \right] - \frac{\partial}{\partial \bar{\mathbf{p}}} \cdot \left[ (\sum_{j=1}^{n_A} \bar{\mathbf{f}}_j - \zeta \bar{\mathbf{v}} - \zeta k_B \bar{T} \frac{\partial \log{\rho}}{\partial \bar{\mathbf{p}}}) \rho \right] \\ 
& = & -\frac{\partial}{\partial \bar{\mathbf{x}}} \cdot \mathbf{J}_{\bar{\mathbf{x}}} - \frac{\partial}{\partial \bar{\mathbf{p}}}  \cdot \mathbf{J}_{\bar{\mathbf{p}}} = - \nabla \cdot \mathbf{J}
\end{eqnarray*}
where $\nabla = \begin{bmatrix} \frac{\partial}{\partial \mathbf{x}}, \frac{\partial}{\partial \mathbf{p}} \end{bmatrix}^T$, the nabla with components restricted to $\bar{\mathbf{x}}$ is denoted by $\frac{\partial}{\partial \bar{\mathbf{x}}}$, the $d$ dimensions of variables across all the $n_A$ atoms are indexed by $i = 1 \ldots n_A$, and we have used $\frac{\partial \rho}{\partial x} = \frac{\partial \log{\rho}}{\partial x} \rho$ \cite{Loos2020}.

The currents of probability are zero in equilibrium and constant (but in general different) in steady state. Detailed balance entails that all probability current components are zero \cite{Loos2020}: $\mathbf{J}_{\bar{\mathbf{x}}} = \mathbf{0}$ and $\mathbf{J}_{\bar{\mathbf{p}}} = \mathbf{0}$.
Then, for $i = 1 \ldots n_A$, we have
\begin{eqnarray}
\mathbf{v}_i &=& \mathbf{0} \label{eq:DB_v} \\ 
\sum_{j=1}^{n_A} \mathbf{f}_{ij} - \zeta \mathbf{v}_i - \zeta k_B T_i \frac{\partial \log{\rho}}{\partial \mathbf{p}_i} &=& \mathbf{0} \label{eq:DB_p}
\end{eqnarray} 
From Eqs.\ \ref{eq:DB_v}, \ref{eq:DB_p} we get $ T_i^{-1} \zeta^{-1} \sum_{j=1}^{n_A} \mathbf{f}_{ij} = k_B \frac{\partial \log{\rho}}{\partial \mathbf{p}_i}$.
This indicates \cite{Loos2020} that $ T_i^{-1} \zeta^{-1} \sum_{j=1}^{n_A} \mathbf{f}_{ij}$ is a conservative or curl-free field: $ \nabla \times (T_i^{-1} \zeta^{-1} \sum_{j=1}^{n_A} \mathbf{f}_{ij}) = 0$.
The curl of a $d$-dimensional vector field $\mathbf{A}$ has $\binom{d}{2} = \frac{d(d-1)}{2}$ components, which (in orthogonal coordinates) comprise the factors $\frac{\partial A_i}{\partial x_j} - \frac{\partial A_j}{\partial x_i}$ for each duplet $i,j = 1 \ldots d : i \neq j$ so setting the curl to null we get $\frac{\partial}{\partial \mathbf{p}_j} T_i^{-1} \sum_{j=1}^{n_A} \mathbf{f}_{ij} = \frac{\partial}{\partial \mathbf{p}_i} T_j^{-1} \sum_{i=1}^{n_A} \mathbf{f}_{ji}$, which leads to $T_i^{-1} \sum_{j=1}^{n_A} \frac{\partial}{\partial \mathbf{p}_j} (-\chi_{ij} \frac{\partial V}{\partial \mathbf{x}_{ij}}) =  T_j^{-1} \sum_{i=1}^{n_A} \frac{\partial}{\partial \mathbf{p}_i} (-\chi_{ji} \frac{\partial V}{\partial \mathbf{x}_{ji}})$.
Since $\frac{\partial V}{\partial \mathbf{x}_{ij}} = -\frac{\partial V}{\partial \mathbf{x}_{ji}}$ and $\chi_{ij}$ are constants, it follows that
\begin{equation*}
\chi_{ij} T_j = \chi_{ji} T_i.
\end{equation*}

\subsection{Mapping non-reciprocal to reciprocal systems \label{app:symmap}}

The interaction coefficients matrix $\chi$ can be broken down (Toeplitz decomposition) into a positive semi-definite symmetric matrix $\chi^S = S = S^T$ and a skew-symmetric matrix $\chi^A = A = -A^T$ with $A_{ii} = 0$ via the sum $\chi = \frac{1}{2}(\chi + \chi^T) + \frac{1}{2}(\chi - \chi^T)= S + A$.
The pairwise asymmetry of interaction coefficients can be characterized with the non-reciprocity parameter $\Xi \geq 0$ or ratio of the non-reciprocal to reciprocal forces \cite{Ivlev2015}.
This can be generalized to a $n_A \times n_A$ dimensional non-reciprocity matrix $\Xi \equiv A S^{-1}$, which is skew-symmetrizable so $\det{\Xi^T} = \det{(-\Xi)} = (-1)^{n_A} \det{\Xi}$.
Then $\mathbf{f}_{ij} = - \chi_{ij} \frac{\partial V(r_{ij})}{\partial \mathbf{x}_{ij}} = - \frac{\partial V^S_{ij}(r_{ij})}{\partial \mathbf{x}_{ij}} (1 + \Xi_{ij})$, where $V^S_{ij} = V S_{ij}$. Note that $\mathrm{1} + \Xi = I + (\mathrm{1}-I) + \Xi = \chi S^{-1} + (\mathrm{1}-I)$, where $\mathrm{1}$ is a matrix of ones.
For $\Xi_{ij} < 1 : \forall i,j$, which is equivalent to Eq.\ \ref{eq:DBcond}, pairwise interactions have a pseudo-Hamiltonian.

Analogously to the two-body problem of classical mechanics, for $n_K=2$ and pairwise interaction laws differing only by a constant, the problem can be rendered reciprocal by symmetrizing the \textit{specieswise} interaction matrix $\chi_{ij} \in \mathbb{R}^{n_K \times n_K}$, with $i,j$ indicating one atom per kind. 
This can be done by rescaling the masses $m_i$ and forces $\mathbf{f}_{ij}$ using $\Delta$ \cite{Ivlev2015} or by rescaling the $\mathbf{x}_i$ and temperatures $T_i$ (if stochastic noise is explicitly included) \cite{Loos2020}. Either way, this enables describing the system exactly as a reciprocal system, via a pseudo-Hamiltonian, with newly rescaled variables $\mathring{m}_i, \mathring{\mathbf{f}}_{ij}, \mathring{T}_i$. 
By the theorem of equipartition of energy, if a degree of freedom makes only a quadratic contribution to the Hamiltonian, then the average energy of the corresponding term in the Hamiltonian is $\frac{1}{2} k_BT$ times the dimensionality of the degree of freedom \cite{Feynman1963,Goldenfeld1992}.\
Hence there is a unique temperature $T$ associated to all atoms in a (symmetrized) system with pseudo-Hamiltonian, so the average kinetic energy of every particle $i = 1 \ldots n_A$ is the same value
\[
\frac{d}{2}k_B \mathring{T} = \frac{1}{2} \mathring{m}_i \langle \mathring{v}_i^2 \rangle,
\]
which after a few manipulations \cite{Ivlev2015,Loos2020} leads to Eq.\ \ref{eq:DBcond}.
\cite{Ivlev2015} provides a measure $\epsilon$ of deviation from pseudo-Hamiltonian systems, which satisfy Eq.\ \ref{eq:DBcond} and may be mapped to reciprocal systems.
For $n_K = 2$ kinds of atom in $d=2$ dimensions  \cite{Ivlev2015} showed that for systems with non-constant interaction coefficients $\epsilon$ is in general non-zero, Eq.\ \ref{eq:DBcond} is not fulfilled so equilibrium is unattainable and in the absence of friction temperatures grow monotonously with time as $T(t) \sim t^{2/3}$ for $t \to \infty$.
However, adding any non-zero friction ($\zeta > 0$) immediately entails that the system approaches a steady state where the temperatures are again related by Eq.\ \ref{eq:DBcond} (as long as the friction coefficient $\zeta$ is shared by all atom kinds) \cite{Ivlev2015}.

However, for $n_K > 2$, where multiple kinds of atom may interact with every kind of atom (so the off-diagonal terms are more than the main diagonal terms) symmetrizing may not be possible if the number of distinct off-diagonal terms in $\chi_{ij}$ is more than the number of atom kinds $n_K$ \cite{Loos2020}.
When symmetrization by mapping onto an equivalent reciprocal system is possible, equilibrium occurs only if the thermal noise terms are set to equal $\mathring{T}_i = \mathring{T}_j : \forall i,j$, which leads again to Eq.\ \ref{eq:DBcond} \cite{Parrondo1996,Loos2020}.
Both (pre- and post- mapping) systems yield identical entropy production \cite{Loos2020}.

\section{BUNCH algorithm implementation details \label{app:BUNCH}}

\subsection{The number of ways to allocate $n$ elements to $k$ clusters \label{app:bellnumber}}

The number of ways to partition a set (universe) of $n_A$ elements (atoms) into $k$ non-empty subsets (clusters) is the Stirling partition number $\left\{ n_A \atop k \right\} := \frac{1}{k!}\sum_{j=0}^k (-1)^{k-j} \binom{k}{j} j^{n_A}$, where $\binom{k}{j} := \frac{k!}{j!(k-j)!}$ is the binomial coefficient.
The Bell number $B_{n_A} := \sum_{k=0}^n \left\{ n_A \atop k \right\}$ is then the total number of partitions of a universe of $n_A $ atoms (not to confuse with the cardinality of the power set, the number of subsets of a set or possible clusters in a universe).
Hence, given a level $i$ with $n_{C_i}$ elements (atoms for $i=0$, otherwise clusters), there are $B_{n_{C_i}}$ possible ways to distribute the $n_{C_i}$ elements over an arbitrary number of $n_{C_{i+1}}$ parent clusters. 
Thus the information content of the cluster dendron (excluding its priors) is
\[
I_B = \sum_{i = 1}^L \log{B_{n_{C_i}}}.
\]

\subsection{Ellipse semi-axes estimation via singular value decomposition \label{app:svd}}

The singular value decomposition (SVD) required by the BUNCH algorithm was computed in JavaScript with the open source library svd-js (v3) \cite{Salvati2019} and in C++ with the C++ template library Eigen (v3.4.0) \cite{Guennebaud2010}.

A $d \times m$ matrix, where $d$ is the dimension of the universe (two) and $m$ the number of children of the cluster, was prepared for each cluster such that its columnwise elements were the two coordinates of the $m$ children centered with respect to the barycenter or centroid of the children and divided by $\sqrt{m-1}$ (using $m-1$ instead of $m$ is Bessel's correction, which yields an unbiased estimate of the variance). 
This guarantees that after performing SVD the resulting eigenvalues appropriately portray the two cluster semi-axes when drawn along the left-singular vectors with size equal to their respective singular values. This is because the singular values are the square roots of the eigenvalues of the covariance matrix of coordinates. Thus the ellipse with semi-axes given by the singular vectors and values indicates the standard deviation of the positions of its children, along the axes of a two-dimensional Gaussian-shaped cluster.
When a cluster has less than three children, its covariance matrix is rank-deficient and SVD yields one or more zero eigenvalues. To prevent the its ellipse from degenerating into a segment or a point, we arbitrarily set the degenerate semi-axes to a small number (the radius of the atoms, 3 pixels).

\subsection{Wandering over parameter space \label{app:roving_params}}

Both implementations enable performing random evolution through two sliders that specify how the system roves in the parameter space of matrix interactions. Random mutation is accomplished through mutation size (a fraction the mutated parameter value) and mutation frequency (probability at each time step).

A straightforward algorithm for selective artificial evolution would proceed as follows:
\begin{enumerate}
\item Pick a random location in the parameter space. 
\item Run the interacting particle system for a long enough time to compute the maximum lifeness over all subdendrons with reasonable accuracy.
\item Transition to a new location in the parameter space that is estimated to have a higher maximum lifeness.
\item Repeat steps 2 and 3 until some criterion (e.g. time or memory limit) is reached.
\end{enumerate}

For step 3, some sort of blind local search optimization algorithm could be used, such as hill climbing or simulated annealing.
Anyhow, the selective evolution algorithm may be overly slow for systems beyond a few hundred atoms with the current typical capability of computers.
The BUNCH algorithm coasts relatively slowly to lower minima on the landscape of surprisal, and more accurate algorithms may be too slow to be practical.

\subsection{Bisect-unite, birth-death, and hierarchy updating routine details \label{app:BUBD}}

\textit{Lazy cluster bisection}: 
There is only one way to merge two clusters, but many to split one.
In the current version of BUNCH, the selection of the candidate two halves (bipartitions) that determine the bisection is random. This is a lazy approach: computationally cheap but short-sighted.
An approach that better optimizes log-evidence would be splitting the ellipse along its minor (or smallest) axis, informed by SVD (cf. Appendix \ref{app:svd}).

\textit{Few-children cluster killing rule}:
In $d$ dimensions, cluster parents with $d$ children or less are degenerate, i.e.\ have measure zero.
So for 2-dimensions, SVD (cf.\ Appendix \ref{app:svd}) on any cluster with 2 children or less will yield at least one zero singular value.
Two-children clusters, which have one zero singular value, are simply modified so that the collapsed (minor) axis is arbitrarily set to a small number (somewhat arbitrarily, the atom radius of 3 pixels, which itself was chosen just for visualization purposes).
Single-child clusters are assigned a isotropic (circular) blob, with a radius that collapses at a fixed rate per iteration: $1-\alpha=0.05$ (where $\alpha$ is an adjustment or learning rate) which entails an exponential rate of collapse of 0.95.
With an adjustment rate of $\alpha = 1$, single-child clusters would collapse immediately in one time step.
By setting $\alpha=.05$, we allow single-child clusters to carry over and possibly to adopt more children in the next time step, thereby avoiding death.
This is because the jittering of children (atoms or subclusters) position prevents blobs from totally collapsing into singularities ---thermal motion adds an irreducible variance or noise to the position of children because temporal derivatives such as velocity are not modeled here.
Note nonetheless that the fine rate at which few-children clusters are killed bears little effect on the conclusions of this article.
Finally, childless clusters and clusters with zero radius (axes) are immediately killed.

\textit{Bottom-up cluster tree updating}:
There must always be a single cluster at the top level: the top cluster. If there is only one level, this is the omnicluster.
If the top cluster is split, then its two halves automatically become the two children of a newly added top cluster, which also results in adding a new level to the hierarchy.
Conversely, if the next-to-top level has only one cluster (due e.g.\ to a recent merge) then it becomes the new top cluster because its former parent is automatically killed, which also results in removing a level from the hierarchy.

In typical hierarchical clustering algorithms, besides the distance metric, a linkage criterion ---which specifies the dissimilarity of sets as a function of the pairwise distances of observations in the sets--- is employed to decide which clusters are merged or split  \cite{Szekely2005}.
In BUNCH, the linkage criterion is effectively the local shift of  the dendron's log-evidence. However, the \textit{selection} of candidates is uniformly random and the log-evidence is only used to \textit{decide} whether to perform or not merges and splits.

\textit{Optimization heuristics}:
Unlike steepest ascent schemes on continuous manifolds, BUNCH does not make use of gradients and attempts to maximize log-evidence via discontinuous leaps while roving the parameter space, which is a hybrid of discrete (e.g. number of levels, clusters at each level, and ways to allocate children to parents) and continuous (cluster positions and covariance matrices) coordinates.
Because the model evidence function is not differentiable (e.g. with respect to cluster births and deaths) BUNCH simply evaluates model evidence at stochastically sampled neighborhood locations to decided whether to take a step, similarly to pattern search or random search approaches \cite{Rastrigin1963}.
Similarly to the EM algorithm and variational Bayesian methods (e.g. DEM \cite{Friston2008}), this is accomplished iteratively: at each time step we bottom-up sweep the layers and at each layer we reallocate children and update sequentially each cluster.
Discrete coordinates naturally enforce the leap sizes (e.g. a child being transfer to other parent, or two cluster parents being merged) while precluding the calculation of directional derivatives.
The ascent step size along continuous coordinates (e.g. cluster centroid speed and bearing and shrinking/growth rate) is determined by a learning rate parameter $\alpha$ that pertains to and bears on the performance of the algorithm, similar to typical gradient descent schemes.

\subsection{Space with boundaries or wrapping around without boundaries \label{app:closedmanifold}}

The BUNCH algorithm can build cluster dendrons for atoms living on closed manifolds that ``wrap around'', such as torii or spheres on two dimensions, as opposed to on open manifolds without boundaries.
However this exerts a disproportionate burden on computational resources and renders the BUNCH algorithm far slower.
This is because all pairwise distance calculations, which comprise the bulk of the most expensive calculations, are compounded by the requirement to find the nearest path between the pair among $2d$ candidates (in a closed manifold, each dimension affords two orientations pointing towards any given location).
Thus computing cluster centroids, the children likelihood or (Mahalanobis) distance to parent centroids, cluster splits and merges (let alone optimal cluster splits; Appendix \ref{app:BUBD}) becomes computationally far more burdensome because centroids are required to maximize model evidence or likelihood over all children spatial coordinates combinations (which is exponential in the number of children), and children membership must also maximize likelihood over all candidate parent clusters (also exponential), and split and merge computations are similarly burdensome.

\subsection{Entities as classes of indistinguishable clusters \label{app:entities}}
 
As mentioned in Section \ref{sec:HMGC}, here an entity denotes an equivalence class that comprises the clusters constituted by the same descendants, in the same way that water refers to the class of any of the indistinguishable molecules of water.
Hence there is a (surjective) map that assigns at least one entity to every conceivable cluster.

This entails that at any time step a cluster is assigned both a cluster index and an entity index.
In practice, this means that two lists, of clusters and entities, must be maintained in main memory. The elements of the list of clusters are appended at birthtime; upon death, a cluster in the list simply tagged as ``dead'' to never be again accessed but otherwise it stays in the list.

Hence the cluster list is akin to a record of all clusters that have been alive at some point. Each element contains an identifying index number, its birthtime, lifespan if alive or deathtime if dead, the number and identity of its cluster children, and current position, axes sizes, and orientation.
Note that clusters can gain and lose children, transform, move and reorient, or be killed (under specific conditions: see Section \ref{sec:BUBD} and Appendix \ref{app:BUBD}).

In contrast, the entity list is akin to a record of all distinct classes of clusters for which at least one instance has been alive at some point. Each element contains an identifying entity number; a \textit{sequence} birthtimes, lifespans, and deathtimes corresponding to the life history of all its instances; and the number and identity of its \textit{entity} children.
Because they are defined by their children (cf. Section \ref{sec:HMGC}), entities cannot gain or lose children be born or killed. However they can transform, move and reorient.

\section{More numerical simulation results \label{app:moresims}}

\subsection{IPS statistics and regression analyses \label{app:statble}}

To analyze how lifeness is affected by IPS parameters we fitted generalized linear models with gamma response distribution and logarithmic link function, separately for the ranges $p \in [-1, 0]$ and $p \in [-3, -1]$, for each of the cluster dendron $l = 1 \ldots nL$ levels, via the R function \texttt{glm} \cite{R2016}.
We used the linear predictors $\langle \bar{\Lambda_l} \rangle_{SS} \sim r_\chi + s_\chi + d_\chi + p$ for $l = 1 \ldots 4$.
$\bar{\Lambda_l}$ is the normalized lifeness averaged over the entities living at level $l$.
The chevrons in $\langle \bar{\Lambda_l} \rangle_{SS}$ indicate that the time average is taken over frames when the system is in NESS. This is approximately defined as the interval starting from the last frame where a change in the cluster dendron height occurred until the end of the simulation run.

% use R's xtable
\begin{table}[H]
\centering
\caption{Regression results for generalized linear models with gamma response distribution, logarithmic link function and linear predictors $\langle \bar{\Lambda_l} \rangle_{SS} \sim r_\chi + s_\chi + d_\chi + p$ for $l = 1 \ldots 4$ and $p \in [-1, 0]$. Cf.\ Section \ref{sec:stroll}.  \label{tab:glm1}}
\begin{tabular}{c|c|llll}
$l$ & coef. & estimate & std.\ error & $t$ value & Pr($> |t|$) \\ 
\hline
&intercept & 1.3419 & 1.4003 & 0.96 & 0.3479 \\ 
&  $r_\chi$ & -1.3688 & 0.7846 & -1.74 & 0.0944 \\ 
1 &  $s_\chi$ & -3.0670 & 2.7667 & -1.11 & 0.2791 \\ 
&  $d_\chi$ & -2.0624 & 2.6104 & -0.79 & 0.4376 \\ 
&  $p$      & 0.0615 & 0.3491 & 0.18 & 0.8617 \\ 
\hline
&intercept & -2.2730 & 1.9464 & -1.17 & 0.2548 \\ 
&  $r_\chi$ & 0.1303 & 1.0906 & 0.12 & 0.9059 \\ 
2 &  $s_\chi$ & 3.6598 & 3.8455 & 0.95 & 0.3511 \\ 
&  $d_\chi$ & 6.0248 & 3.6283 & 1.66 & 0.1104 \\ 
&  $p$ & -0.0821 & 0.4853 & -0.17 & 0.8672 \\ 
\hline
&intercept & 0.9017 & 3.0525 & 0.30 & 0.7703 \\ 
&  $r_\chi$ & 0.8044 & 1.7103 & 0.47 & 0.6425 \\ 
3 &  $s_\chi$ & 1.8236 & 6.0308 & 0.30 & 0.7651 \\ 
&  $d_\chi$ & 2.6120 & 5.6902 & 0.46 & 0.6505 \\ 
&  $p$ & 2.0219 & 0.7611 & 2.66 & 0.0141* \\ 
\hline
&intercept & 1.7712 & 1.1632 & 1.52 & 0.1415 \\ 
&  $r_\chi$ & 1.3168 & 0.6517 & 2.02 & 0.0551 \\ 
4 &  $s_\chi$ & -0.5456 & 2.2981 & -0.24 & 0.8144 \\ 
&  $d_\chi$ & -1.7924 & 2.1683 & -0.83 & 0.4169 \\ 
&  $p$ & 0.7693 & 0.2900 & 2.65 & 0.0142* \\ 
\hline
\end{tabular}
\end{table}

\begin{table}[H]
\centering
\caption{Regression results for generalized linear models with gamma response distribution, logarithmic link function and linear predictors $\langle \mathring{\Lambda_l} \rangle_{SS} \sim r_\chi + s_\chi + d_\chi + p$ for $l = 1 \ldots 4$ and $p \in [-1, 0]$. Cf.\ Section \ref{sec:stroll}. * indicates p-values less than .05. \label{tab:glm2}}
\begin{tabular}{c|c|llll}
$l$ & coef. & estimate & std.\ error & $t$ value & Pr($> |t|$) \\ 
\hline
& intercept & 3.5481 & 2.0411 & 1.74 & .0955 \\ 
&  $r_\chi$ & 0.2910 & 1.1436 & 0.25 & .8014 \\ 
1 &  $s_\chi$ & -3.1205 & 4.0327 & -0.77 & .4469 \\ 
&  $d_\chi$ & -1.9543 & 3.8049 & -0.51 & .6124 \\ 
&  $p$ & 1.3910 & 0.5089 & 2.73 & .0118* \\ 
\hline
& intercept & -1.5577 & 2.1797 & -0.71 & .4820 \\ 
& $r_\chi$ & 0.7120 & 1.2213 & 0.58 & .5656 \\ 
2 & $s_\chi$ & 2.7038 & 4.3065 & 0.63 & .5363 \\ 
& $d_\chi$ & 4.6829 & 4.0633 & 1.15 & .2610 \\ 
& $p$ & 0.1525 & 0.5435 & 0.28 & .7815 \\ 
\hline
& intercept & 0.5603 & 1.3141 & 0.43 & .6738 \\ 
&  $r_\chi$ & -0.5869 & 0.7363 & -0.80 & .4336 \\ 
3 &  $s_\chi$ & -1.9848 & 2.5963 & -0.76 & .4524 \\ 
& $d_\chi$ & -1.2209 & 2.4497 & -0.50 & .6230 \\ 
&  $p$ & 0.0276 & 0.3277 & 0.08 & .9337 \\ 
\hline
&  intercept & -2.5169 & 9.0260 & -0.28 & .7851 \\ 
&  $r_\chi$ & -1.7151 & 1.2809 & -1.34 & .2054 \\ 
4 & $s_\chi$ & 5.3731 & 18.6351 & 0.29 & .7780 \\ 
&  $d_\chi$ & 4.8546 & 14.3074 & 0.34 & .7402 \\ 
&  $p$ & 0.7298 & 0.6496 & 1.12 & .2832 \\ 
\hline
\end{tabular}
\end{table}

We also analyzed how the surprisal $I_W$ relates to IPS parameters by fitting generalized linear models with gamma response distribution and logarithmic link function, separately for the ranges $p \in [-1, 0]$ and $p \in [-3, -1]$. We similarly used the linear predictors $I_W \sim r_\chi + s_\chi + d_\chi + p$ (Table \ref{tab:glm3}).

\begin{table}[H]
\centering
\caption{Regression results for generalized linear models with gamma response distribution, logarithmic link function and linear predictors $ I_W \sim r_\chi + s_\chi + d_\chi + p$ for $l = 1 \ldots 4$ and $p \in [-1, 0]$, $p \in [-3, -1]$. Cf.\ Section \ref{sec:stroll}.  \label{tab:glm3}}
\begin{tabular}{c|c|llll}
$p$ range & coef. & estimate & std.\ error & $t$ value & Pr($> |t|$) \\ 
\hline
& intercept & 1.5642 & 2.9272 & 0.53 & .5982 \\ 
& $r_\chi$ & 4.3555 & 1.6401 & 2.66 & .0141* \\ 
$[-1,0]$ & $s_\chi$ & 16.3750 & 5.7833 & 2.83 & .0095* \\ 
& $d_\chi$ & 11.7766 & 5.4566 & 2.16 & .0416* \\ 
& $p$ & 1.7453 & 0.7298 & 2.39 & .0254* \\ 
\hline
& intercept & -42.8505 & 12.6174 & -3.40 & .0193* \\ 
& $r_\chi$ & 3.4406 & 2.7074 & 1.27 & .2597 \\ 
$[-3,-1]$ & $s_\chi$ & 100.6104 & 25.2240 & 3.99 & .0104* \\ 
& $d_\chi$ & 88.6311 & 21.3404 & 4.15 & .0089* \\ 
& $p$ & -1.6767 & 0.7455 & -2.25 & .0744 \\ 
\hline
\end{tabular}
\end{table}

%\csvautotabular{wps.csv}
% or www.tablesgenerator.com/latex_tables to import wps.csv
\begin{table}[!ht]
\centering
\caption{Simulation run parameters and statistics for 41 different IPS worlds. See Section \ref{sec:simul}. 
World names follow the convention specified in Section \ref{sec:p09_30}.
\label{tab:simstats1}}
\begin{tabular}{l|llllllllll}
\hline
   & $p$   & $I_W$  & $r_\chi$   & $s_\chi$  & $d_\chi$   & $\langle \bar{\Lambda_1} \rangle_{SS}$   & $\langle \bar{\Lambda_2} \rangle_{SS}$   & $\langle \bar{\Lambda_3} \rangle_{SS}$ & $\langle \bar{\Lambda_4} \rangle_{SS}$  & $n_L$ \\
\hline
(-1,+.08,.37)  & -1  & 5.37  & 0.077  & 0.373  & 0.168  & 2.845  & 1.38  & 0.887  & 1.108  & 4  \\
(-1,0,1)       & -1     & 12.32  & 0      & 1     & -0.321 & 0.549  & 0.446  & 0.314 & NA    & 3      \\
(-1,-.04,0)    & -1     & 6.91   & -0.04  & 0     & 0.357  & 46.88  & 15.94  & 0.724 & 0.847 & 4      \\
(-1,-.17,0)    & -1     & 7.24   & -0.17  & 0     & 0.414  & 2.738  & 0.554  & 0.803 & 1.22  & 4      \\
(-1,-.15,.63)  & -1     & 7.77   & -0.151 & 0.637 & -0.171 & 2.672  & 0.703  & 0.661 & 0.9   & 4      \\
(-1,+.14,.55)  & -1     & 9.47   & 0.14   & 0.545 & -0.06  & 0.9062 & 0.9368 & 1.265 & NA    & 3      \\
(-1,+.23,0)    & -1     & 13.67  & 0.23   & 0     & 0.4    & 80.55  & 45.2   & 42.24 & NA    & 3      \\
(-1,-.07,.50)  & -1     & 10.94  & -0.07  & 0.5   & -0.002 & 0.638  & 0.216  & 0.288 & 0.332 & 4      \\
(-1,-.00,.45)  & -1     & 8.65   & -0.002 & 0.45  & 0.08   & 0.652  & 0.193  & 0.308 & 0.581 & 4      \\
(-1,-.33,.39)  & -1     & 5.674  & -0.334 & 0.393 & 0.154  & 0.598  & 0.538  & 1.066 & 1.919 & 4      \\
(-1,+.09,.30)  & -1     & 9.24   & 0.094  & 0.3   & 0.265  & 1.227  & 0.408  & 0.375 & 0.386 & 4      \\
(-1,-.27,.23)  & -1     & 6.99   & -0.272 & 0.236 & 0.365  & 0.983  & 0.438  & 0.541 & 0.537 & 4      \\
(-2,-.07,.41)  & -2     & 12.15  & -0.075 & 0.411 & 0.114  & 0.147  & 0.127  & 0.164 & NA    & 3      \\
(-1.5,-.20,.42) & -1.5   & 7.51   & -0.197 & 0.417 & 0.088  & 3.65   & 2.79   & 0.349 & 0.314 & 4      \\
(-0.5,+.08,.41) & -0.5   & 7.77   & 0.084  & 0.414 & 0.117  & 11.71  & 2.46   & 1.27  & NA    & 3      \\
(-0.5,-.20,.53) & -0.5   & 7.38   & -0.198 & 0.529 & -0.028 & 0.842  & 0.559  & 0.785 & 2.441 & 4      \\
(-0.33,-.11,.46) & -0.333 & 8.86   & -0.112 & 0.461 & 0.048  & 3.328  & 0.297  & 0.396 & 0.526 & 4      \\
(-0.1,-.17,.49)  & -0.1   & 9.828  & -0.17  & 0.494 & 0.006  & 13.76  & 0.696  & 0.563 & NA    & 3      \\
(-0.75,-.17,.60) & -0.75  & 7.788  & -0.172 & 0.593 & -0.118 & 3.63   & 1.034  & 1.297 & NA    & 3      \\
(-1.25,-.14,.43) & -1.25  & 11.027 & -0.142 & 0.431 & 0.098  & 17.34  & 4.63   & 1.15  & NA    & 3      \\
(-0.66,-.11,.57) & -0.666 & 7.686  & -0.11  & 0.571 & -0.103 & 1.334  & 1.278  & 0.697 & 0.909 & 4      \\
(-0.25,-.10,.44) & -0.25  & 9.874  & -0.095 & 0.44  & 0.077  & 0.366  & 0.418  & 0.549 & NA    & 3      \\
(-1.75,-.24,.23) & -1.75  & 12.2   & -0.241 & 0.226 & 0.35   & 2.181  & 2.846  & 1.128 & NA    & 3      \\
(0,+.17,.39)     & 0      & 10.28  & 0.172  & 0.385 & 0.132  & 2.113  & 0.272  & 0.28  & NA    & 3      \\
(-2,+.24,.45)    & -2     & 11.724 & 0.243  & 0.451 & 0.063  & 6.348  & 3.49   & 2.187 & 0.707 & 4      \\
(-0.25,-.11,.27) & -0.25  & 9.439  & -0.115 & 0.27  & 0.318  & 14.46  & 6.322  & 1.488 & NA    & 3      \\
(-1.33,+.08,.20) & -1.33  & 8.665  & 0.077  & 0.205 & 0.315  & 0.489  & 0.31   & 0.156 & 0.164 & 4      \\
(-0.5,-.06,.51)  & -0.5   & 9.808  & -0.065 & 0.51  & -0.014 & 5.073  & 0.537  & 0.538 & NA    & 3      \\
(-0.2,+.04,.29)  & -0.2   & 9.138  & 0.037  & 0.29  & 0.246  & 2.677  & 0.709  & 0.858 & 1.688 & 4      \\
(-0.3,-.11,.48)  & -0.3   & 8.228  & -0.114 & 0.483 & 0.023  & 5.773  & 1.128  & 0.592 & 0.787 & 4      \\
(-3,+.04,.59)  & -3     & 10.48  & 0.042  & 0.59  & -0.127 & 6.851  & 3.971  & 2.05  & 1.138 & 4      \\
(0,+.17,.38)   & 0      & 11.32  & 0.165  & 0.383 & 0.148  & 0.276  & 0.619  & 0.841 & NA    & 3      \\
(-0.4,-.03,.49) & -0.4   & 7.6    & -0.029 & 0.478 & 0.028  & 0.736  & 0.353  & 0.573 & 0.756 & 4      \\
(-0.9,+.34,.29) & -0.9   & 10.45  & 0.343  & 0.29  & 0.29   & 4.69   & 4.09   & 0.318 & NA    & 3      \\
(-0.7,-.19,.56)  & -0.7   & 6.888  & -0.193 & 0.555 & -0.084 & 4.72   & 0.102  & 0.165 & 0.179 & 4      \\
(-0.8,+.05,.54) & -0.8   & 7.954  & 0.046  & 0.545 & -0.041 & 3.69   & 0.361  & 0.475 & 0.509 & 4      \\
(-1.6,-.17,.44) & -1.6   & 12.3   & -0.167 & 0.44  & 0.084  & 1.5    & 1.47   & 1.83  & NA    & 3      \\
(-1.4,-.03,.56)  & -1.4   & 8.904  & -0.031 & 0.556 & -0.069 & 24.3   & 18.6  & 1.22  & 1.12  & 4      \\
(-1.9,-.15,.34)  & -1.9   & 11.86  & -0.151 & 0.338 & 0.199  & 0.816  & 0.762  & 0.679 & NA    & 3      \\
(-0.5,+.02,.38)  & -0.5   & 7.624  & 0.022  & 0.378 & 0.128  & 3      & 0.529  & 0.551 & 0.765 & 4      \\
(-0.25,+.18,.36) & -0.25  & 8.337  & 0.178  & 0.364 & 0.162  & 5.35   & 0.927  & 0.446 & 0.658 & 4     \\
\hline
\end{tabular}
\end{table}

\begin{table}[!ht]
\centering
\caption{Simulation run parameters and statistics for 41 different IPS worlds. See Section \ref{sec:simul}. 
World names follow the convention specified in Section \ref{sec:p09_30}.
$t_l$ indicates the frame number at which the cluster dendron first attained the height $l$.
The total simulation time is expressed in frames ($t_{run}(\text{f})$) and seconds ($t_{run}(\text{s})$).
\label{tab:simstats2}}
\begin{tabular}{l|lllllllllll}
\hline
    & $t_1$    & $t_2$   & $t_3$   & $t_4$ & $t_{run}(\text{f})$   & $t_{B_0}$  & $t_{run}(\text{s})$  & $\langle \mathring{\Lambda_1} \rangle_{SS}$   & $\langle \mathring{\Lambda_2} \rangle_{SS}$   & $\langle \mathring{\Lambda_3} \rangle_{SS}$ & $\langle \mathring{\Lambda_4} \rangle_{SS}$  \\               
\hline               
(-1,+.08,.37)  & 5103  & 5103  & 5121  & 5280  & 5542  & 439  & 138721 & 1.108 & 0.887  & 1.38   & 2.845 \\
(-1,0,1)       & 5535  & 5535  & 9888  & NA    & 10731 & 5196 & 240.7  & 0.314 & 0.446  & 0.549  & 2.845 \\
(-1,-.04,0)    & 1304  & 1307  & 1384  & 1676  & 1778  & 474  & 68320  & 0.847 & 0.724  & 15.94  & 46.88 \\
(-1,-.17,0)    & 17570 & 17570 & 17677 & 17807 & 18091 & 521  & 108885 & 1.22  & 0.803  & 0.554  & 2.738 \\
(-1,-.15,.63)  & 4996  & 4996  & 5059  & 5222  & 5657  & 661  & 116805 & 0.9   & 0.661  & 0.703  & 2.672 \\
(-1,+.14,.55)  & 3252  & 3252  & 3281  & NA    & 4055  & 932  & 345.7  & 1.265 & 0.9368 & 0.9062 & 2.845 \\
(-1,+.23,0)    & 4471  & 4471  & 10075 & NA    & 16899 & 803  & 23478  & 42.24 & 45.2   & 80.55  & 2.845 \\
(-1,-.07,.50)  & 11560 & 11560 & 11601 & 12439 & 13502 & 1942 & 3248   & 0.332 & 0.288  & 0.216  & 0.638 \\
(-1,-.00,.45)  & 608   & 608   & 681   & 1151  & 2246  & 1638 & 8356   & 0.581 & 0.308  & 0.193  & 0.652 \\
(-1,-.33,.39)  & 2520  & 2520  & 2538  & 2679  & 2982  & 462  & 3075   & 1.919 & 1.066  & 0.538  & 0.598 \\
(-1,+.09,.30)  & 711   & 711   & 812   & 1475  & 1549  & 838  & 392.7  & 0.386 & 0.375  & 0.408  & 1.227 \\
(-1,-.27,.23)  & 1715  & 1715  & 1738  & 1995  & 2610  & 895  & 820.7  & 0.537 & 0.541  & 0.438  & 0.983 \\
(-2,-.07,.41)  & 623   & 623   & 4085  & NA    & 10173 & 9550 & 18546  & 0.164 & 0.127  & 0.147  & 2.845 \\
(-1.5,-.20,.42)& 3546  & 3546  & 3556  & 4042  & 5629  & 2083 & 12661  & 0.314 & 0.349  & 2.79   & 3.65  \\
(-0.5,+.08,.41)& 2001  & 2001  & 2065  & NA    & 2733  & 732  & 177.6  & 1.27  & 2.46   & 11.71  & 2.845 \\
(-0.5,-.20,.53)& 1924  & 1924  & 1996  & 2044  & 2700  & 776  & 369.6  & 2.441 & 0.785  & 0.559  & 0.842 \\
(-0.33,-.11,.46) & 2201  & 2201  & 2376  & 2778  & 3145  & 944  & 465.9  & 0.526 & 0.396  & 0.297  & 3.328 \\
(-0.1,-.17,.49) & 1725  & 1725  & 1968  & NA    & 2668  & 943  & 192.9  & 0.563 & 0.696  & 13.76  & 2.845 \\
(-0.75,-.17,.60) & 1189  & 1189  & 1217  & NA    & 1969  & 780  & 302.7  & 1.297 & 1.034  & 3.63   & 2.845 \\
(-1.25,-.14,.43) & 1326  & 1326  & 1366  & NA    & 2080  & 754  & 129.6  & 1.15  & 4.63   & 17.34  & 2.845 \\
(-0.66,-.11,.57) & 2496  & 2496  & 2540  & 2820  & 3036  & 540  & 121.9  & 0.909 & 0.697  & 1.278  & 1.334 \\
(-0.25,-.10,.44) & 907   & 907   & 1222  & NA    & 1699  & 792  & 133.29 & 0.549 & 0.418  & 0.366  & 2.845 \\
(-1.75,-.24,.23) & 2867  & 2867  & 2911  & NA    & 4331  & 1464 & 93.4   & 1.128 & 2.846  & 2.181  & 2.845 \\
(0,+.17,.39)   & 1855  & 1855  & 2634  & NA    & 3152  & 1297 & 122.5  & 0.28  & 0.272  & 2.113  & 2.845 \\
(-2,+.24,.45)  & 1894  & 1894  & 2247  & 4861  & 5600  & 3706 & 113.1  & 0.707 & 2.187  & 3.49   & 6.348 \\
(-0.25,-.11,.27) & 718   & 718   & 739   & NA    & 1358  & 640  & 145.4  & 1.488 & 6.322  & 14.46  & 2.845 \\
(-1.33,+.08,.20) & 1181  & 1181  & 2445  & 2864  & 3180  & 1999 & 167.5  & 0.164 & 0.156  & 0.31   & 0.489 \\
(-0.5,-.06,.51) & 3070  & 3070  & 3410  & NA    & 3855  & 785  & 128.9  & 0.538 & 0.537  & 5.073  & 2.845 \\
(-0.2,+.04,.29)& 4991  & 4991  & 5014  & 5162  & 5565  & 574  & 121    & 1.688 & 0.858  & 0.709  & 2.677 \\
(-0.3,-.11,.48)& 765   & 765   & 807   & 1079  & 1413  & 648  & 178.6  & 0.787 & 0.592  & 1.128  & 5.773 \\
(-3,+.04,.59)  & 40653 & 40653 & 41838 & 49118 & 50526 & 9873 & 350.8  & 1.138 & 2.05   & 3.971  & 6.851 \\
(0,+.17,.38)   & 890   & 890   & 1018  & NA    & 1623  & 733  & 233.2  & 0.841 & 0.619  & 0.276  & 2.845 \\
(-0.4,-.03,.49)  & 1131  & 1131  & 1218  & 1520  & 1717  & 586  & 229.7  & 0.756 & 0.573  & 0.353  & 0.736 \\
(-0.9,+.34,.29)  & 1604  & 1604  & 2236  & NA    & 2790  & 1186 & 248    & 0.318 & 4.09   & 4.69   & 2.845 \\
(-0.7,-.19,.56)  & 1136  & 1136  & 2402  & 2796  & 2974  & 1838 & 272.7  & 0.179 & 0.165  & 0.102  & 4.72  \\
(-0.8,+.05,.54)  & 1224  & 1224  & 1248  & 1832  & 1915  & 691  & 275.2  & 0.509 & 0.475  & 0.361  & 3.69  \\
(-1.6,-.17,.44)  & 2057  & 2057  & 2406  & NA    & 4055  & 1998 & 132    & 1.83  & 1.47   & 1.5    & 2.845 \\
(-1.4,-.03,.56)  & 2983  & 2983  & 3007  & 3105  & 3554  & 571  & 177.3  & 1.12  & 1.22   & 18.6   & 24.3  \\
(-1.9,-.15,.34)  & 6583  & 6583  & 8926  & NA    & 9621  & 3038 & 95.4   & 0.679 & 0.762  & 0.816  & 2.845 \\
(-0.5,+.02,.38)  & 890   & 890   & 914   & 1285  & 1581  & 691  & 182.6  & 0.765 & 0.551  & 0.529  & 3     \\
(-0.25,+.18,.36) & 1268  & 1268  & 1288  & 1715  & 2192  & 924  & 556.8  & 0.658 & 0.446  & 0.927  & 5.35 \\
\hline
\end{tabular}
\end{table}

\subsection{World (-1,-.17,.00) and World (-1,0,1) \label{app:n1n1}}

\begin{figure}
\centering
\begin{tabular}{cc}
\includegraphics[width=.5\textwidth]{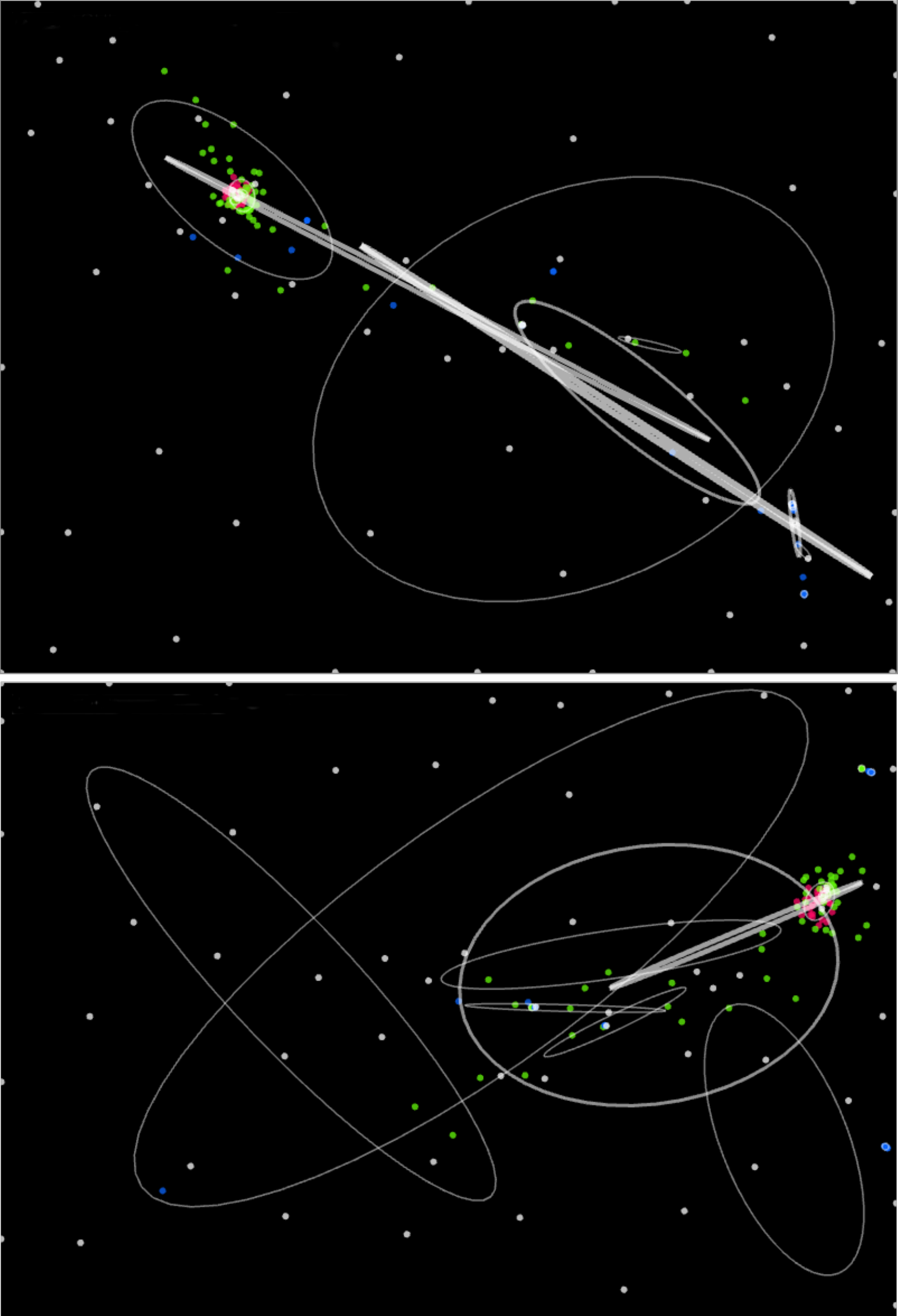}
&
\includegraphics[width=.48\textwidth]{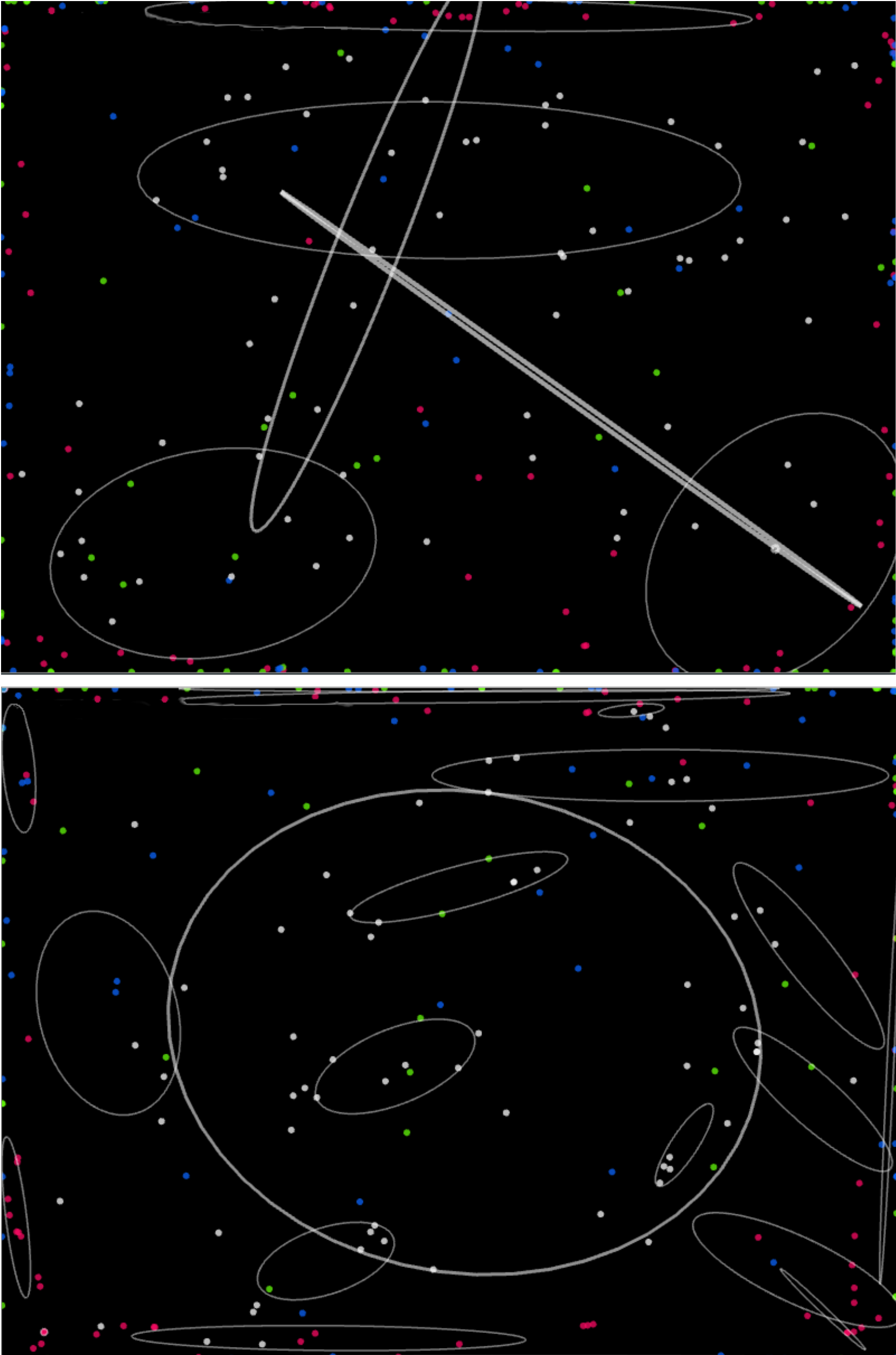}
\end{tabular}
\caption{Two arbitrarily chosen frames (17820 and 18007) of World $(-1,-.17,.00)$ and another two (10120 and 12846) of World $(-1,0,1)$. BUNCH started at frames 11560 and 5535 respectively. Cf.\ Fig.\ \ref{fig:p09_30_f1406+1541}. \label{fig:n17_00_and_0_1_screenshots}}
\end{figure}

\begin{figure*}
\centering
\begin{tabular}{c}
\includegraphics[width=1\textwidth]{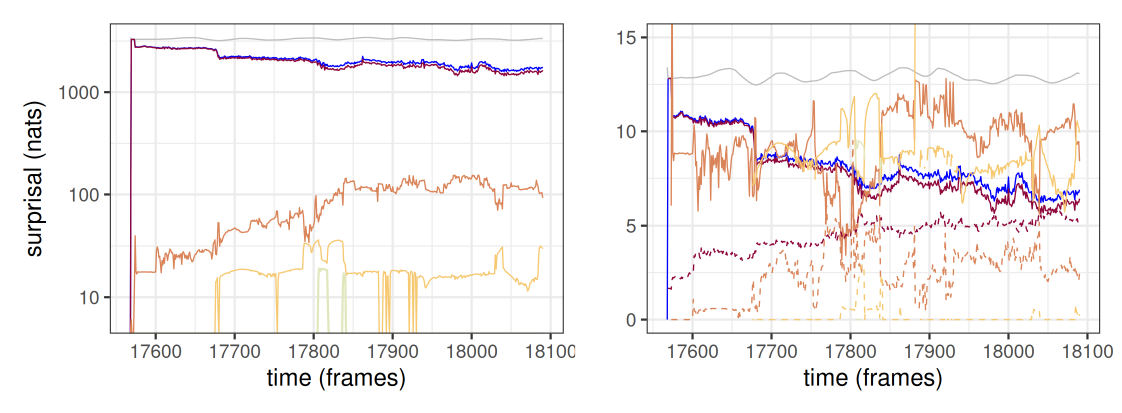}
\\
\includegraphics[width=1\textwidth]{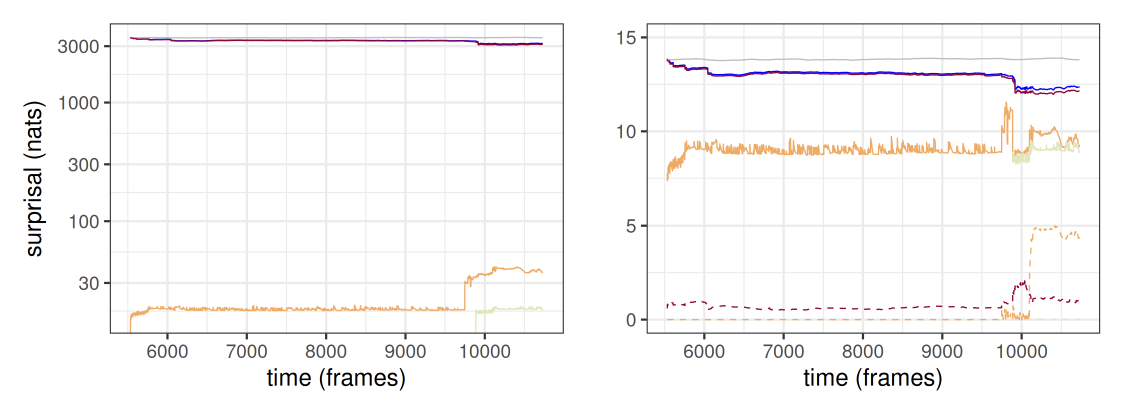}
\end{tabular}
\caption{The surprisal of entities at each dendron level for (top) World $(-1,-.17,.00)$ and (bottom) World $(-1,0,1)$. Cf.\ Fig.\ \ref{fig:p09_30_su_L_sum_mean}. \label{fig:n17_0_and_0_1_su_L_sum_mean}}
\end{figure*}

\begin{figure}
\centering
\begin{tabular}{cc}
\includegraphics[width=.5\textwidth]{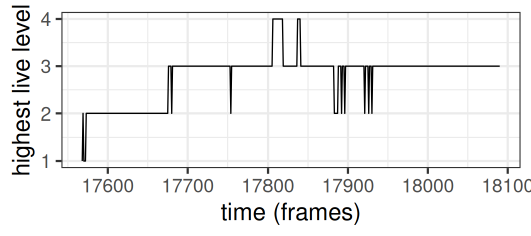}
&
\includegraphics[width=.5\textwidth]{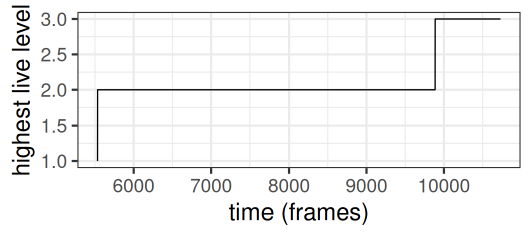}
\end{tabular}
\caption{Evolution of the current maximum dendron level for (left) Worlds $(-1,-.17,.00)$ and (right) $(-1,0,1)$. Cf.\ Fig.\ \ref{fig:p09_30_hystlev}. \label{fig:n17_0_and_0_1_hystlev}}
\end{figure}

\begin{figure}
\centering
\begin{tabular}{cc}
\includegraphics[width=.5\textwidth]{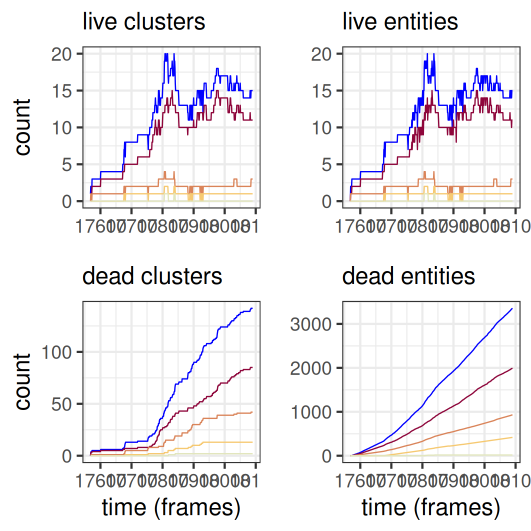}
&
\includegraphics[width=.5\textwidth]{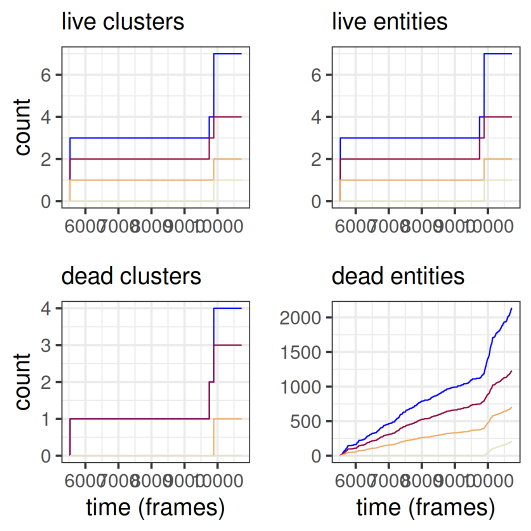}
\end{tabular}
\caption{Number of live and dead clusters and entities at each level for (left) World $(-1,-.17,.00)$ and (right) World $(-1,0,1)$. Cf.\ Fig.\ \ref{fig:p09_30_LivDed_C_E}. \label{fig:n17_0_and_0_1_LivDed_C_E}}
\end{figure}

\begin{figure}
\centering
\begin{tabular}{cc}
\includegraphics[width=.5\textwidth]{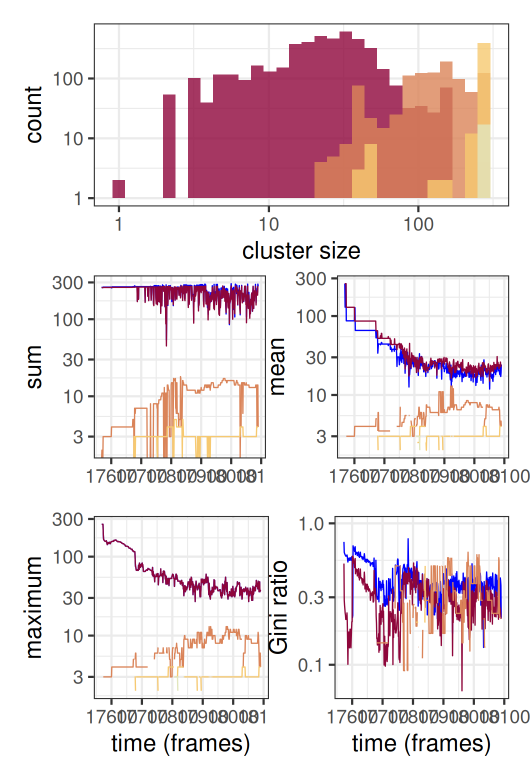}
&
\includegraphics[width=.5\textwidth]{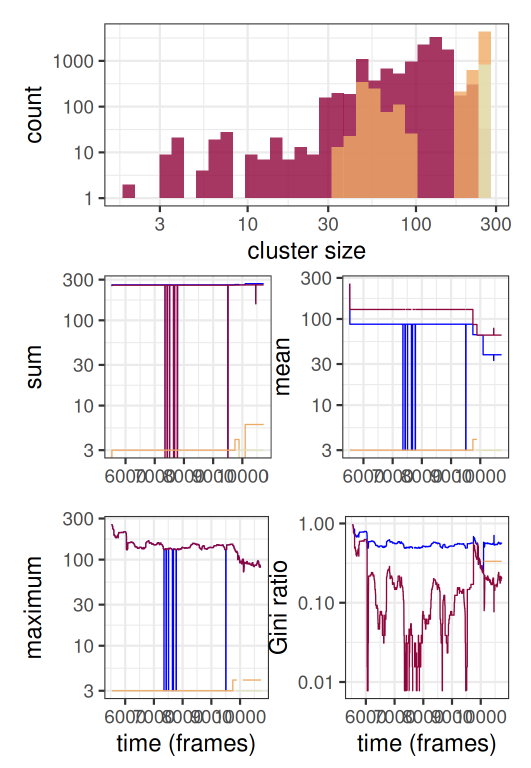}
\end{tabular}
\caption{Cluster size histogram and time-series of statistics, for each level, for (left) World $(-1,-.17,.00)$ and (right) World $(-1,0,1)$. Cf.\ Fig.\ \ref{fig:p09_30_CluSiz_hist_statime}. \label{fig:n17_0_and_0_1_CluSiz_hist_statime}}
\end{figure}

\begin{figure*}
\begin{tabular}{c}
\includegraphics[width=1\textwidth]{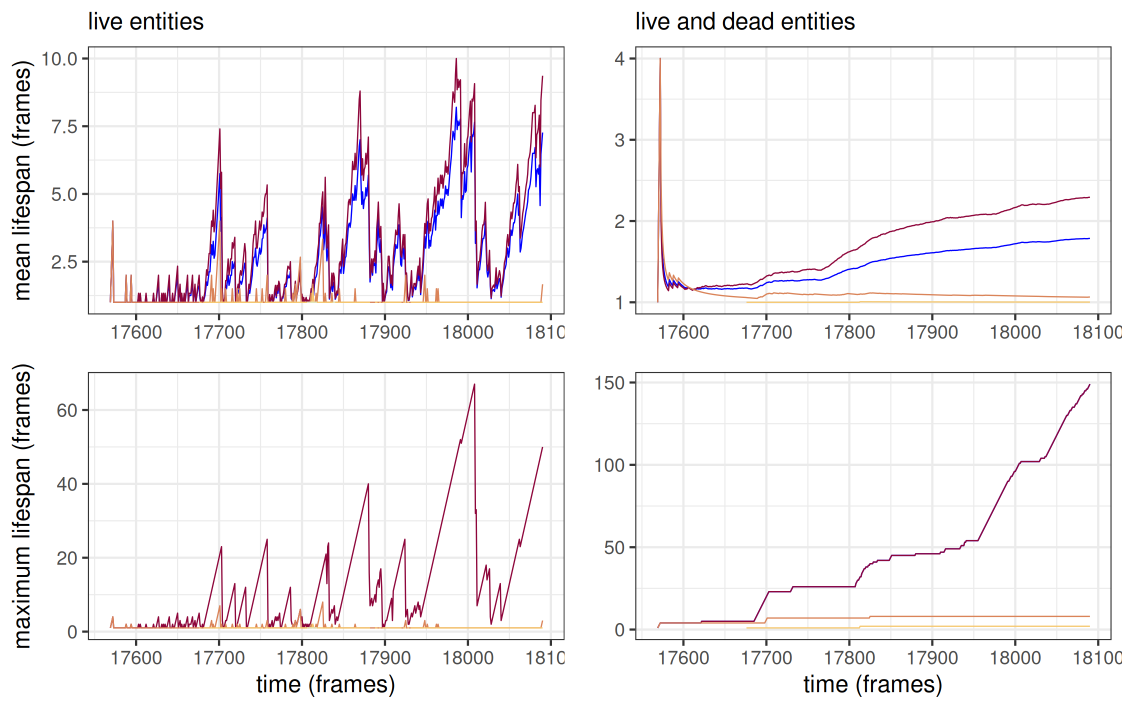}
\\
\includegraphics[width=1\textwidth]{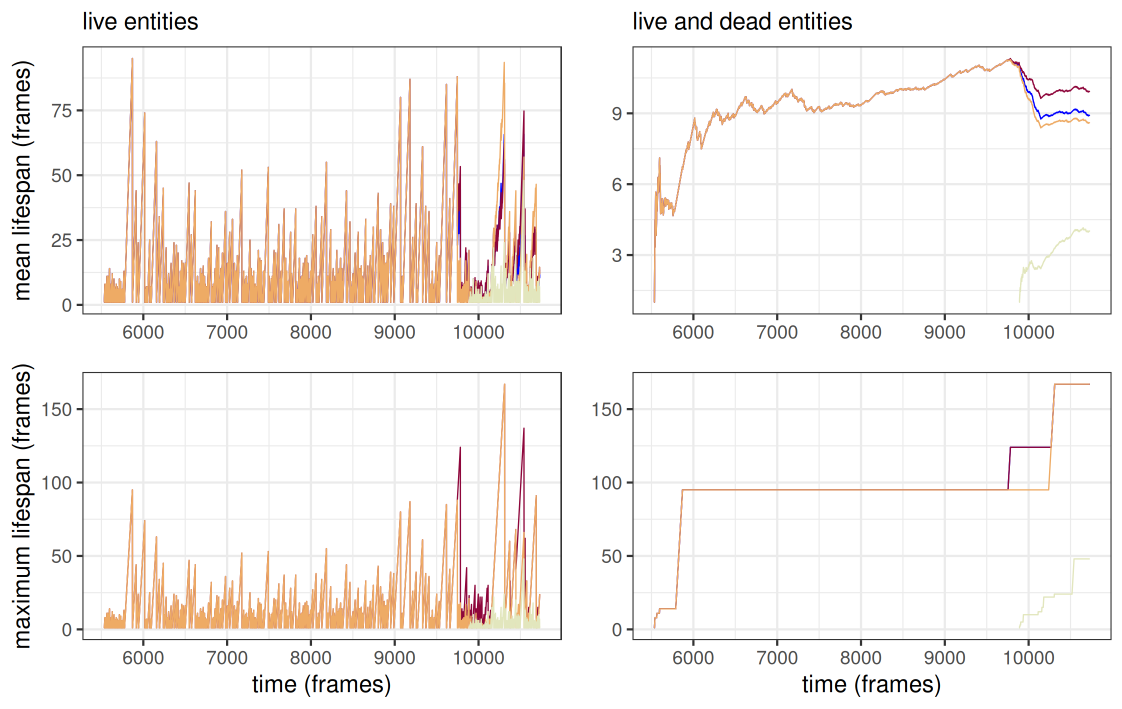}
\end{tabular}
\caption{Current mean and maximum lifespans for only live and both live and dead entities for (left) World $(-1,-.17,.00)$ and (right) World $(-1,0,1)$. Cf.\ Fig.\ \ref{fig:p09_30_L_cur}. \label{fig:n17_0_and_0_1_L_cur}}
\end{figure*}

\begin{figure}
\centering
\begin{tabular}{cc}
\includegraphics[width=.5\textwidth]{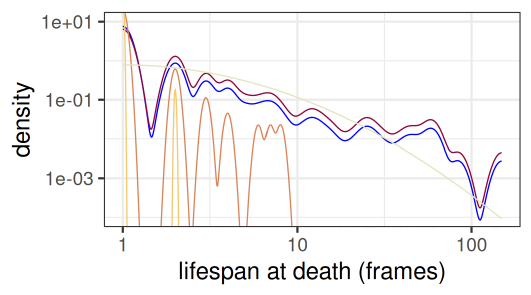}
&
\includegraphics[width=.5\textwidth]{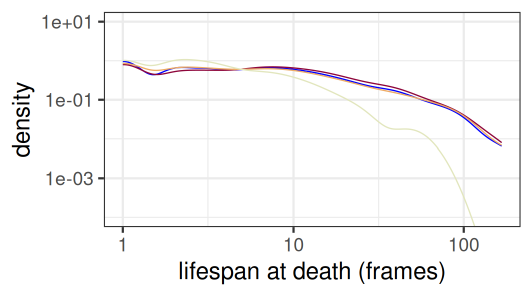}
\end{tabular}
\caption{Distribution of lifespans across all entities at the end of the simulation on log-log scale for (left) World $(-1,-.17,.00)$ and (right) World $(-1,0,1)$. Cf.\ Fig.\ \ref{fig:p09_30_LTcum_dens}. \label{fig:n17_0_and_0_1_LTcum_dens}}
\end{figure}

\begin{figure}
\begin{tabular}{c}
\includegraphics[width=1\textwidth]{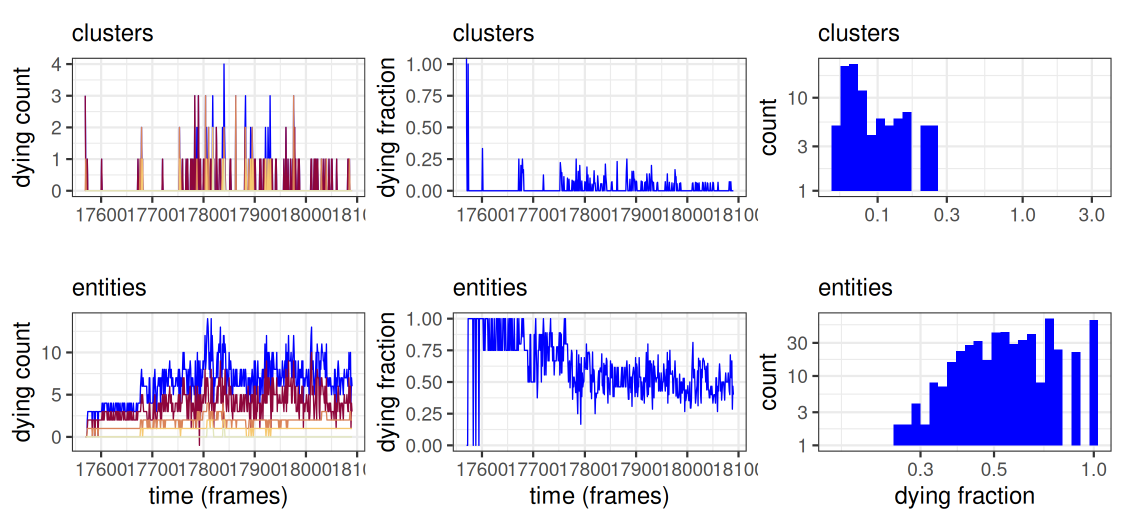}
\\
\includegraphics[width=1\textwidth]{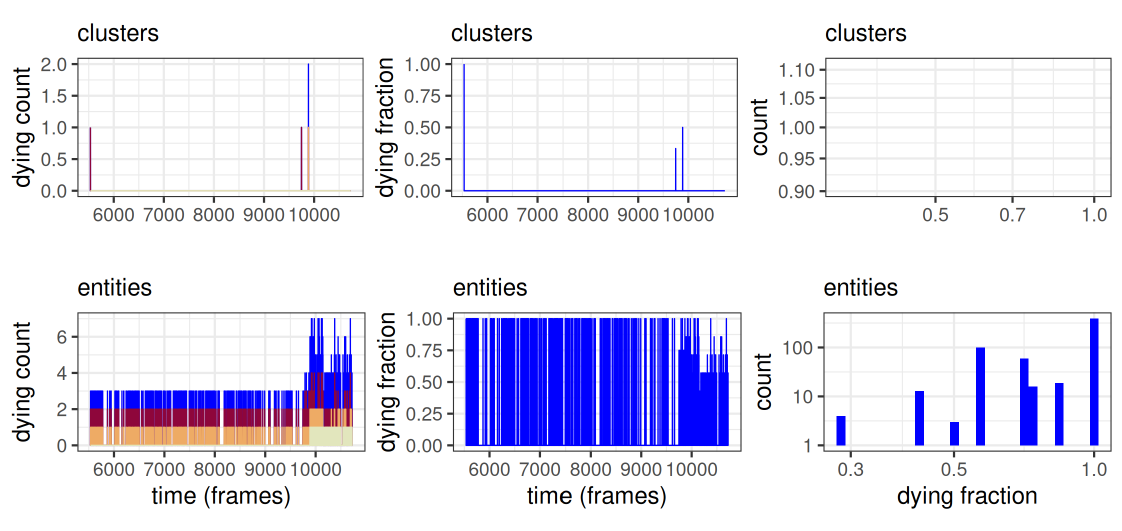}
\end{tabular}
\caption{Count and fraction of dying clusters and entities for (top) World $(-1,-.17,.00)$ and (bottom) World $(-1,0,1)$. Cf.\ Fig.\ \ref{fig:p09_30_Dying_C_E}. \label{fig:n17_0_and_0_1_Dying_C_E}}
\end{figure}

\begin{figure}
\begin{tabular}{c}
\includegraphics[width=.75\textwidth]{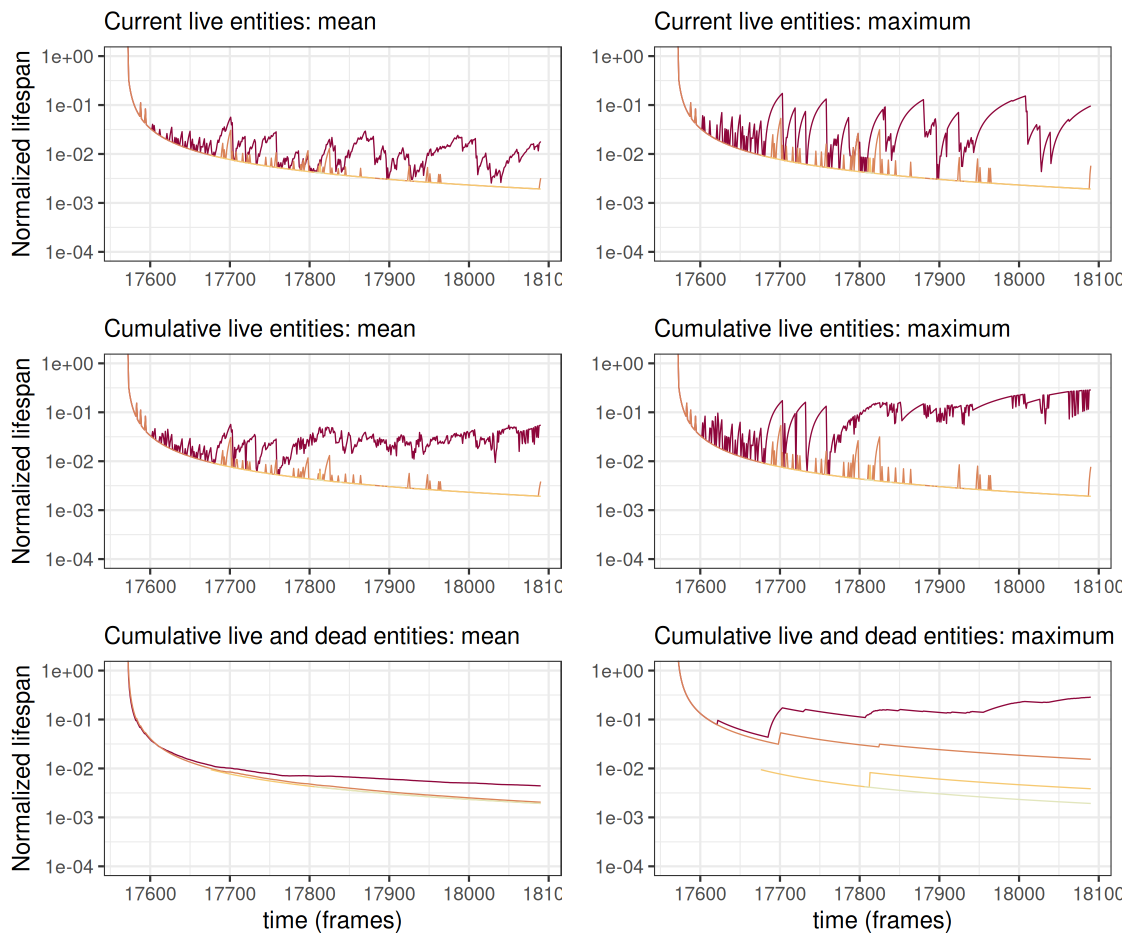}
\\
\includegraphics[width=.75\textwidth]{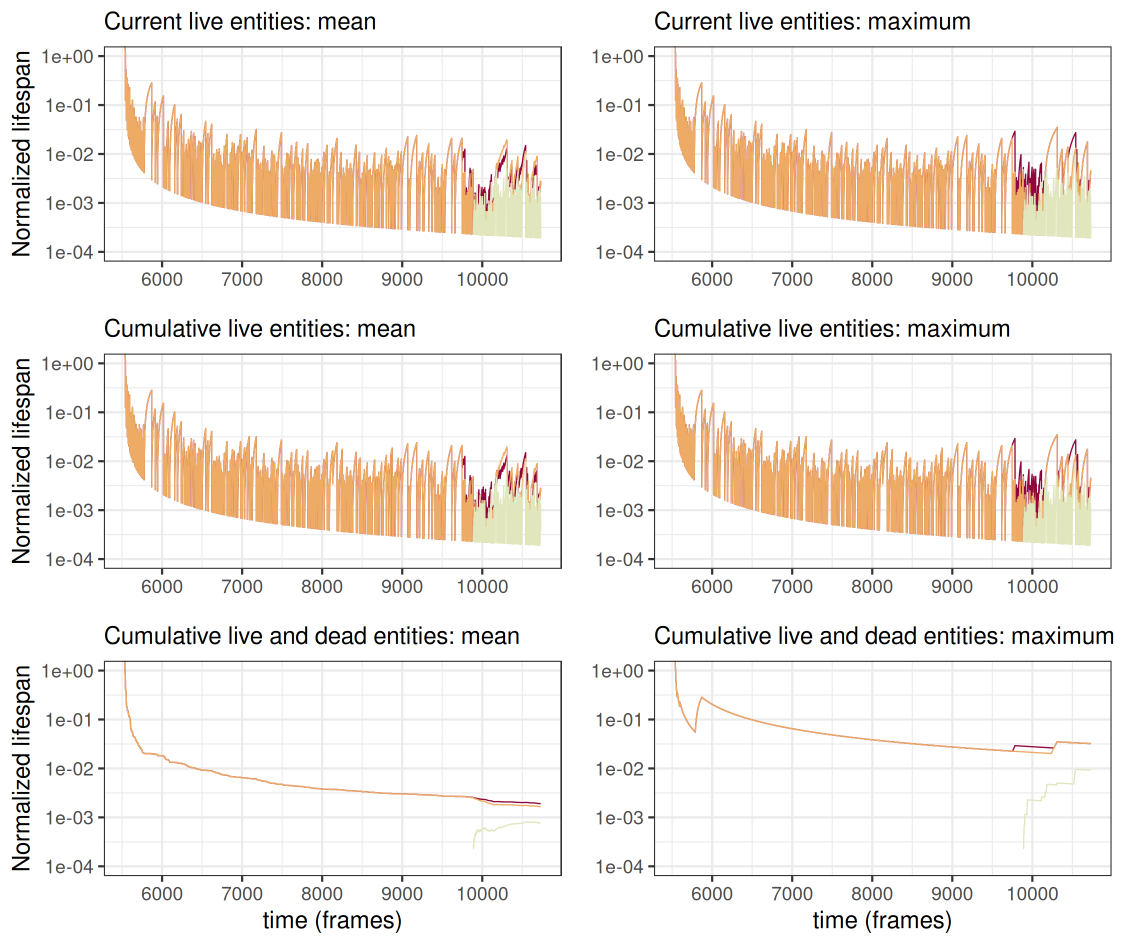}
\end{tabular}
\caption{Normalized live current, live cumulative and all cumulative lifespans statistics for (top) World $(-1,-.17,.00)$ and (bottom) World $(-1,0,1)$. Cf.\ Fig.\ \ref{fig:p09_30_LN_cur} \label{fig:n17_0_and_0_1_LN_cur}}
\end{figure}

\begin{figure}
\centering
\begin{tabular}{cc}
\includegraphics[width=.5\textwidth]{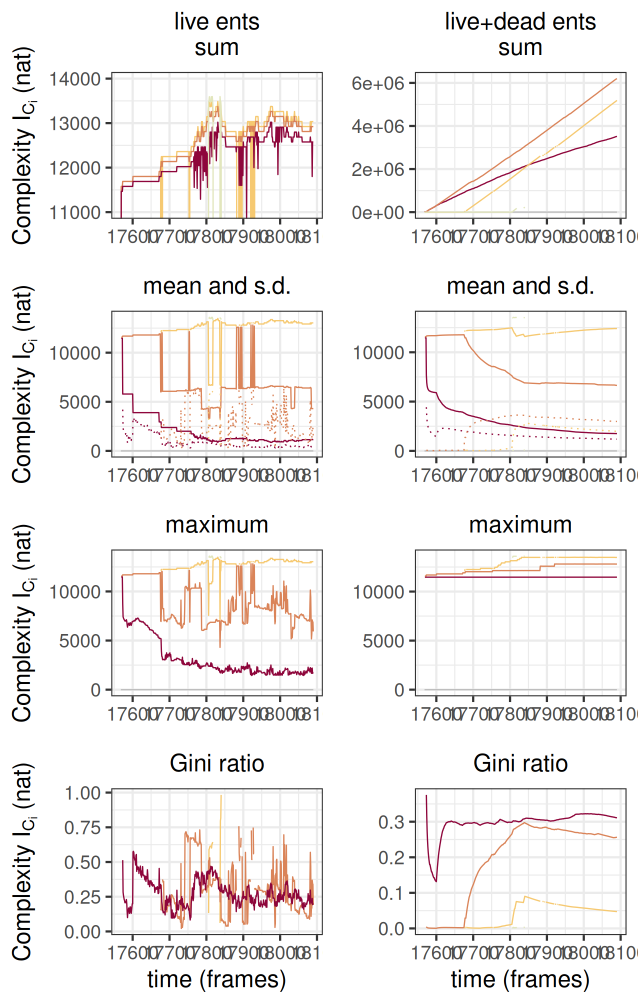}
&
\includegraphics[width=.5\textwidth]{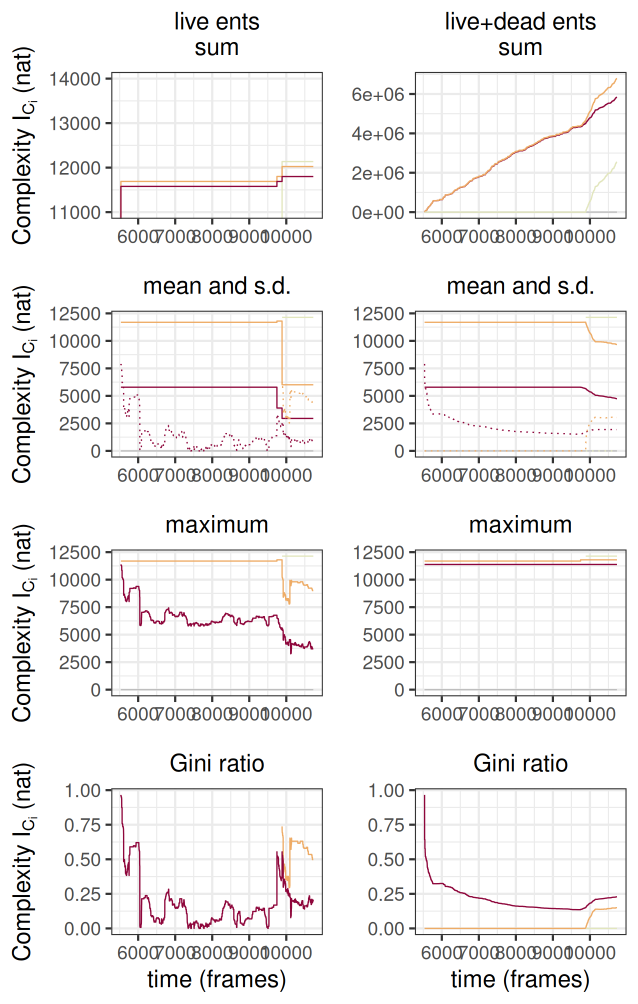}
\end{tabular}
\caption{Live and all entities complexity level-wise statistics for (left) World $(-1,-.17,.00)$ and (right) World $(-1,0,1)$. Cf.\ Fig.\ \ref{fig:p09_30_LivDed_Ic}. \label{fig:n17_0_and_0_1_LivDed_Ic}}
\end{figure}

\begin{figure}
\centering
\begin{tabular}{c}
\includegraphics[width=.75\textwidth]{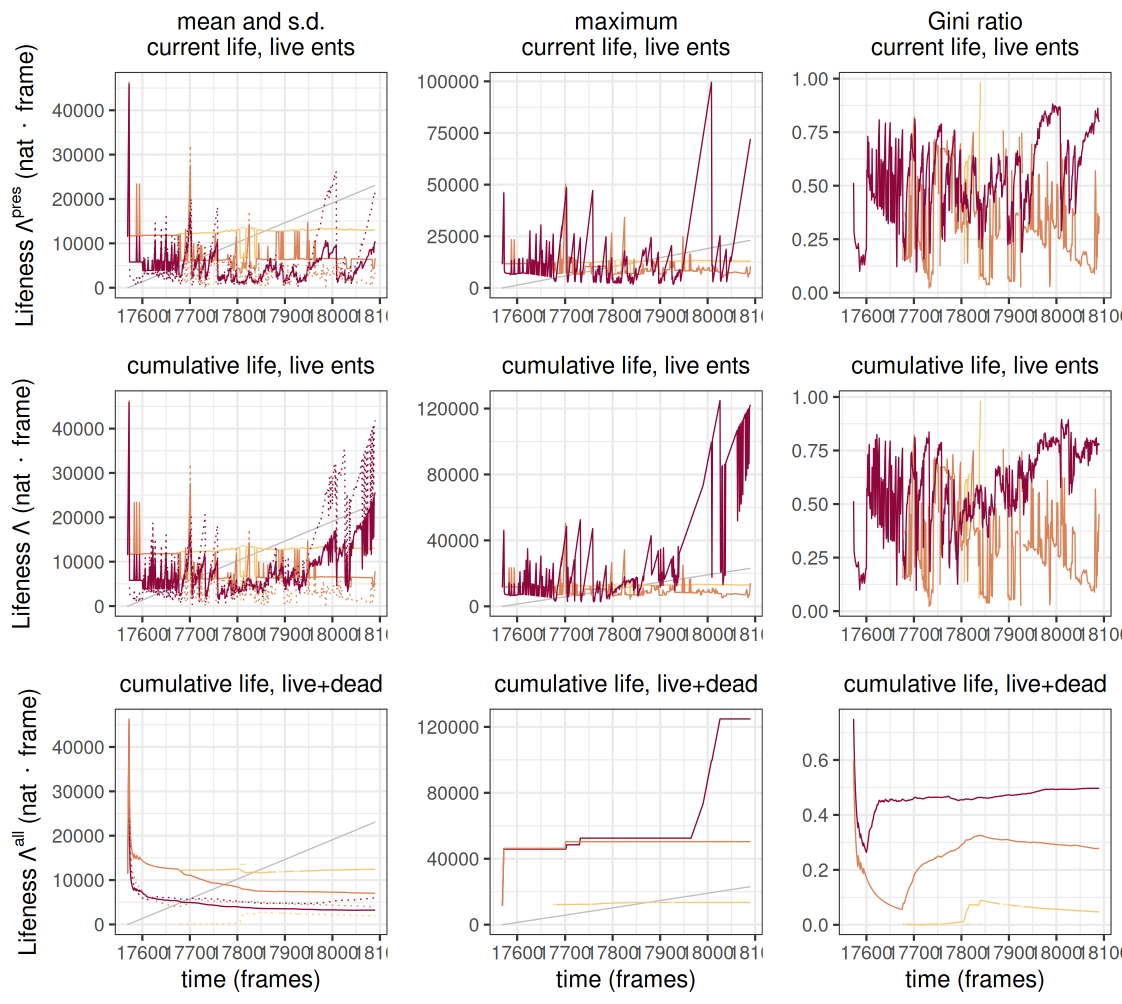}
\\
\includegraphics[width=.75\textwidth]{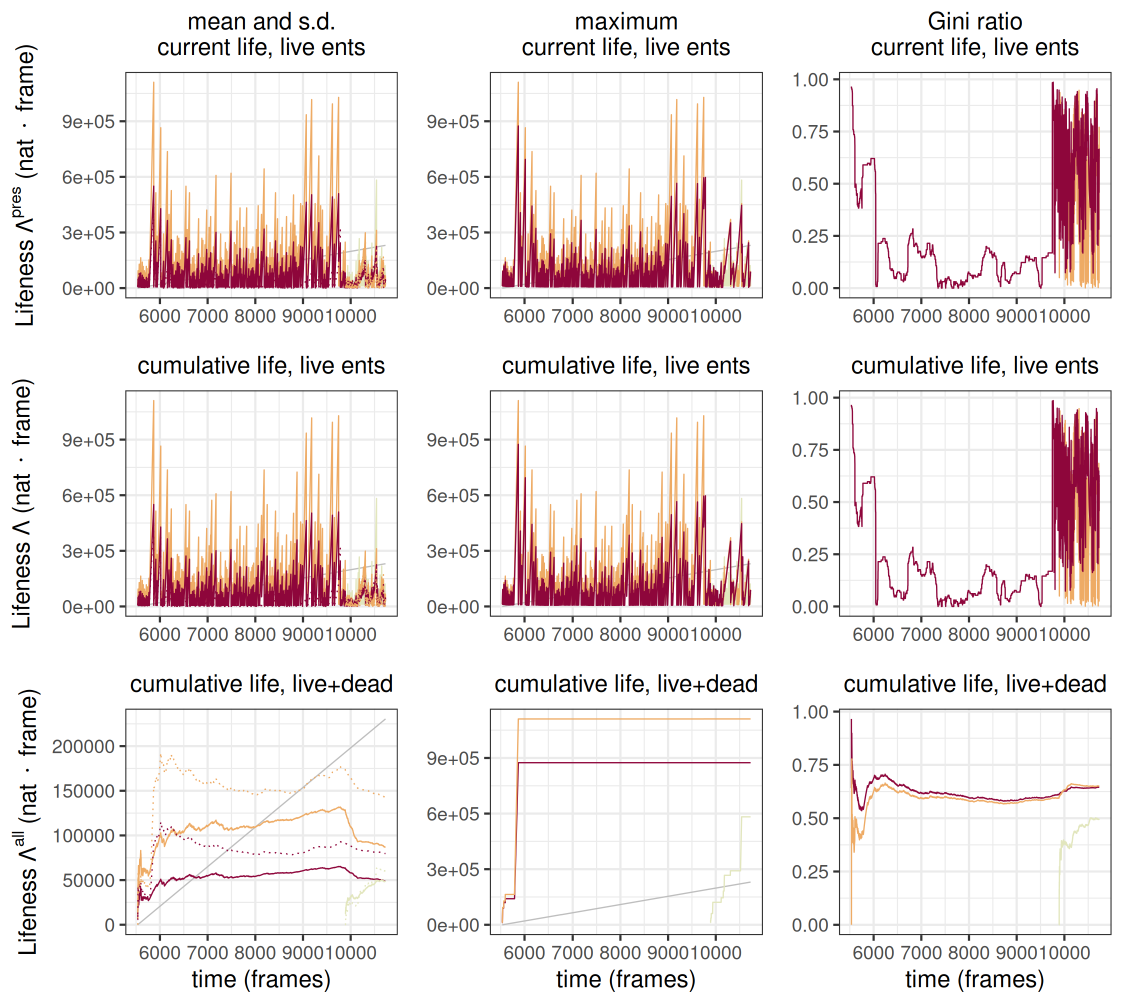}
\end{tabular}
\caption{Lifeness for (top) World $(-1,-.17,.00)$ and (bottom) World $(-1,0,1)$. Cf.\ Fig.\ \ref{fig:p09_30_Lifeness}.  \label{fig:n17_0_and_0_1_Lifeness}}
\end{figure}

\begin{figure}
\begin{tabular}{c}
\includegraphics[width=1\textwidth]{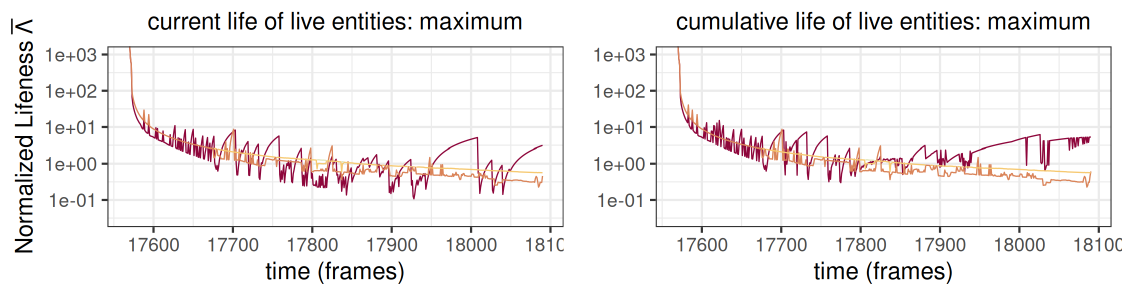}
\\
\includegraphics[width=1\textwidth]{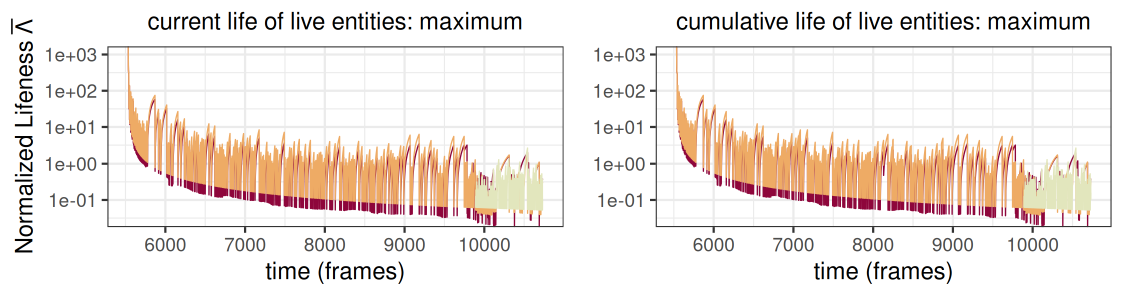}
\end{tabular}
\caption{Normalized maximum lifeness for the current life of live entities for (top) World $(-1,-.17,.00)$ and (bottom) World $(-1,0,1)$. Cf.\ Fig.\ \ref{fig:p09_30_LifenessN}. \label{fig:n17_0_and_0_1_LifenessN}}
\end{figure}

World (-1,-.17,.00) has predominantly attractive and purely symmetric interactions $\chi_{(-1,.09,.30)} \approx \begin{bmatrix} 
-.52 & -.94 & .29 & -.38 \\
-.94 & .44 & -.46 & .12 \\
.29 & -.46 & -.22 & -.18 \\
-.38 & .12 & -.18 & .70 \end{bmatrix}$, whereas World (-1,0,1) has purely skew-symmetric interactions $\chi_{(-1,0,1)} \approx \begin{bmatrix} 
.00 & .67 &  .35 & -.36 \\
-.67 & .00 & -.69 & .35 \\
-.35  & .69 & .00 & -.16 \\
.36  & -.35 & .16 & .00 \end{bmatrix}$. Both systems' interaction forces decrease as a power law with exponent -1. 
The time unit is set to frames (simulation time steps) instead of seconds because as the simulation progresses and the data structures storing relevant variables grow, the BUNCH algorithm slows down, burdened by writing and reading operations.
Figs. \ref{fig:n17_00_and_0_1_screenshots}, \ref{fig:n17_0_and_0_1_su_L_sum_mean}, \ref{fig:n17_0_and_0_1_hystlev},\ref{fig:n17_0_and_0_1_LivDed_C_E}, \ref{fig:n17_0_and_0_1_CluSiz_hist_statime}, \ref{fig:n17_0_and_0_1_L_cur}, \ref{fig:n17_0_and_0_1_LTcum_dens}, \ref{fig:n17_0_and_0_1_Dying_C_E}, \ref{fig:n17_0_and_0_1_LN_cur}, \ref{fig:n17_0_and_0_1_LivDed_Ic}, \ref{fig:n17_0_and_0_1_Lifeness}, \ref{fig:n17_0_and_0_1_LifenessN} are homologous to those of Section \ref{sec:p09_30}.

\subsection{World (-.33,-.11,.46) and World (-2,.24,.45) \label{app:n.33n2}}

\begin{figure}
\centering
\begin{tabular}{cc}
\includegraphics[width=.5\textwidth]{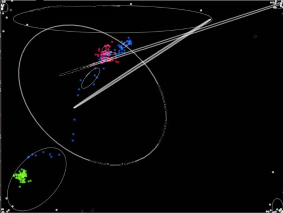}
&
\includegraphics[width=.48\textwidth]{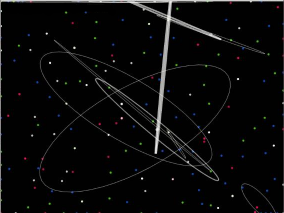}
\end{tabular}
\caption{Frame 2848 of World $(-.333,-.11,.46)$ and frame 5555 of World $(-2,.24,.45)$. BUNCH started at frames 2201 and 1894. Cf.\ Fig.\ \ref{fig:p09_30_f1406+1541}. \label{fig:n11_46_n.33_and_p24_45_n2_screenshots}}
\end{figure}

\begin{figure*}
\centering
\begin{tabular}{c}
\includegraphics[width=1\textwidth]{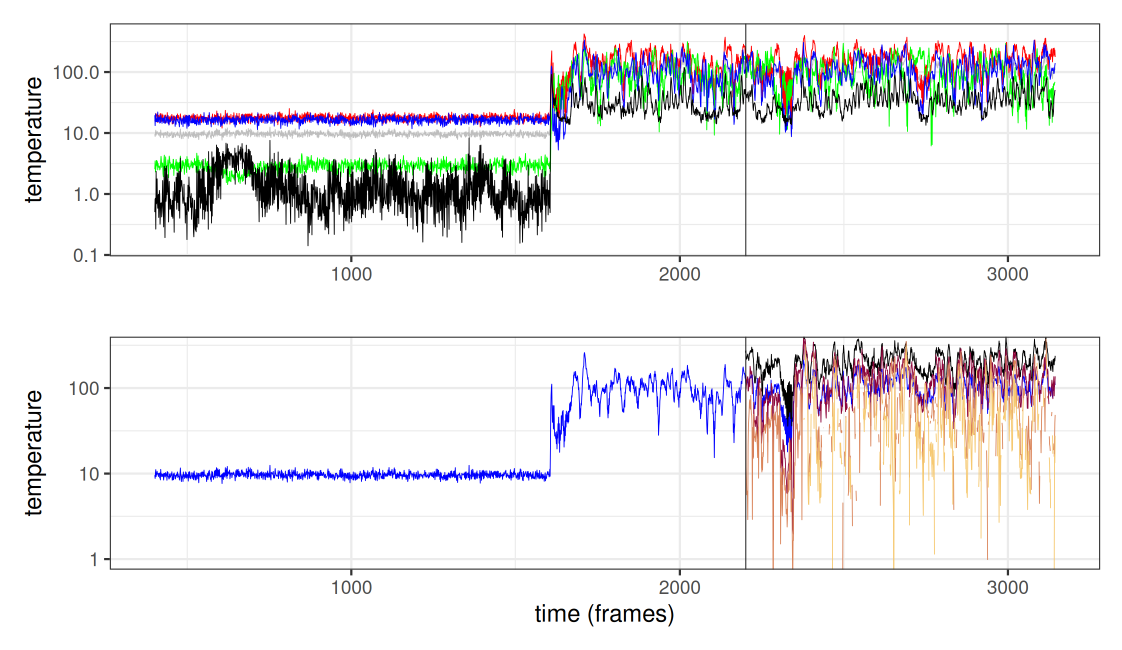}
\\
\includegraphics[width=1\textwidth]{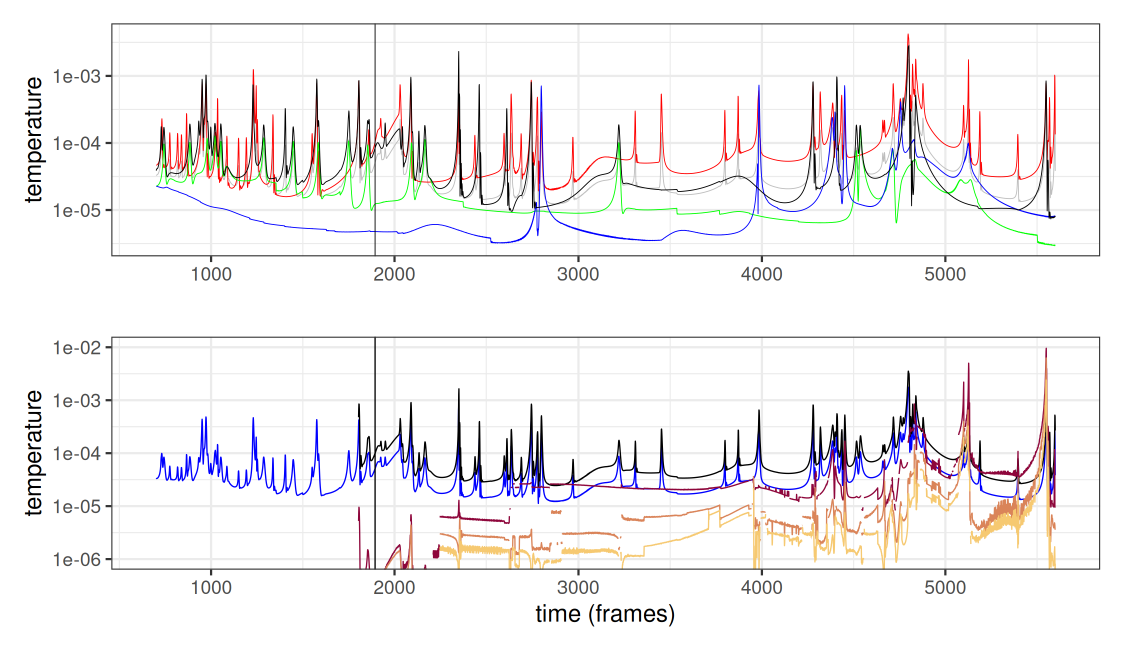}
\end{tabular}
\caption{The temperature for World $(-.333,-.11,.46)$ and $(-2,.24,.45)$. Cf.\ Fig.\ \ref{fig:p09_30_T_K_L}. \label{fig:n11_46_n.33_and_p24_45_n2_T_K_L}}
\end{figure*}

\begin{figure*}
\centering
\begin{tabular}{c}
\includegraphics[width=1\textwidth]{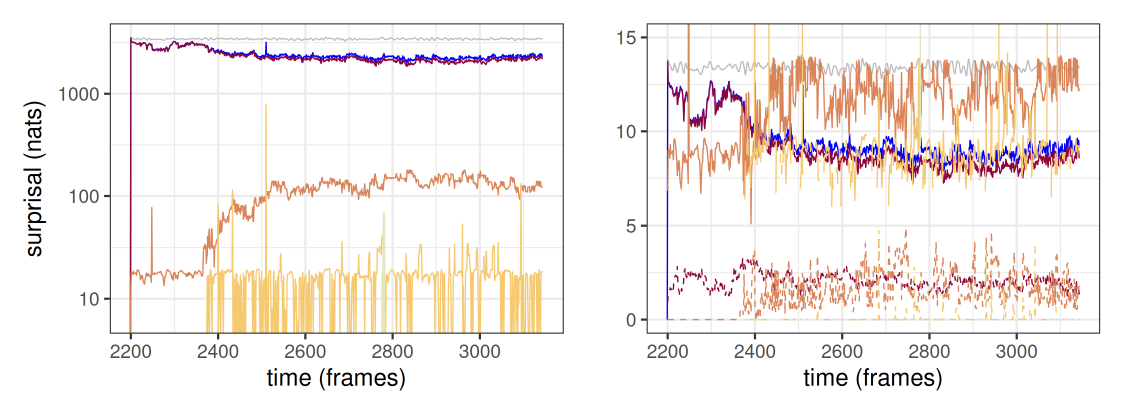}
\\
\includegraphics[width=1\textwidth]{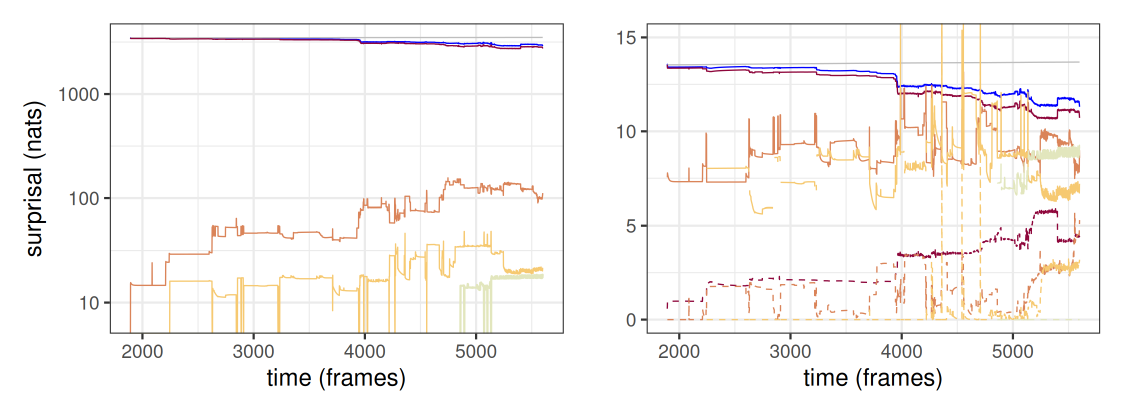}
\end{tabular}
\caption{The surprisal of entities at each dendron level for (top) World $(-.333,-.11,.46)$ and (bottom) World $(-2,.24,.45)$. Cf.\ Fig.\ \ref{fig:p09_30_su_L_sum_mean}. \label{fig:n11_46_n.33_and_p24_45_n2_su_L_sum_mean}}
\end{figure*}

\begin{figure}
\centering
\begin{tabular}{cc}
\includegraphics[width=.5\textwidth]{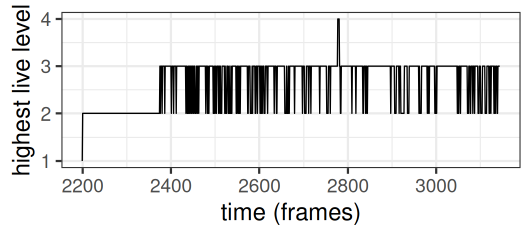}
&
\includegraphics[width=.5\textwidth]{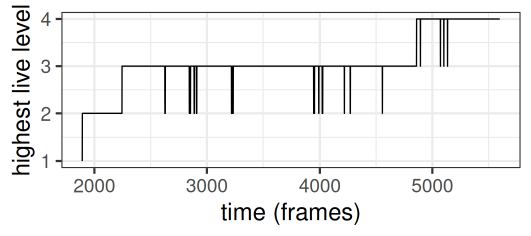}
\end{tabular}
\caption{Evolution of the current maximum dendron level for (left) Worlds $(-.333,-.11,.46)$ and (right) $(-2,.24,.45)$. Cf.\ Fig.\ \ref{fig:p09_30_hystlev}. \label{fig:n11_46_n.33_and_p24_45_n2_hystlev}}
\end{figure}

\begin{figure}
\centering
\begin{tabular}{cc}
\includegraphics[width=.5\textwidth]{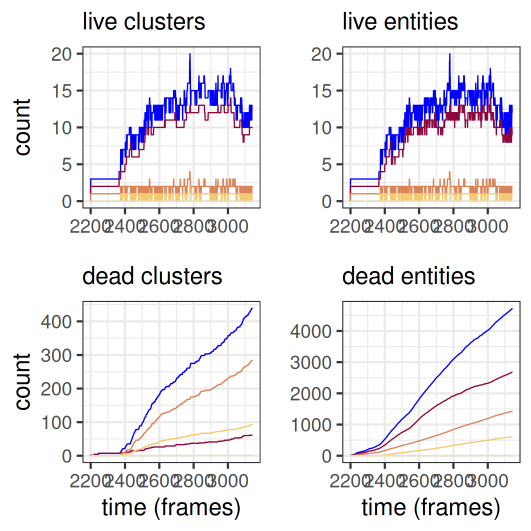}
&
\includegraphics[width=.5\textwidth]{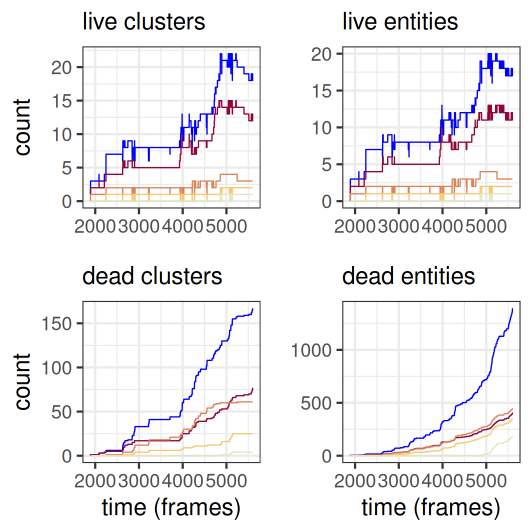}
\end{tabular}
\caption{Number of live and dead clusters and entities at each level for (left) World $(-.333,-.11,.46)$ and (right) World $(-2,.24,.45)$. Cf.\ Fig.\ \ref{fig:p09_30_LivDed_C_E}. \label{fig:n11_46_n.33_and_p24_45_n2_LivDed_C_E}}
\end{figure}

\begin{figure}
\centering
\begin{tabular}{cc}
\includegraphics[width=.5\textwidth]{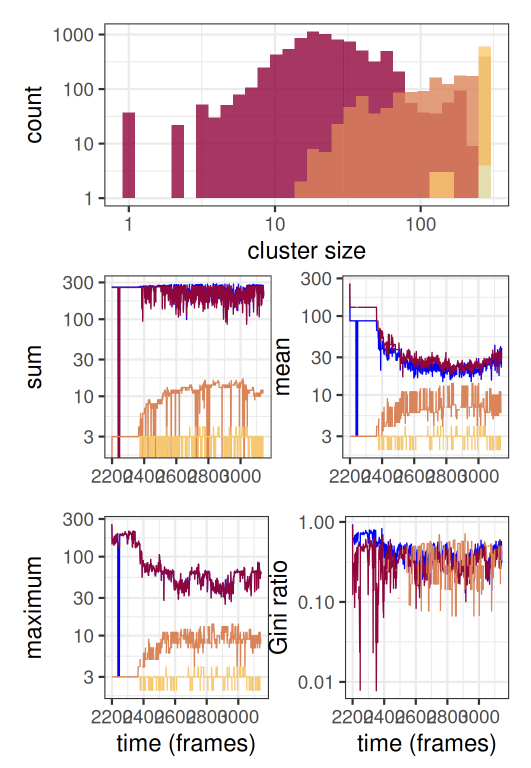}
&
\includegraphics[width=.5\textwidth]{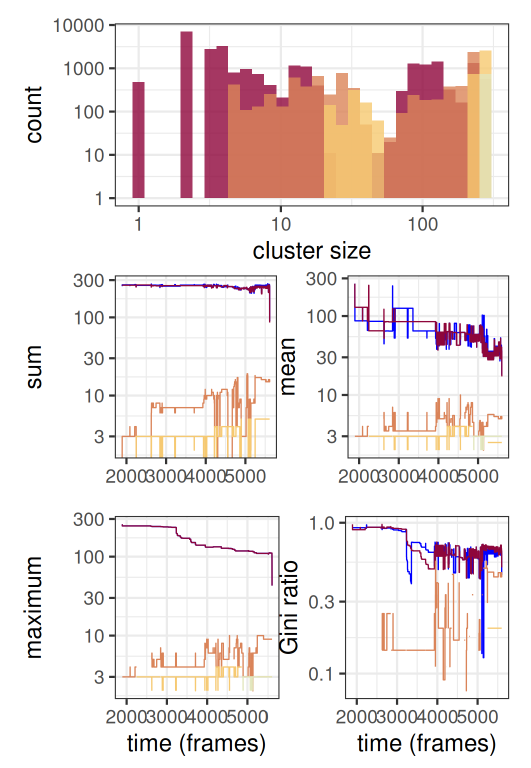}
\end{tabular}
\caption{Cluster size histogram and time-series of statistics, for each level, for (left) World $(-.333,-.11,.46)$ and (right) World $(-2,.24,.45)$. Cf.\ Fig.\ \ref{fig:p09_30_CluSiz_hist_statime}. \label{fig:n11_46_n.33_and_p24_45_n2_CluSiz_hist_statime}}
\end{figure}

\begin{figure*}
\begin{tabular}{c}
\includegraphics[width=1\textwidth]{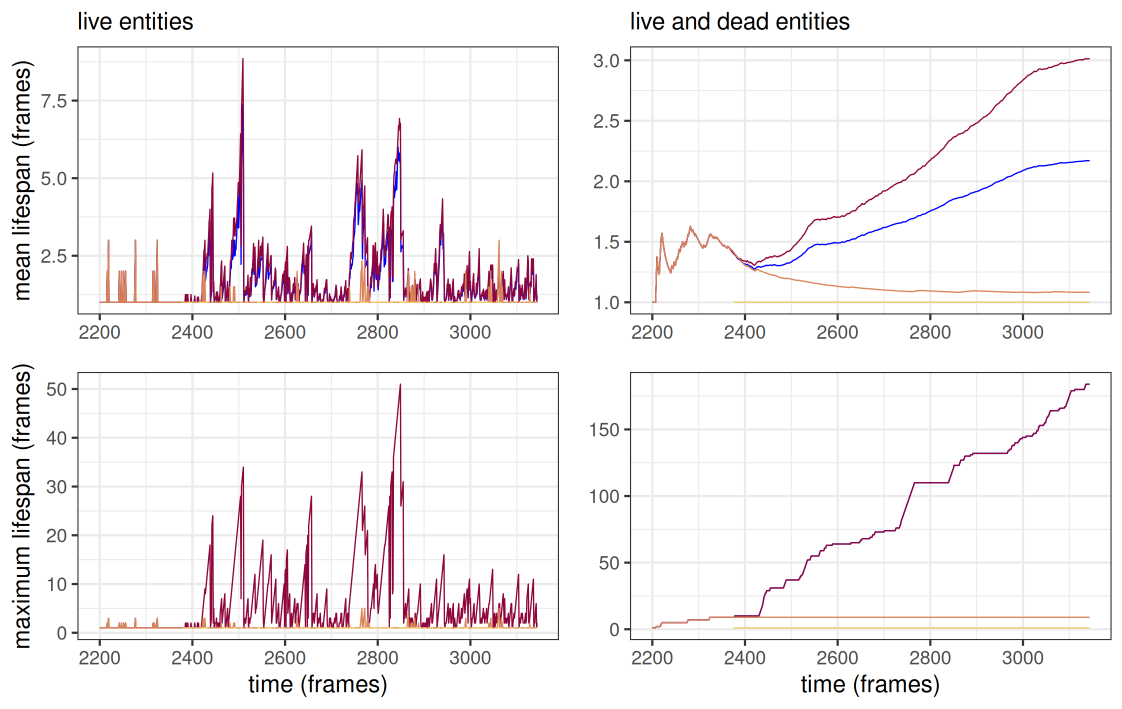}
\\
\includegraphics[width=1\textwidth]{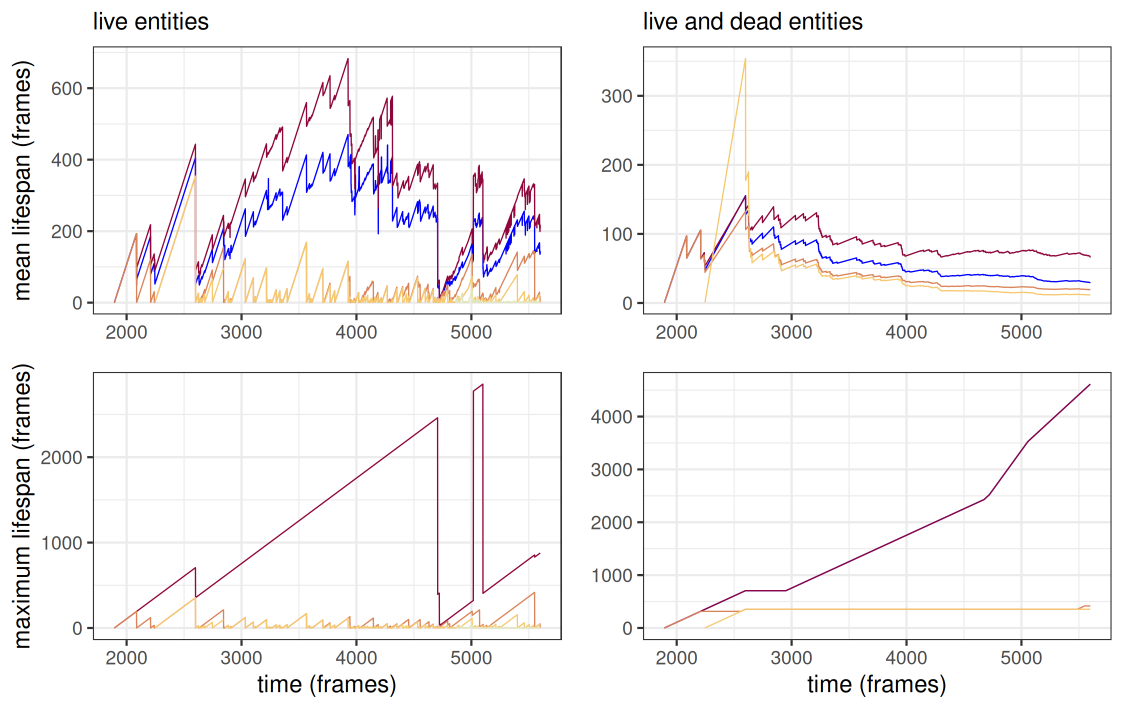}
\end{tabular}
\caption{Current mean and maximum lifespans for only live and both live and dead entities for (left) World $(-.333,-.11,.46)$ and (right) World $(-2,.24,.45)$. Cf.\ Fig.\ \ref{fig:p09_30_L_cur}. \label{fig:n11_46_n.33_and_p24_45_n2_L_cur}}
\end{figure*}

\begin{figure}
\centering
\begin{tabular}{cc}
\includegraphics[width=.5\textwidth]{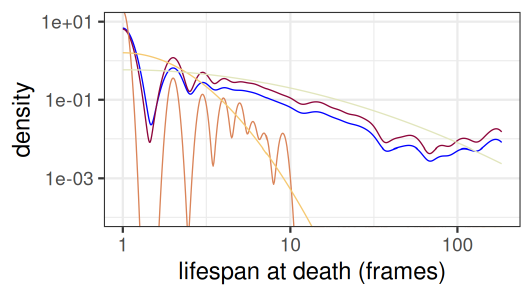}
&
\includegraphics[width=.5\textwidth]{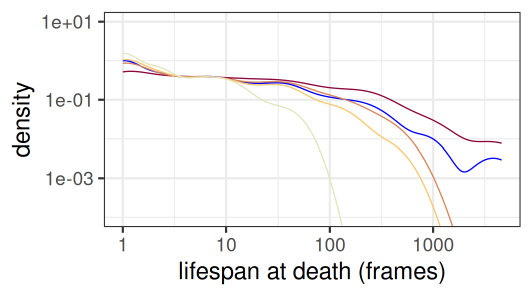}
\end{tabular}
\caption{Distribution of lifespans across all entities at the end of the simulation on log-log scale for (left) World $(-.333,-.11,.46)$ and (right) World $(-2,.24,.45)$. Cf.\ Fig.\ \ref{fig:p09_30_LTcum_dens}. \label{fig:n11_46_n.33_and_p24_45_n2_LTcum_dens}}
\end{figure}

\begin{figure}
\begin{tabular}{c}
\includegraphics[width=1\textwidth]{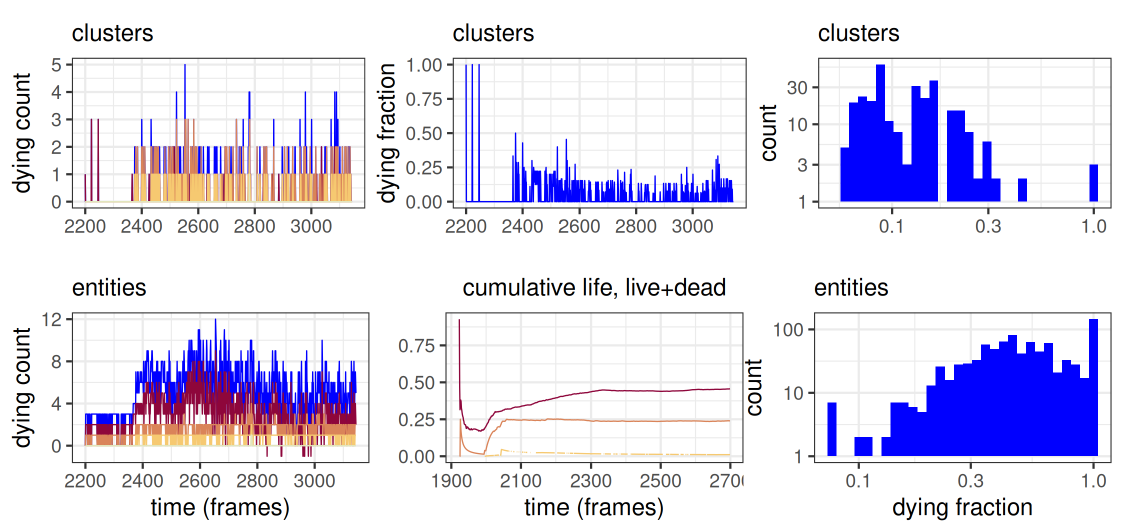}
\\
\includegraphics[width=1\textwidth]{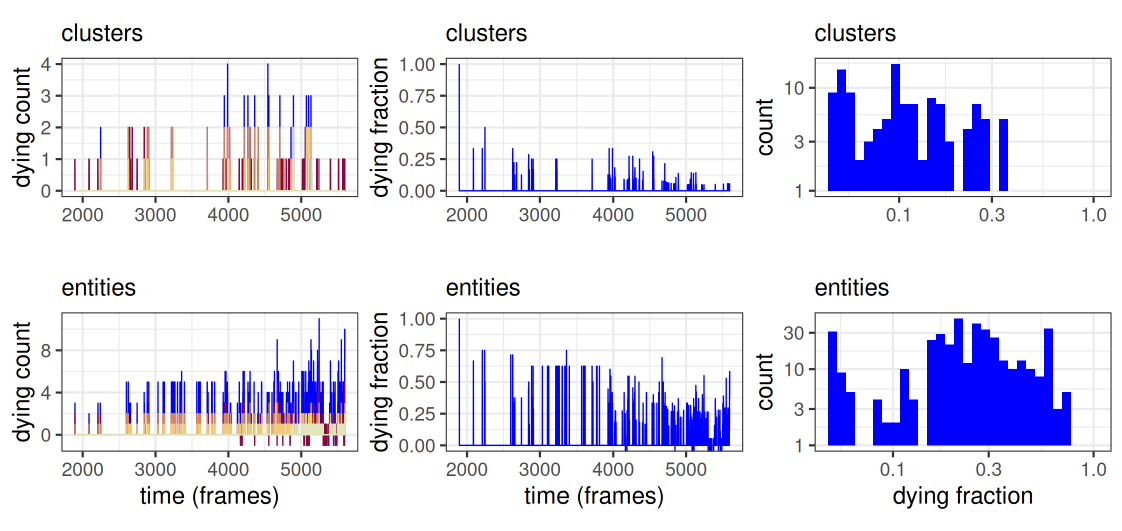}
\end{tabular}
\caption{Count and fraction of dying clusters and entities for (top) World $(-.333,-.11,.46)$ and (bottom) World $(-2,.24,.45)$. Cf.\ Fig.\ \ref{fig:p09_30_Dying_C_E}. \label{fig:n11_46_n.33_and_p24_45_n2_Dying_C_E}}
\end{figure}

\begin{figure}
\centering
\begin{tabular}{c}
\includegraphics[width=.75\textwidth]{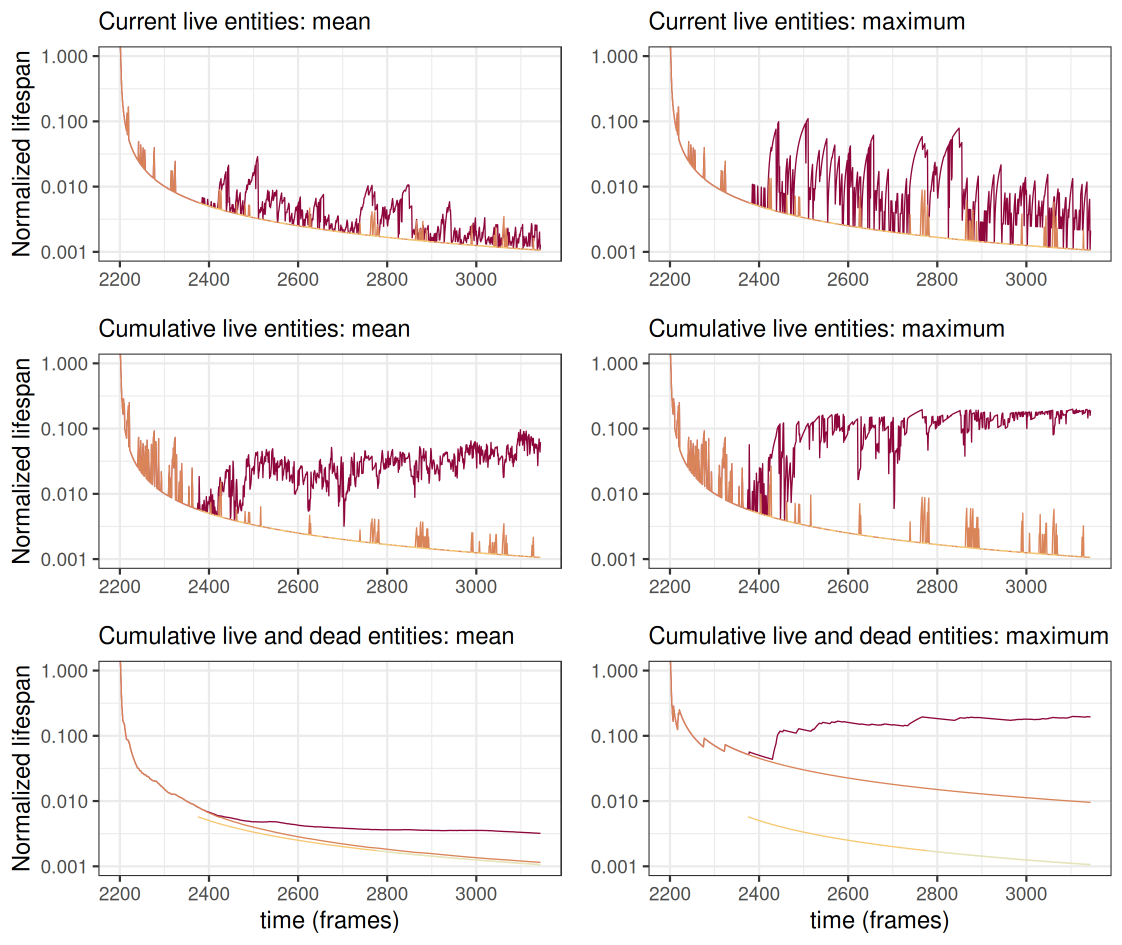}
\\
\includegraphics[width=.75\textwidth]{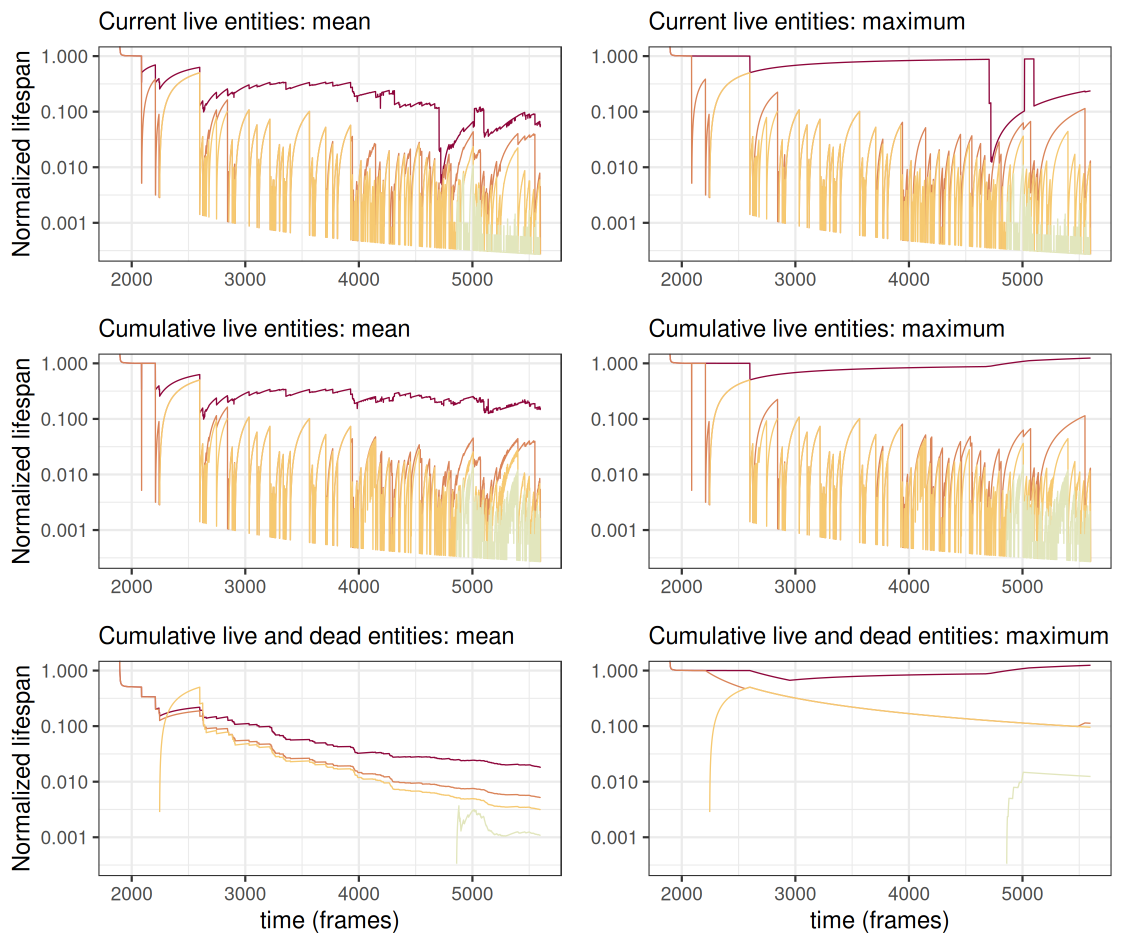}
\end{tabular}
\caption{Normalized live current, live cumulative and all cumulative lifespans statistics for (top) World $(-.333,-.11,.46)$ and (bottom) World $(-2,.24,.45)$. Cf.\ Fig.\ \ref{fig:p09_30_LN_cur} \label{fig:n11_46_n.33_and_p24_45_n2_LN_cur}}
\end{figure}

\begin{figure}
\centering
\begin{tabular}{cc}
\includegraphics[width=.5\textwidth]{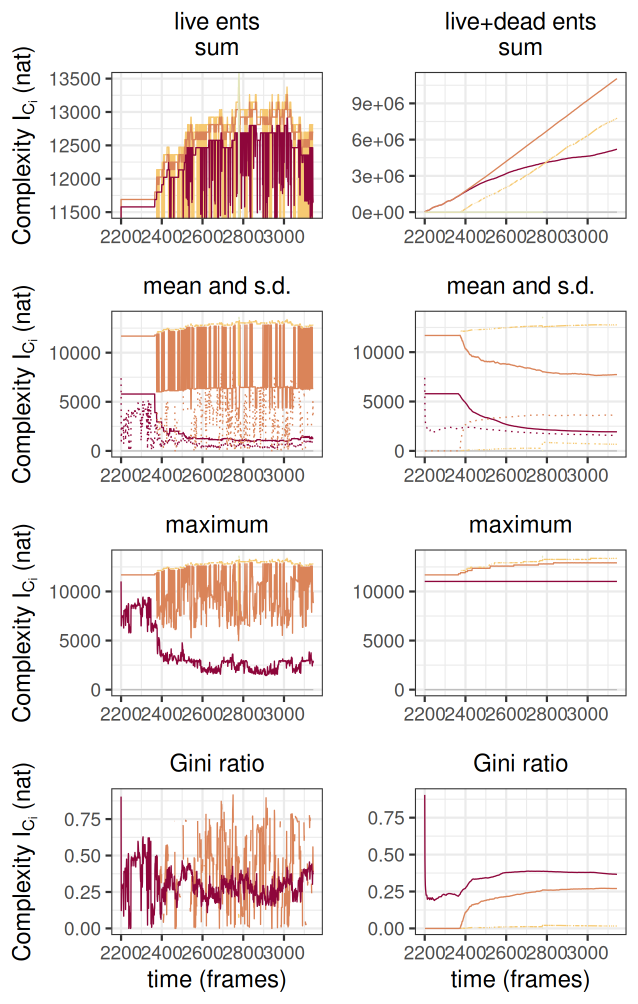}
&
\includegraphics[width=.5\textwidth]{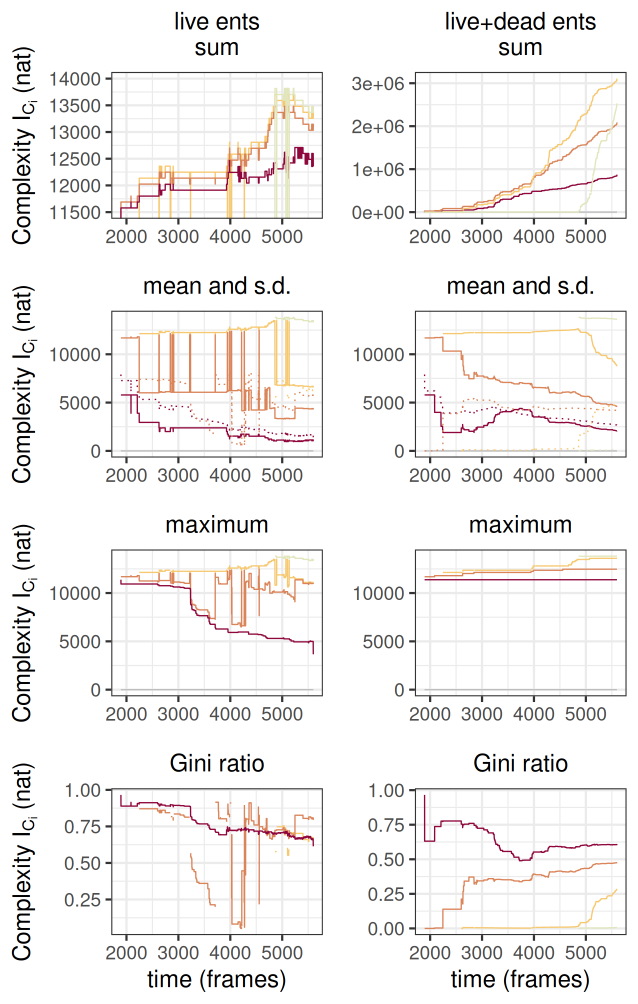}
\end{tabular}
\caption{Live and all entities complexity level-wise statistics for (left) World $(-.333,-.11,.46)$ and (right) World $(-2,.24,.45)$. Cf.\ Fig.\ \ref{fig:p09_30_LivDed_Ic}. \label{fig:n11_46_n.33_and_p24_45_n2_LivDed_Ic}}
\end{figure}

Both World (-.33,-.11,.46) and World (-2,.24,.45) have predominantly attractive and repulsive respectively and moderately asymmetric interactions $\chi_{(-.33,-.11,.46)} \approx \begin{bmatrix}
-.94 & .33 &-.60 & .62 \\
 .10 &-.63 &-.34 &-.37 \\
-.24 &-.54 & .25 & .46 \\
-.63 & .91 &-.70 & .52 \end{bmatrix}$ and $\chi_{(-2,.24,.45)} \approx \begin{bmatrix}
-.39 & .62 & .77 &-.76 \\
 .95 & .44 & .65 &-.50 \\
-.56 &-.65 & .67 & .54 \\
 .80 & .33 &-.02 &-.32 \end{bmatrix}$. The former system's interaction forces decrease as a power law with exponent -.333, whereas the latter with exponent -2.
Figs. \ref{fig:n11_46_n.33_and_p24_45_n2_screenshots},
\ref{fig:n11_46_n.33_and_p24_45_n2_T_K_L}, 
\ref{fig:n11_46_n.33_and_p24_45_n2_su_L_sum_mean}, \ref{fig:n11_46_n.33_and_p24_45_n2_hystlev},\ref{fig:n11_46_n.33_and_p24_45_n2_LivDed_C_E}, \ref{fig:n11_46_n.33_and_p24_45_n2_CluSiz_hist_statime}, \ref{fig:n11_46_n.33_and_p24_45_n2_L_cur}, \ref{fig:n11_46_n.33_and_p24_45_n2_LTcum_dens}, \ref{fig:n11_46_n.33_and_p24_45_n2_Dying_C_E}, \ref{fig:n11_46_n.33_and_p24_45_n2_LN_cur}, \ref{fig:n11_46_n.33_and_p24_45_n2_LivDed_Ic}, \ref{fig:n11_46_n.33_and_p24_45_n2_Lifeness}, \ref{fig:n11_46_n.33_and_p24_45_n2_LifenessN} are as in Appendix \ref{app:n1n1}.

\begin{figure}
\centering
\begin{tabular}{c}
\includegraphics[width=.75\textwidth]{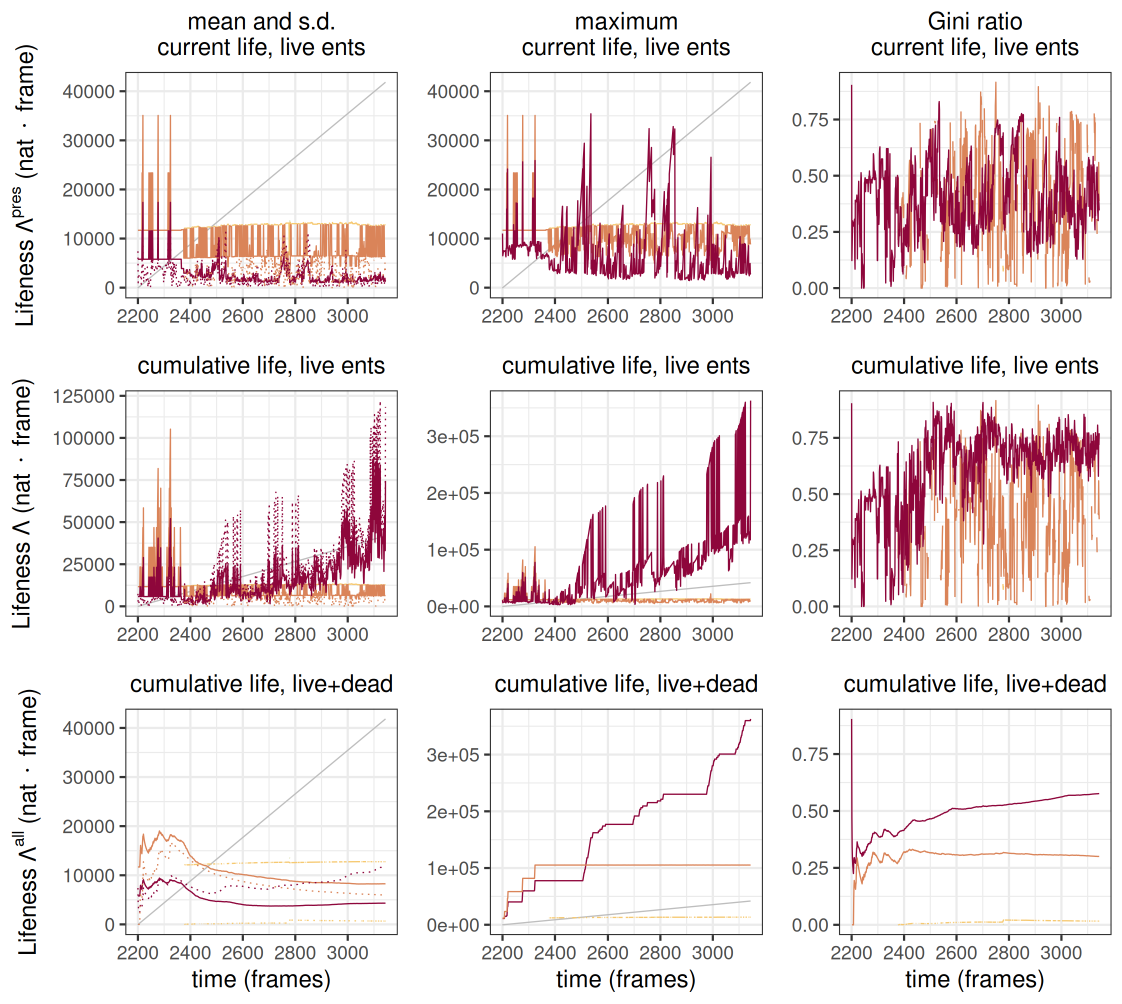}
\\
\includegraphics[width=.75\textwidth]{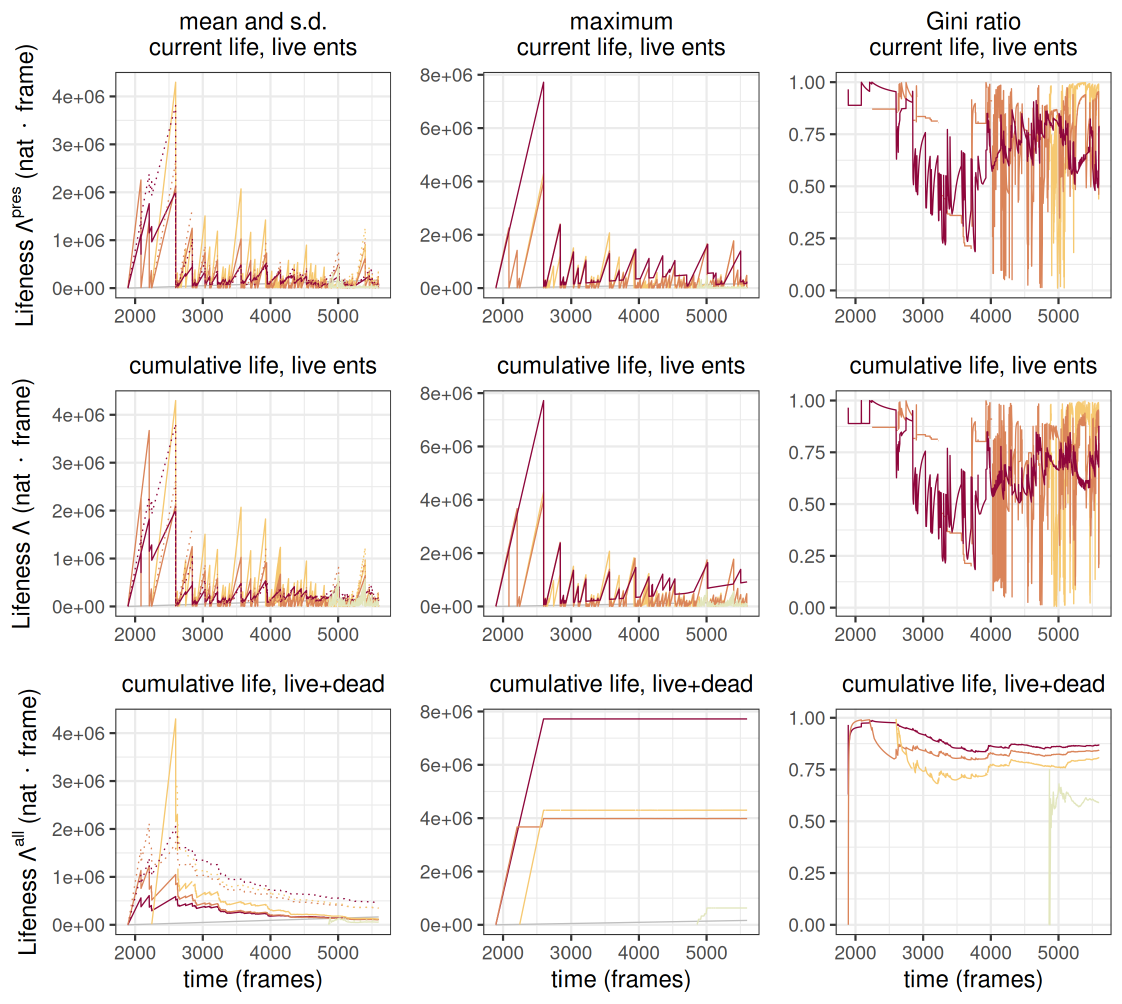}
\end{tabular}
\caption{Lifeness for (top) World $(-.333,-.11,.46)$ and (bottom) World $(-2,.24,.45)$. Cf.\ Fig.\ \ref{fig:p09_30_Lifeness}.  \label{fig:n11_46_n.33_and_p24_45_n2_Lifeness}}
\end{figure}

\begin{figure}
\begin{tabular}{c}
\includegraphics[width=1\textwidth]{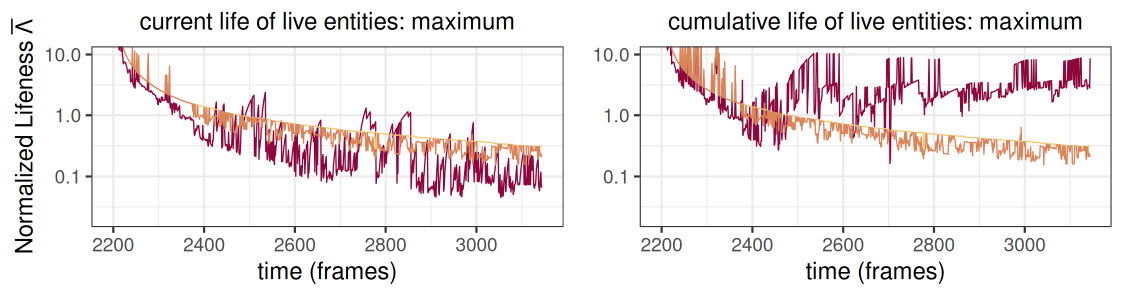}
\\
\includegraphics[width=1\textwidth]{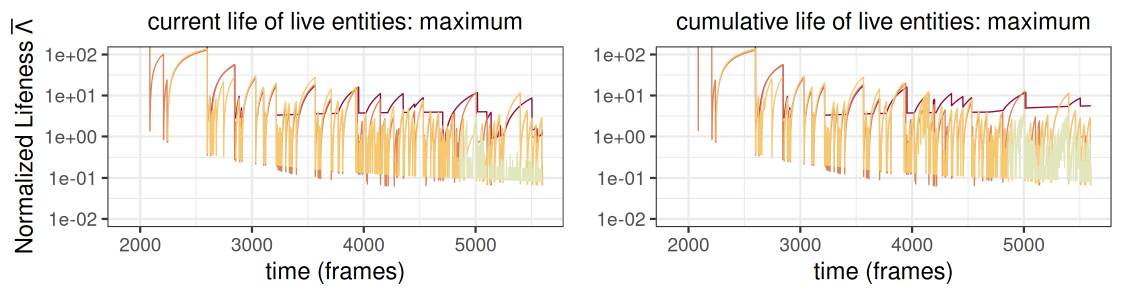}
\end{tabular}
\caption{Normalized maximum lifeness for the current life of live entities for (top) World $(-.333,-.11,.46)$ and (bottom) World $(-2,.24,.45)$. Cf.\ Fig.\ \ref{fig:p09_30_LifenessN}. \label{fig:n11_46_n.33_and_p24_45_n2_LifenessN}}
\end{figure}

\end{document}